%% file: article_vamos_m5_c.tex
\documentclass[twocolumn,showpacs,showkeys,preprintnumbers,amsmath,amssymb,superscriptaddress,unsortedaddress]{revtex4}

\usepackage{graphicx}
\usepackage{setspace}
\usepackage{dcolumn}
\usepackage{amssymb}
\usepackage{color}
\usepackage{mathrsfs}
\usepackage{longtable}
\setlongtables
\usepackage{amsmath}

\begin{document}

\title{Isotopic Yield Distributions of Transfer- and Fusion-Induced Fission from $^{238}$U+$^{12}$C Reactions in Inverse Kinematics}


\author{M.~Caama\~no}
\email[Email address: ]{manuel.fresco@usc.es}
\affiliation{GANIL, CEA/DSM-CNRS/IN2P3, BP 55027, F-14076 Caen Cedex 5, France}
\affiliation{Universidade de Santiago de Compostela, E-15706 Santiago de Compostela, Spain}
\author{O.~Delaune}
\email[Email address: ]{olivier.delaune@cea.fr}
\altaffiliation[Present address:]{CEA DAM DIF, F-91297 Arpajon, France}
\affiliation{GANIL, CEA/DSM-CNRS/IN2P3, BP 55027, F-14076 Caen Cedex 5, France}
\author{F.~Farget}
\email[Email address: ]{fanny.farget@ganil.fr}
\affiliation{GANIL, CEA/DSM-CNRS/IN2P3, BP 55027, F-14076 Caen Cedex 5, France}
\author{X.~Derkx}
\altaffiliation[Present address:]{Universit\"at Mainz, D-55128 Mainz, Germany}
\affiliation{GANIL, CEA/DSM-CNRS/IN2P3, BP 55027, F-14076 Caen Cedex 5, France}
\author{K.-H.~Schmidt}
\affiliation{GANIL, CEA/DSM-CNRS/IN2P3, BP 55027, F-14076 Caen Cedex 5, France}
\author{L.~Audouin}
\affiliation{IPN Orsay, IN2P3/CNRS-UPS, F-91406 Orsay Cedex, France}
\author{C.-O.~Bacri}
\affiliation{IPN Orsay, IN2P3/CNRS-UPS, F-91406 Orsay Cedex, France}
\author{G.~Barreau}
\affiliation{CENBG, IN2P3/CNRS-UB1, F-33175 Gradignan Cedex, France}
\author{J.~Benlliure}
\affiliation{Universidade de Santiago de Compostela, E-15706 Santiago de Compostela, Spain}
\author{E.~Casarejos}
\affiliation{Universidade de Vigo, E-36310, Spain}
\author{A.~Chbihi}
\affiliation{GANIL, CEA/DSM-CNRS/IN2P3, BP 55027, F-14076 Caen Cedex 5, France}
\author{B.~Fern\'andez-Dom\'inguez}
\affiliation{University of Liverpool, Liverpool L69 7ZE, United Kingdom}
\author{L.~Gaudefroy}
\affiliation{CEA/DAM \^Ile-de-France, BP 12 91680 Bruy\`eres-le-Ch\^atel, France}
\author{C.~Golabek}
\affiliation{GANIL, CEA/DSM-CNRS/IN2P3, BP 55027, F-14076 Caen Cedex 5, France}
\author{B.~Jurado}
\affiliation{CENBG, IN2P3/CNRS-UB1, F-33175 Gradignan Cedex, France}
\author{A.~Lemasson}
\affiliation{GANIL, CEA/DSM-CNRS/IN2P3, BP 55027, F-14076 Caen Cedex 5, France}
\author{A.~Navin}
\affiliation{GANIL, CEA/DSM-CNRS/IN2P3, BP 55027, F-14076 Caen Cedex 5, France}
\author{M.~Rejmund}
\affiliation{GANIL, CEA/DSM-CNRS/IN2P3, BP 55027, F-14076 Caen Cedex 5, France}
\author{T.~Roger}
\affiliation{GANIL, CEA/DSM-CNRS/IN2P3, BP 55027, F-14076 Caen Cedex 5, France}
\author{A.~Shrivastava}
\affiliation{GANIL, CEA/DSM-CNRS/IN2P3, BP 55027, F-14076 Caen Cedex 5, France}
\author{C.~Schmitt}
\affiliation{GANIL, CEA/DSM-CNRS/IN2P3, BP 55027, F-14076 Caen Cedex 5, France}

\begin{abstract}
A novel method to access the complete identification in atomic number $Z$ and mass $A$ of fragments produced in low-energy fission of actinides is presented. This method, based on the use of multi-nucleon transfer and fusion reactions in inverse kinematics, is applied in this work to reactions between a $^{238}$U beam and a $^{12}$C target to produce and induce fission of moderately excited actinides. The fission fragments are detected and fully identified with the VAMOS spectrometer of GANIL, allowing the measurement of fragment yields of several hundreds of isotopes in a range between $A\sim 80$ and $\sim 160$, and from $Z\sim 30$ to $\sim 64$. For the first time, complete isotopic yield distributions of fragments from well-defined fissioning systems are available. Together with the precise measurement of the fragment emission angles and velocities, this technique gives further insight into the nuclear-fission process.

\end{abstract}


\maketitle

\section{Introduction}
The fission of the atomic nucleus is an especially rich process in which intrinsic excitation transforms into a large amplitude deformation until the nucleus splits. The dissipative nature of the nuclear matter is revealed in the coupling of the intrinsic excitation and the collective degrees of freedom, and affects different aspects of the fission observables. In addition, during the collective motion, the single-particle shell structure of nucleons appears to strongly influence the shape of the fissioning nucleus if the excitation energy remains moderate. Because of this complexity and despite several decades of theoretical works, the accurate description of the fission process remains out of reach. 

The experimental measurements are focused on observables from both the incoming and outgoing channels. The characteristics of the fissioning system (excitation energy, angular momentum, etc.) and the probability of reaction determine the entrance channel, while the measurement of the fragments mass, charge, or energy distributions~\cite{ThH08,ItB07,BaS11,boc89,lan80,boe97,ste97}, particle~\cite{NiN98} and $\gamma$ evaporation~\cite{TeH94,Nif74}, etc. give information about the evolution of the process. The study of these observables gives access to features of the fission process that are included in the present models and aims at understanding the underlying physical mechanisms in order to predict the properties of fissioning systems not experimentally available.

Predicting the fission properties of exotic nuclei is of importance for the design of new-generation nuclear power plants and for the incineration of nuclear waste. An accurate knowledge of the fission probabilities in the relevant channels and the properties of the products of the fission reactions are essential to predict the incineration rate of actinides, the evolution of the neutron flux, and the decay heat with time. It is also of importance for various applications in fundamental research, such as the production of neutron-rich radioactive beams \cite{spiral2} and the modeling of nucleosynthesis in explosion of supernovae~\cite{BeM08}.

In the present work, a new experimental approach that allows the simultaneous measurement of the complete mass- and atomic-number distributions of fission fragments is presented. This technique is based on the use of inverse kinematics, which improves the resolution on the atomic-number measurement due to the kinematical boost, with a magnetic spectrometer, in order to provide the mass identification of the fragments. This method has been used widely to measure the fission fragments and the evaporation residues in spallation reactions~\cite{FRS1,FRS2}, from which results on the influence of viscosity on the dynamics of the fission at high excitation energy were derived~\cite{BeA02,PeB07}. However, in these reactions, the fissioning system is not well defined and its determination is based on models describing the spallation reaction. To overcome this drawback, the present technique employs multi-nucleon transfer reactions to produce and identify different fissioning systems at moderate excitation energy.

\subsection{The need of fragment yields}
Fission-fragment yield distributions are one of the key observables for the modeling of the fission process. At low excitation energy, shell structure and pairing correlations mark the main features of the fragment distributions. It was experimentally found that the fission-fragment mass distribution of most actinides in the range of masses from 230 to 256 are characterized by two- or even three-humped shapes, with an average mass of the heavy fragment of $A\sim 140$ \cite{fly72,itk00}. This feature has been understood as a superposition of three independent fission modes \cite{tur51,bro90} corresponding to different paths in the potential-energy surface of the deforming nucleus \cite{bro90,pas71,pas88}: one symmetric and two asymmetric fission channels, the asymmetric modes being commonly associated with spherical and deformed neutron shell gaps appearing at neutron number $N = 82$ and $N\sim 88$, respectively \cite{itk00,ben98}. The measurement of atomic-number distributions of fragments from low-energy fission of a broad range of neutron-deficient actinides and pre-actinides revealed that the heavy-fragment distributions are centered on an average atomic number of $Z\sim 54$, independently of the fissioning system \cite{sch01}. A detailed analysis of these data found the two asymmetric modes centered on $Z\sim 53$ and $Z\sim55$, for all identified actinides, including neutron-rich ones~\cite{boc08}. These results are somehow surprising as no shell-gap structure is expected close to these proton numbers~\cite{wilkins,str68,rag84}. In addition, according to the neutron-shell influence described above, a change of the average atomic number of the heavy fragment with the mass of the fissioning system would be expected under the assumption of \textit{unchanged charge density}. The controversial interpretations of the independent measurements of fragment mass- and atomic-number distributions sets a challenge to the theoretical description of fission, namely, on what drives the evolution of the potential energy of the deforming nucleus. 

Another important characteristic of low-energy fission in even-$Z$ nuclei is the enhanced production of fragments with an even number of protons. This has been interpreted as a signature that completely-paired proton configurations are preserved up to the scission point \cite{boc89,hen94}. The difference between even and odd element yields may be related to the intrinsic excitation energy gained by the fissioning nucleus on its way to the scission point~\cite{nif82,rej00} and hence to the viscosity of cold nuclear matter. Experimentally, this even-odd staggering has been observed to depend strongly on the fissility of the fissioning nucleus, its excitation energy~\cite{pom93,pers97}, and the fragment distribution asymmetry \cite{sid89,hen94,roc04,ste97}. Recently, a systematic study of the available data including odd-$Z$ fission in systems showed that the well-established evolution of the dissipated energy with the fissioning system accounts only for a part of the even-odd structure \cite{caa11}. In fact, the amplitude of the even-odd staggering in the atomic-number distributions increases systematically with the asymmetry of the fission-fragment split, with this increase depending on the fissioning system. These observations can be explained with a new interpretation of the even-odd staggering, based on new concept of energy sorting mechanism in the fission process~\cite{ScJ13}.

These are two examples of features of fission that can be traced back with the study of fragment mass- and atomic-number distributions. Systematic measurements of fission characteristics over a broad range of fissioning systems are very important to constrain the different models. Nonetheless, as different quantities were so far measured with different techniques and reaction mechanisms (fission-fragment mass in direct kinematics, or atomic number in the case of inverse-kinematics experiments), the simultaneous measurement of both mass- and atomic-number distributions of fission fragments is required to go further. Unfortunately, information on full isotopic-fragment distributions is scarce; it mainly consists of thermal neutron-induced fission of a limited number of actinides \cite{sie76,lan80,sch84,qua88,sid89,dje89,boc89,hen94,tse01}, and is restricted to the light fragments, due to the low kinetic energy of fission fragments that induces important straggling and ionic charge-states fluctuations in the ionization detectors, preventing for the atomic-number identification of the heavy fragments. The isotopic distribution of the heaviest fragments can be determined with radio-chemical techniques \cite{nis97,tri04} or $\beta$-delayed $\gamma$ spectroscopy~\cite{BaS11}, but with poor precision or in a reduced range of the total production. The limitation in the investigation of the isotopic distributions of heavy fission fragments is critical for the interpretation of the role that nuclear shell structure plays in the process. Interesting developments in the domain of fission yields have been performed using a Penning trap, where isotopic distributions of proton-induced fission fragments could be measured \cite{pent1}. However, the yields obtained in this work suffer from different extraction efficiencies from the ion guide for the different elements and from an evolving extraction efficiency. Further developments to calibrate the method are still under study \cite{pent2}.

In the following, a new experimental approach is described. In-flight measurements of the fragment production is free from the above-described experimental drawbacks. Combined with a spectrometer, inverse kinematics allows to access the simultaneous measurement of the complete mass and atomic-number distributions of fission-fragments. In addition, transfer reaction are chosen to produce and induce fission on fissioning systems well-defined in atomic number, mass number, and excitation energy.  Also, these reaction channels allow to widen the systematic study of neutron-deficient actinides at GSI~\cite{sch01}, as heavy and neutron-rich actinides are produced. 

\section{Experimental setup}

The present experiment made use of transfer reactions between a $^{238}$U beam at 6.1 AMeV and a 100 $\mu$g/cm$^2$ $^{12}$C target to induce fission in a set of different actinides, ranging from U to Cm. An advantage of this method lies on the possibility of studying fission of short-lived species that would be otherwise impossible to handle as individual beams or targets. The use of inverse kinematics facilitates the measurement and detection of the products both from the transfer reactions and the subsequent fission: the fission fragments are confined in a cone of about 25$^\circ$, while the recoils from the transfer reaction reach greater angles. In addition, the measurement of the properties of the recoil product allows the identification of the fissioning system, and to deduce its angle, kinetic energy, and excitation energy \cite{der10}.

The beam energy was chosen as a compromise between the fission production, the geometry of detection, and the opening of other reaction channels. At this energy, around 10\% above the Coulomb barrier, a total fission cross-section of 300 mb was previously measured for $^{12}$C+$^{232}$Th, with a total transfer-induced fission probability one order of magnitude lower \cite{bis97}. A similar cross-section is expected for the present reaction. Different fissioning systems are produced from the interaction of the U beam and the carbon target: inelastic collisions provide a range of excitation energy to the $^{238}$U beam, while transfer reactions produce a collection actinides, always in a range of excitation energy below 30 MeV. In addition, fusion reactions produce a compound system of $^{250}$Cf with $\sim$45 MeV of excitation energy.

\begin{figure}[t]
\begin{center}
\includegraphics[width=8cm]{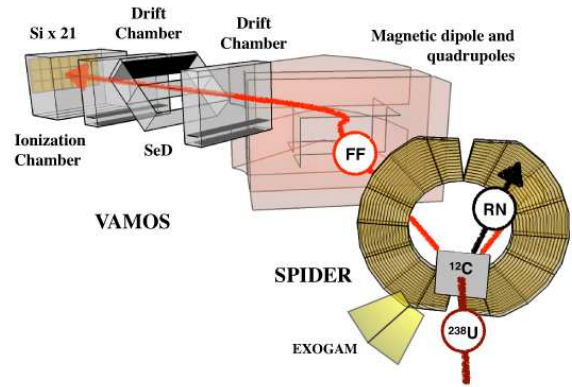}
 \caption{\label{setup} Schematic layout of the experimental set-up. The interaction between a $^{238}$U beam and a $^{12}$C target produces a recoil (RN) and a fissioning system. The former is detected and identified in SPIDER. One of the fragments (FF) emitted by the fissioning system is detected and identified in VAMOS. Few EXOGAM detectors are placed around the target to detect $\gamma$ emission from the products. See text for details.}
\end{center}
\end{figure}

The experimental setup is described in Figure~\ref{setup}. It consisted in two stages of detection: an annular segmented silicon detector (SPIDER) was devoted to reconstruct the transfer reactions from the kinematics of the detected target-like recoils, while the variable mode spectrometer VAMOS~\cite{savajols,pul08} was used to reconstruct the fission reaction from the kinematic properties of the fragments. SPIDER (Silicon Particle Identification DEtector Ring) comprised two annular double-sided silicon detectors, manufactured by Micron \cite{mic12}, separated by 4 mm. Both detectors are segmented into 16 rings of 1.5 mm width on one side and 16 radial sectors on the other one, allowing the measurement of the recoil angle. The ensemble had a diameter of 96 mm and an inner hole of 48 mm. The recoils were through the first detector, 65 $\mu$m thick, which measured their energy-loss $\Delta E$, and they were stopped in the second 1 mm thick detector, where their residual energy $E_{res}$ was measured. SPIDER was centered on the beam axis, and placed 32 mm after the target. The angular coverage was between 35$^\circ$ and 55$^\circ$, at the limit of the grazing angle ($\theta^{lab}_{grazing}(^{12}$C$)$ = 35.7$^\circ$) where the maximum transfer cross-section is expected \cite{bis97}.

The fragments produced in the fission of the beam-like products are emitted in a forward cone that passes through the inner hole of SPIDER. The spectrometer VAMOS was rotated at an angle of 20$^\circ$ with respect to the beam axis, in order to cover most of the fission production. Fission fragments emitted within the acceptance of the spectrometer arrived to the detection set-up at the focal plane after being deflected by the large dipole of the spectrometer. The detection set comprises a Secondary Electron Detector (SeD) for the time of flight measurement \cite{dro02}, two drift chambers for position and angle determination, and an ionization chamber and a wall of twenty-one silicon detectors to perform energy-loss and residual energy measurements. Further details about the detection ensemble at VAMOS can be found in \cite{pul08}. The combination of these measurements results in the identification of the charge state, atomic and mass numbers, and the reconstruction of the emission angles ($\theta_L$ and $\phi_L$) and energy of the fragment. In addition, part of the EXOGAM array~\cite{SiA00} was set around the target station to detect in-flight $\gamma$-emission from fission fragments. The identification of both the fissioning system and the fragments is described in the following section.

\section{Data Analysis}
The analysis of the data collected in the present experiment can be split, at first, into two sections: the first one includes the characterization of the binary reaction, and thus the fissioning system, by SPIDER, and the identification of the fragments and their fission kinematics in the VAMOS spectrometer. In a second stage, the correlation between the fissioning system and the resulting fragments allows for a complete characterization of the binary and subsequent fission reactions.

\subsection{Fissioning System Identification}
\label{section_SPIDER}
The different transfer channels are selected by means of the target-recoil identification, performed by $\Delta$E-E$_{res}$ measurements in the SPIDER telescope. Unfortunately the resolution achieved with SPIDER has not been as good as expected. The detector response suffered from the high counting rates (up to 40~000 per second) of high-energy scattered $^{12}$C target nuclei (up to 200~MeV). The consequence was an important increase of the current in the detectors (up to several $\mu$A) which implied a reduction of the applied high-voltage. Therefore, the detectors were not permanently fully depleted and the response dropped as a function of time. Besides, the high counting rate increased the temperature which also deteriorates the resolution in energy. Moreover, the side of the $\Delta$E detector close to the target received a high electromagnetic flux, $\delta$ electrons in particular, from the interaction of the highly charged $^{238}$U beam with the target. In addition, the beam showed spatial instabilities of several millimeters that worsened the angular resolution. This effect had also an impact on the particle energy calibration, performed with the elastic scattering of the target. A detailed description of these problems and the software solutions applied are described in Ref.~\cite{der10}. Most of these issues were addressed in a second iteration of the experiment, where the isotopic identification of the fissioning system was achieved and the reconstruction of the transfer channels was improved \cite{carme}.

\begin{figure}[t]
\includegraphics[width=8cm]{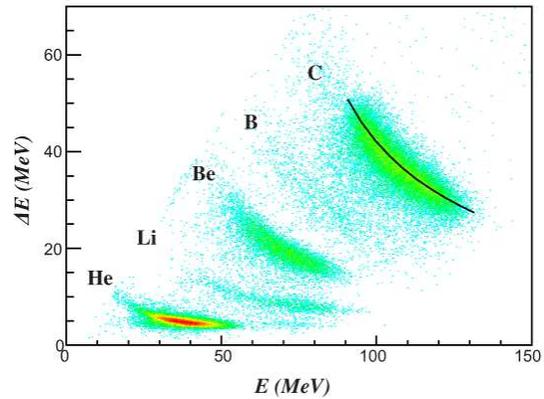}
 \caption{\label{fig:matrix}$\Delta$E-E identification matrix from the SPIDER telescope gated by the detection of a fission fragment in the VAMOS spectrometer. Five recoil (compound) nuclei were identified: (C) carbon (uranium), (B) boron (neptunium), (Be) beryllium (plutonium), (Li) lithium (americium), and (He) helium (curium). The line represents the calculated $\Delta$E-E relation for $^{12}$C.}
 \end{figure}

Figure~\ref{fig:matrix} shows the resulting identification matrix. Five different target recoils are identified from helium to carbon. Due to the above-mentioned experimental difficulties, isotopic separation is not achieved. In addition, some of the channels exhibit further difficulties: Li transfer channels are identified through the identification of lithium recoils, but they can also be wrongly assign to the pile-up of two $^4$He particles in the same ring and same sector of the detector. The low statistics recorded do not allow a correct evaluation of this contamination. Consequently, this channel is not considered in the following analysis. Similarly, the measured helium particles can be produced in different reaction channels: they can be the result of a direct $^8$Be transfer, of the target-recoil $^8$Be decay after a $^4$He transfer, or they can be evaporated after a fusion reaction. Due to this complexity in the interpretation of the data, they are neither considered in the following analysis. Also, it is worth noting that one-proton transfer from the U beam to the C target, leading to the production of nitrogen recoils, is not observed within the limits of the collected statistics.
 
The atomic number $Z_{fs}$ of the fissioning system was defined to be $Z_{fs}=Z_{beam}+Z_{target}-Z_{recoil}$. Due to the limited energy dissipated in the collision (see Sec. \ref{transfer_induced} and Fig. \ref{figex}), the evaporation of charge particles was not considered. Also, the excitation of the target-like recoil that could lead to the evaporation of a neutron was considered to be negligible based on the assumption of a thermal equilibrium between the two tranfer partners. In any case, further studies on the same reaction showed that the excitation of the target-like recoil remains limited to the first escited states, well bellow the emission threshold \cite{carme}. To determine the most probable isotope for each proton channel, a systematic behaviour based on the ground-state $Q$-values of the transfer was used: for different isotopes of the same element, the production cross-section exponentially decreases with the corresponding $Q$-value~\cite{kar82}. It can be written as:
\begin{equation}
\label{eq:qsyst}
\frac{\sigma_{(Z,i)}}{\sigma_{(Z,j)}} = \frac{\exp{Q_{(Z,i)}}}{\exp{Q_{(Z,j)}}}
\end{equation}
where $\sigma_{(Z,x)}$ is the cross-section of a particular transfer channel for the corresponding recoil element and $Q_{(Z,x)}$, its ground-state reaction heat. Following this prescription, the uranium production corresponds essentially to $^{238}$U from inelastic scattering and 20 \% of $^{237}$U. The neptunium channel is dominated by $^{239}$Np, where the contribution of $^{240}$Np is less than 1\%. The plutonium channel has a stronger but still limited contamination. The main produced isotope is $^{240}$Pu with a minor contribution of $^{241}$Pu around 20\%. 

\subsection{\label{fragid}Fragment Identification}
\label{section_ID}
The fission-fragment identification is done with the measurement of the mass ($A$), the mass over atomic charge ratio ($A/q$), and the atomic number ($Z$) of each particle detected in VAMOS. These quantities are, in principle, independent on each other; however, their measurement employs common observables, such as energy or time, producing an entanglement that benefits from feedback between them in order to achieve the best results.

In the present analysis, $A/q$ is measured from the magnetic rigidity, $B\rho$ in Tm, the velocity, $\beta$ in units of $c$, and the Lorentz factor $\gamma$ of the particle:
\begin{equation}
A/q=\frac{{\rm B}\rho}{3.107\cdot \beta\gamma}
\end{equation}
In VAMOS, a detailed simulation of the particle trajectories inside the magnetic fields of the quadrupoles and dipole allows to deduce the $B\rho$, scattered angle, and path of the particles from the measurement of the position and angle in the drift chambers placed after the dipole \cite{pul08}. The velocity is determined by the Time of Flight ($ToF$) measured between the radio frequency of the cyclotron and the SeD detector, and the reconstructed flight path ($D$) of the particle:
\begin{equation}
\beta=\frac{D}{ToF\cdot c}
\end{equation}
The resolution of the time-of-flight measurement is mainly limited by the dispersion of velocity of the beam and the quality of the high radio-frequency signal of the cyclotron. This was of the order of $\sim 900$ ps through all the experiment, compared to the $\sim 130$ ps of the SeD detector. With a $ToF$ between 152 and 280 ns, the final resolution is of the order of 0.4 \%. The velocity resolution is also affected by the reconstruction procedure to determine the length of the path inside the spectrometer, which is of the order of 0.6 \%~\cite{pul08}.

The mass, $A$, of the fragments is measured using the total energy, $E_{tot}$, and the velocity:
\begin{equation}
A=\frac{E_{tot}}{u\cdot(\gamma-1)}
\label{aeq}
\end{equation}

where $u$ is the unified atomic mass unit. The fact that $A$ and $q$ are integer numbers makes that only a specific set of ($A$, $A/q$) values is possible; this property was used to improve the measurement of the $ToF$, and to correct aberrations of the $B\rho$ reconstruction that appear for extreme angles. Other corrections include the estimation of $ToF$ variation due to the slowing down of the particles when traversing the first drift chamber before reaching the SeD detector. Once the $ToF$, and thus the velocity, was improved, the energy measurement was also corrected through the relation between the velocity and $A$, and with the energy loss in windows and other dead layers of matter along the particle trajectory.

\begin{figure}[ht]
\begin{center}
\includegraphics[width=8cm]{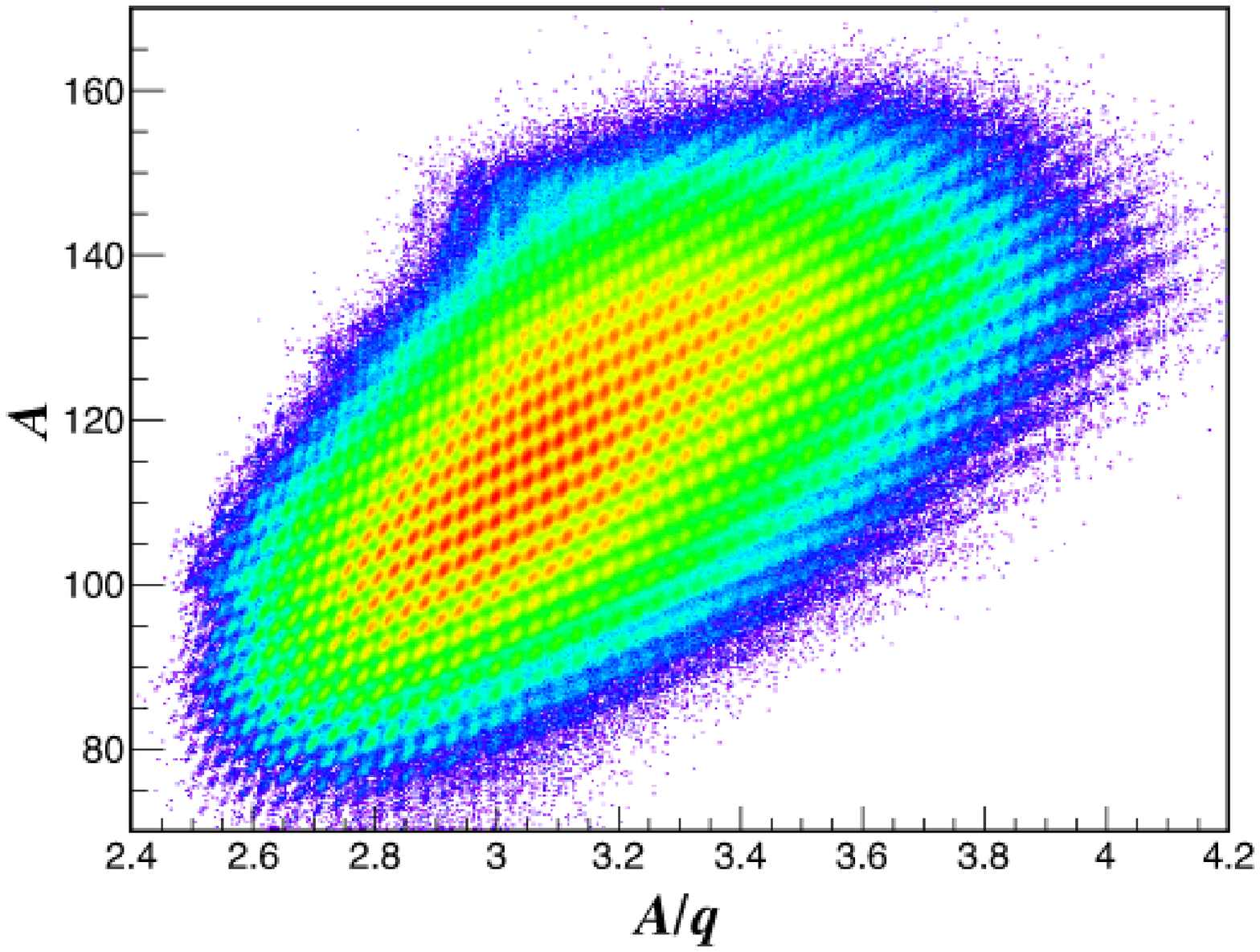}\\
\includegraphics[width=8cm]{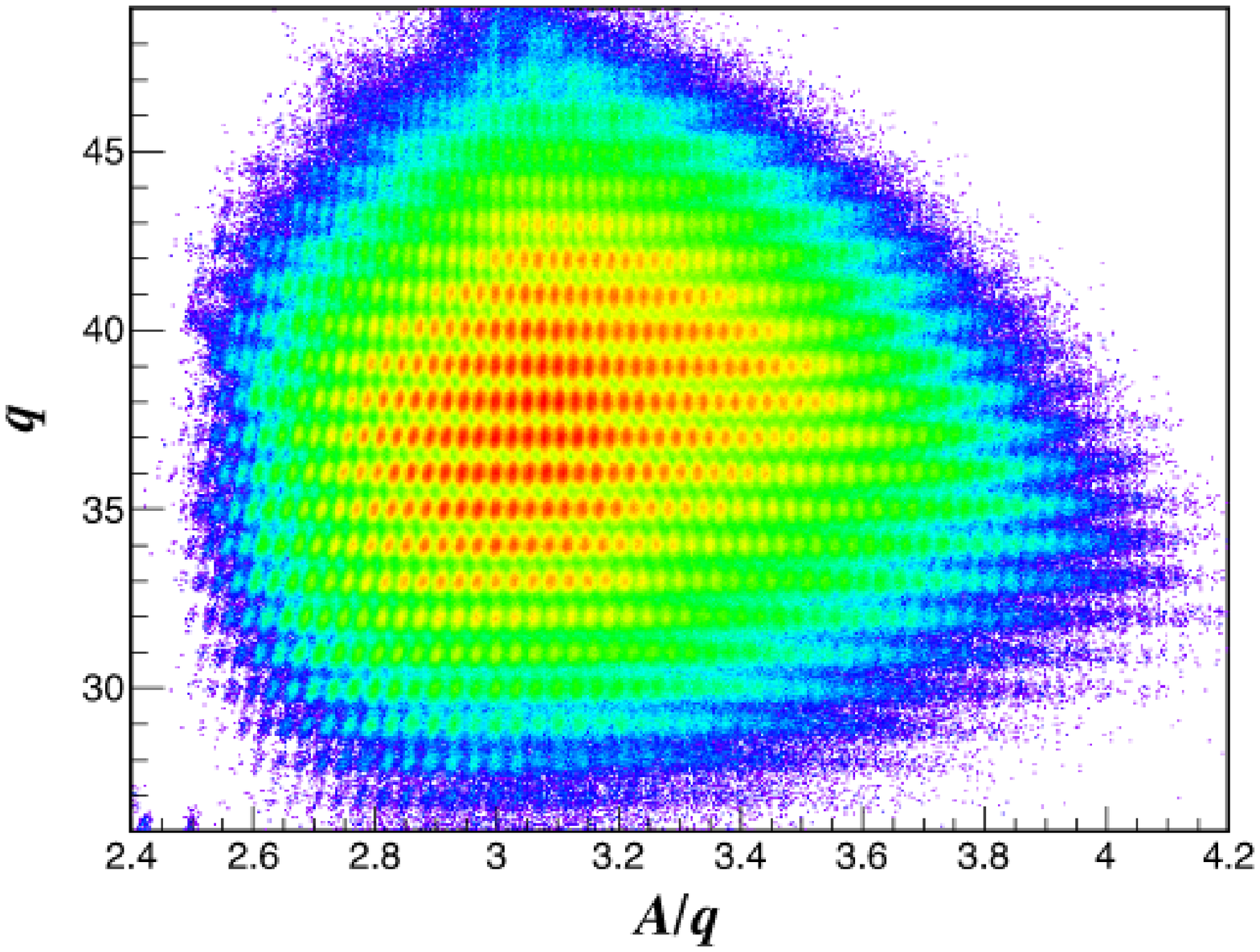}
\caption[Fragment Identification on $A$, $A/q$, and $q$]{\label{fig1} Identified fission fragments. Upper panel: $A$ as a function of $A/q$. Lower panel: $q_{meas}=A/(A/q)$ as a function of $A/q$.}
\end{center}
\end{figure}

Upper panel of Fig. \ref{fig1} shows the resulting $A$ as a function of $A/q$. The calculated $q$ for each system with respect to $A/q$ are shown on the lower panel of the same figure. In the present experimental setup, the final $A$ identification benefits from the better resolution in $A/q$: fragments with the same charge state form diagonal lines with a slope equal to $q$, the final $A$ identification is obtained multiplying the $A/q$ of systems contained in the same diagonal by the corresponding $q$.

Charge states between $q=29$ and 45 are identified with a resolution of $\approx 0.7$ full width at half maximum (FWHM). As explained before, the determination of $q$ allows the mass identification through $A=q\times (A/q)$ between $A=80$ and $A=150$, with a resolution below FWHM$_A\lesssim 0.8$.

\begin{figure}[ht]
\begin{center}
\includegraphics[width=8cm]{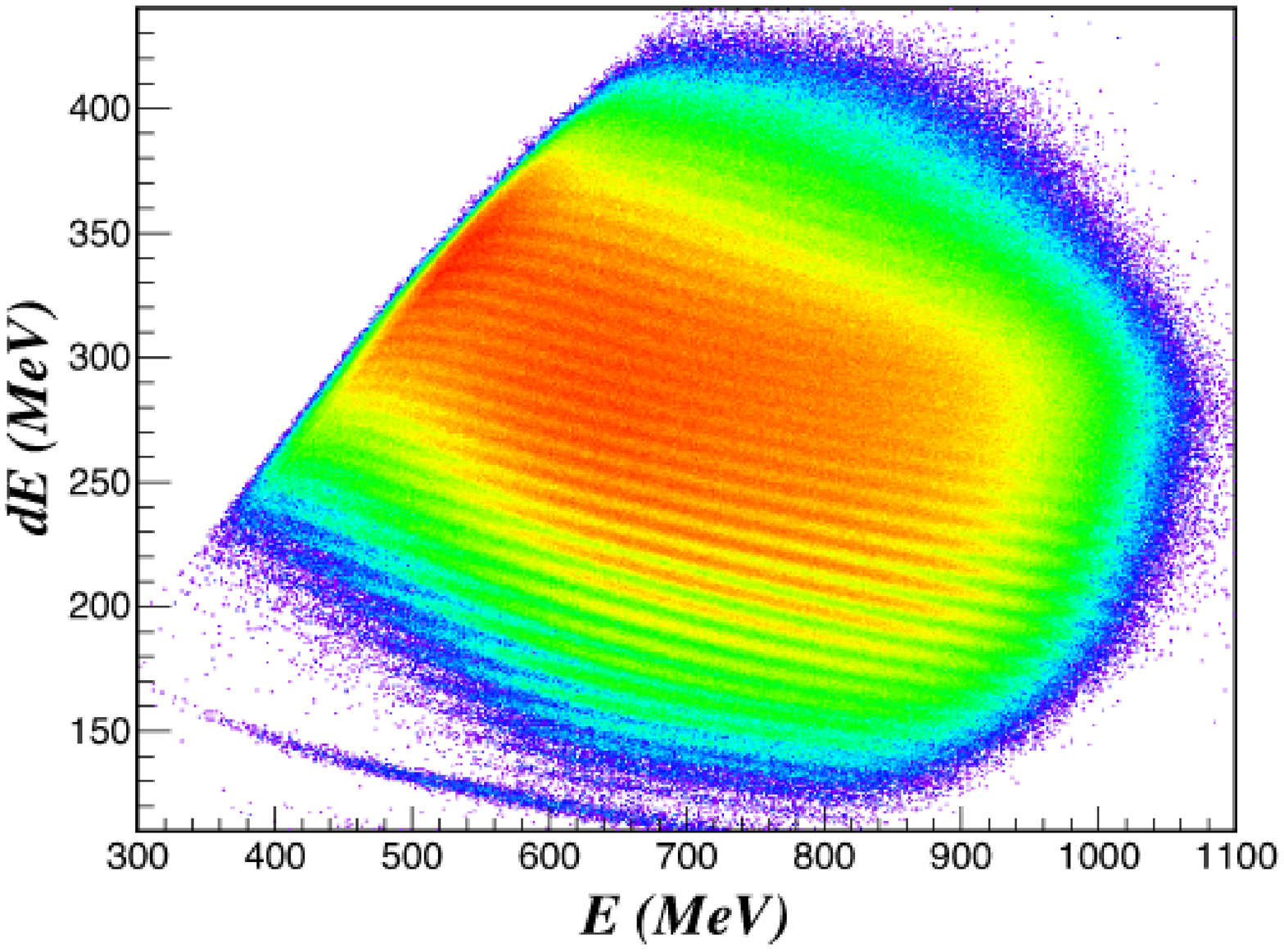}
\includegraphics[width=8cm]{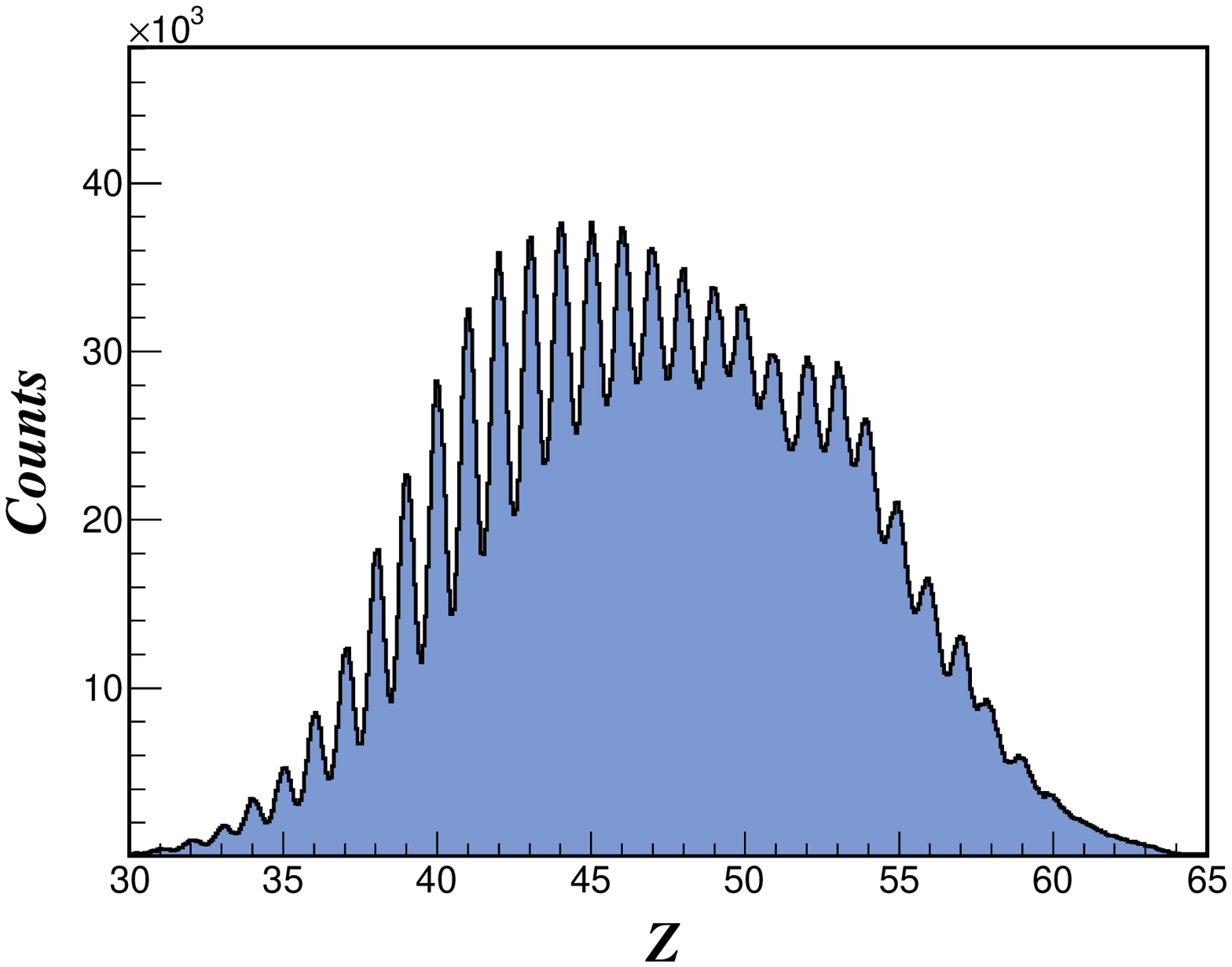}
\caption{\label{fig9} Upper panel: Identification of $Z$ via the correlation between the energy loss $\Delta E$ in the ionization chamber and the total energy $E_{tot}$, measured in the ionization chamber and the wall of silicon detectors. Each line is produced by isotopes of the same element. Lower panel: projection along the ridges of maximum statistics. See text for details. }
\end{center}
\end{figure}

The atomic-number $Z$ identification of the fission fragments is done by means of the correlation between the energy loss $\Delta E$ measured with the ionization chamber and the total energy $E_{tot}$ measured with the silicon detectors and the ionization chamber. The pressure of the ionization chamber gas (C$_4$H$_{10}$) was varied during the experiment, and it was found that the best resolution corresponds to an energy-loss of about 30\% of the fragment total kinetic energy. Figure \ref{fig9} shows the $\Delta E-E_{tot}$ correlation for the ensemble of detected particles with each line corresponding to a single $Z$. The ridges of maximum-statistics for each $\Delta E-E_{tot}$ line were defined with an ensemble of 40 points. The space between two consecutive points along the same ridge was interpolated by lines. Each event was then computed to determine its distance to the closest ridge. The given distance gave a spectrum in $Z$ which is shown in the bottom panel of Fig. \ref{fig9}. The atomic-number resolution is limited by the broad atomic-charge changes that happen at this energy during the course of the ions within the gas. The resolution obtained in the experiment is around FWHM$_Z/Z\approx 1.5~\% $ along the $Z$ distribution.

\begin{figure}[ht]
\centering
 \includegraphics[width=5cm]{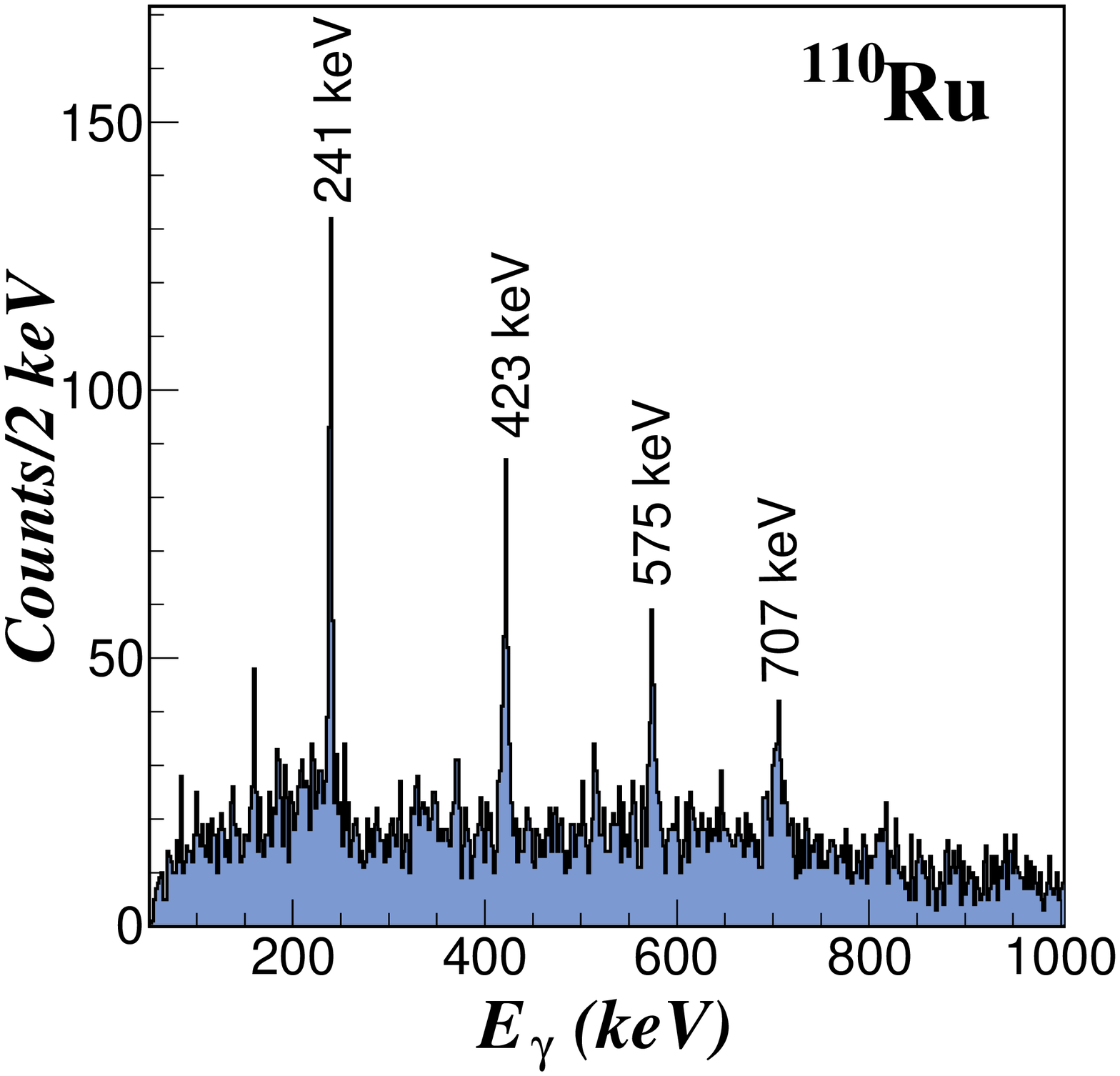}
~~
 \includegraphics[width=2cm]{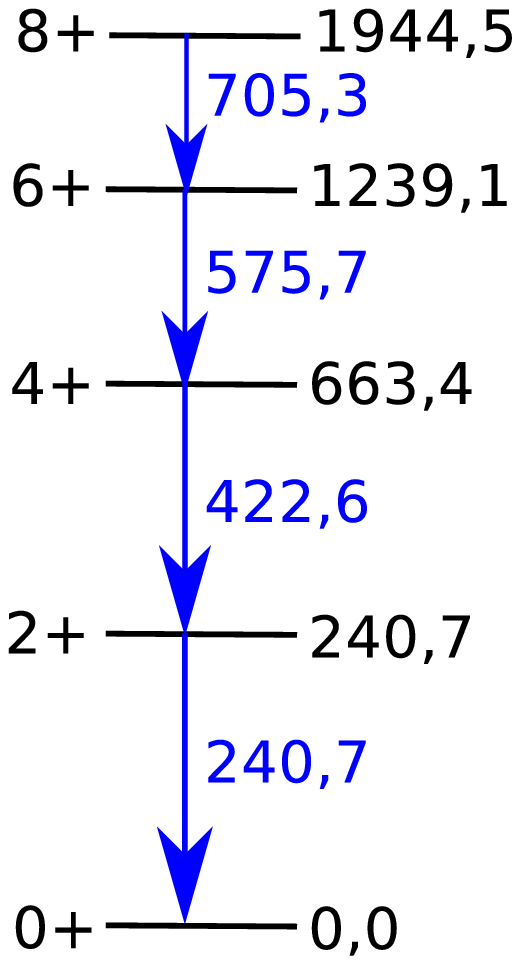}
~~~~~~~~~
\\
 \includegraphics[width=5cm]{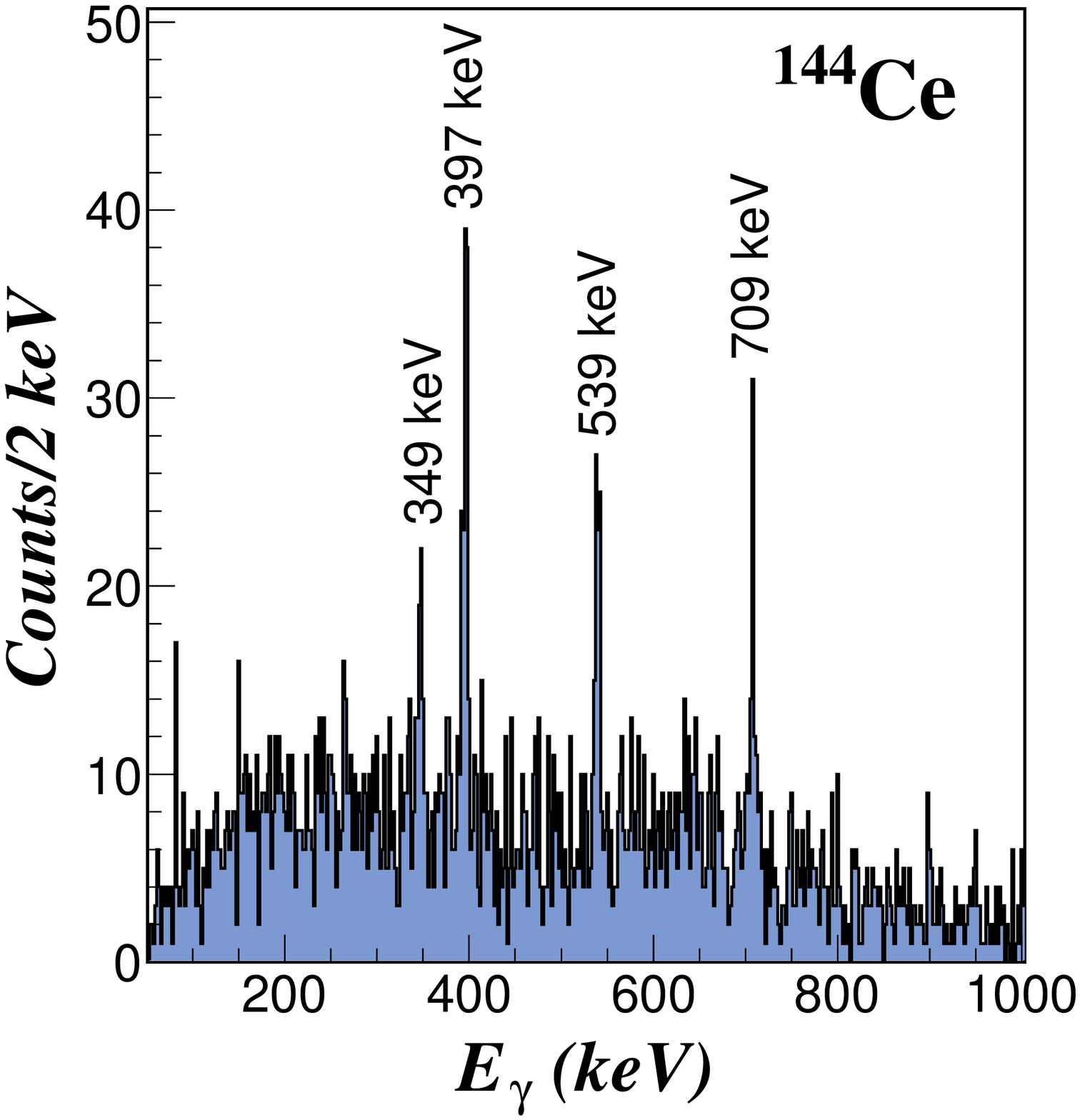}
 \includegraphics[width=3.5cm]{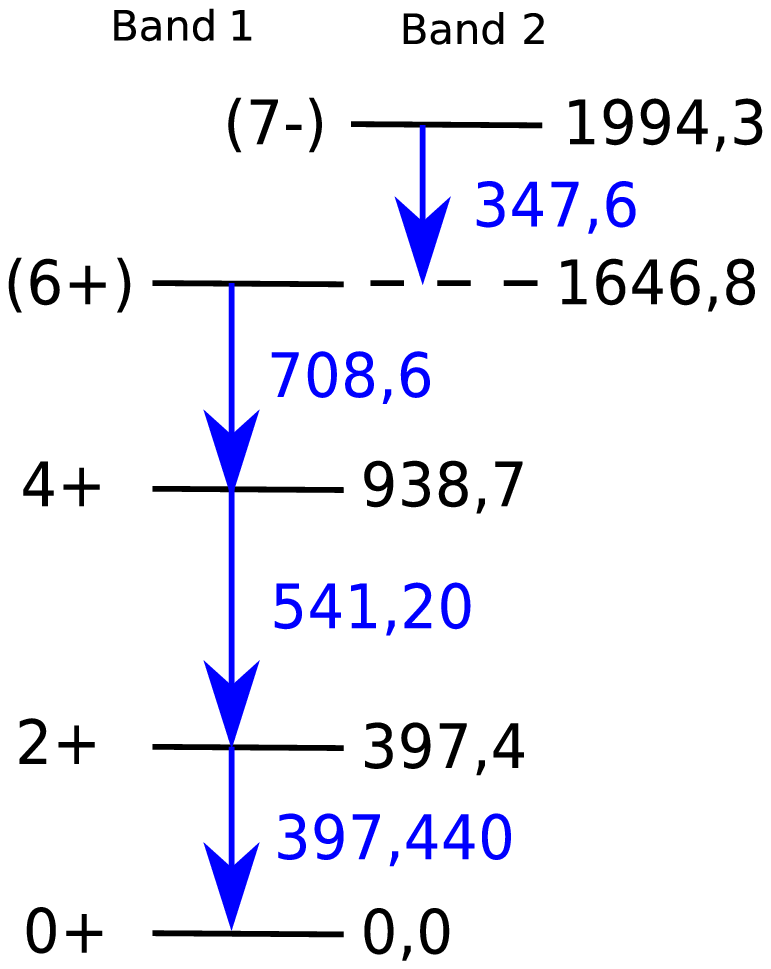}
 \\
\caption{\label{figg} Single $\gamma$-ray spectra obtained in coincidence with the specified fragments.}
\end{figure}

\subsubsection{Confirmation of the fragment identification}

Two clover-type detectors of the EXOGAM array~\cite{SiA00} were set around the target station to detect in-flight $\gamma$-emission from fission-fragments. This allowed not only for a cross-check of the fission-fragment identification, but also for the extraction of new results on the level scheme of $^{134}$Xe~\cite{ara09}. Figure \ref{figg} shows the single spectra of two well-produced fission fragments, together with their level scheme, to show the quality of the $\gamma$-ray spectra obtained with the spectrometer selection. 
The absence of $\gamma$-ray transitions from neighboring nuclei confirms the quality of the selection and the stability of the calibration throughout the experiment. 

\begin{figure*}[ht]
\begin{center}
\includegraphics[width=8cm]{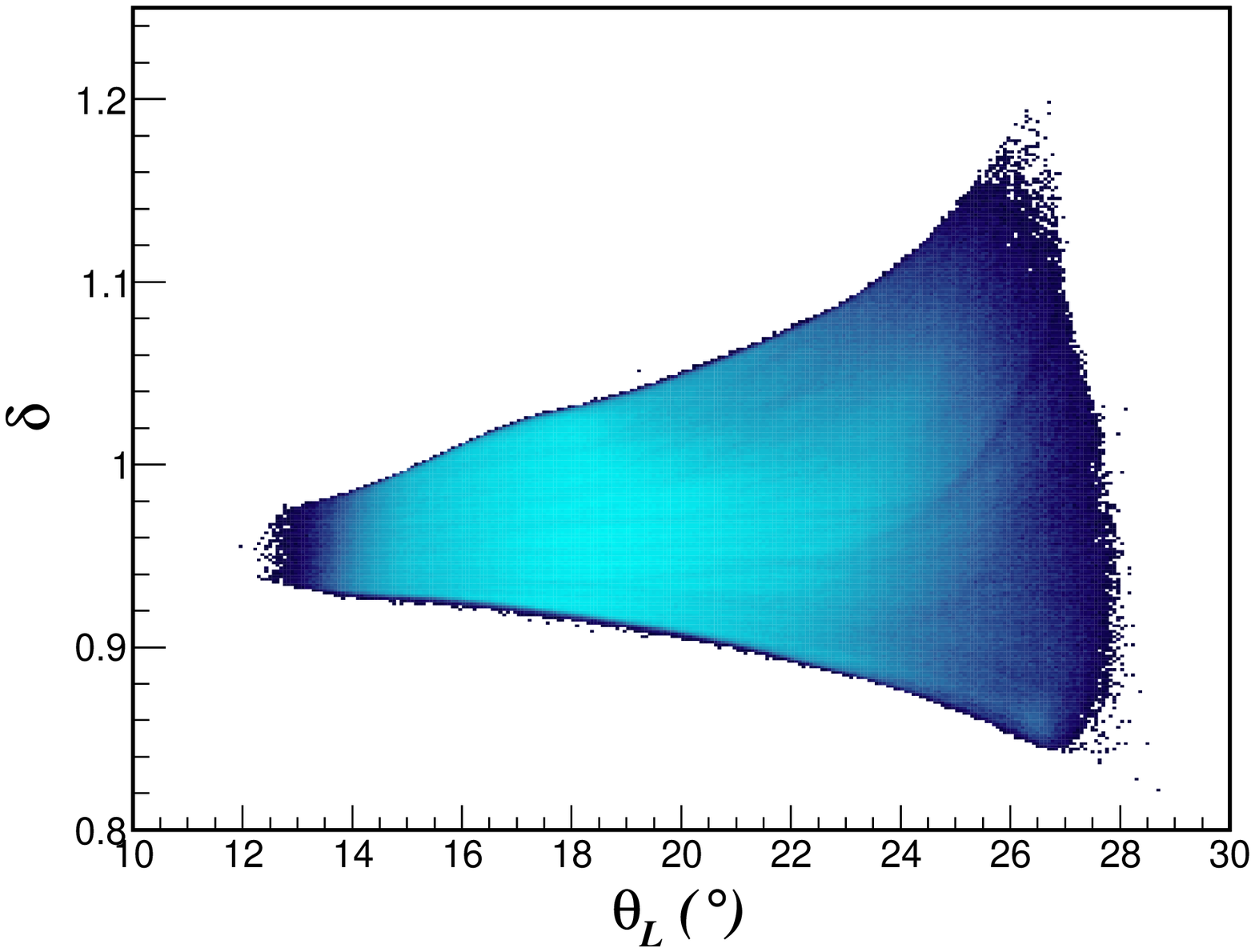}
\includegraphics[width=8cm]{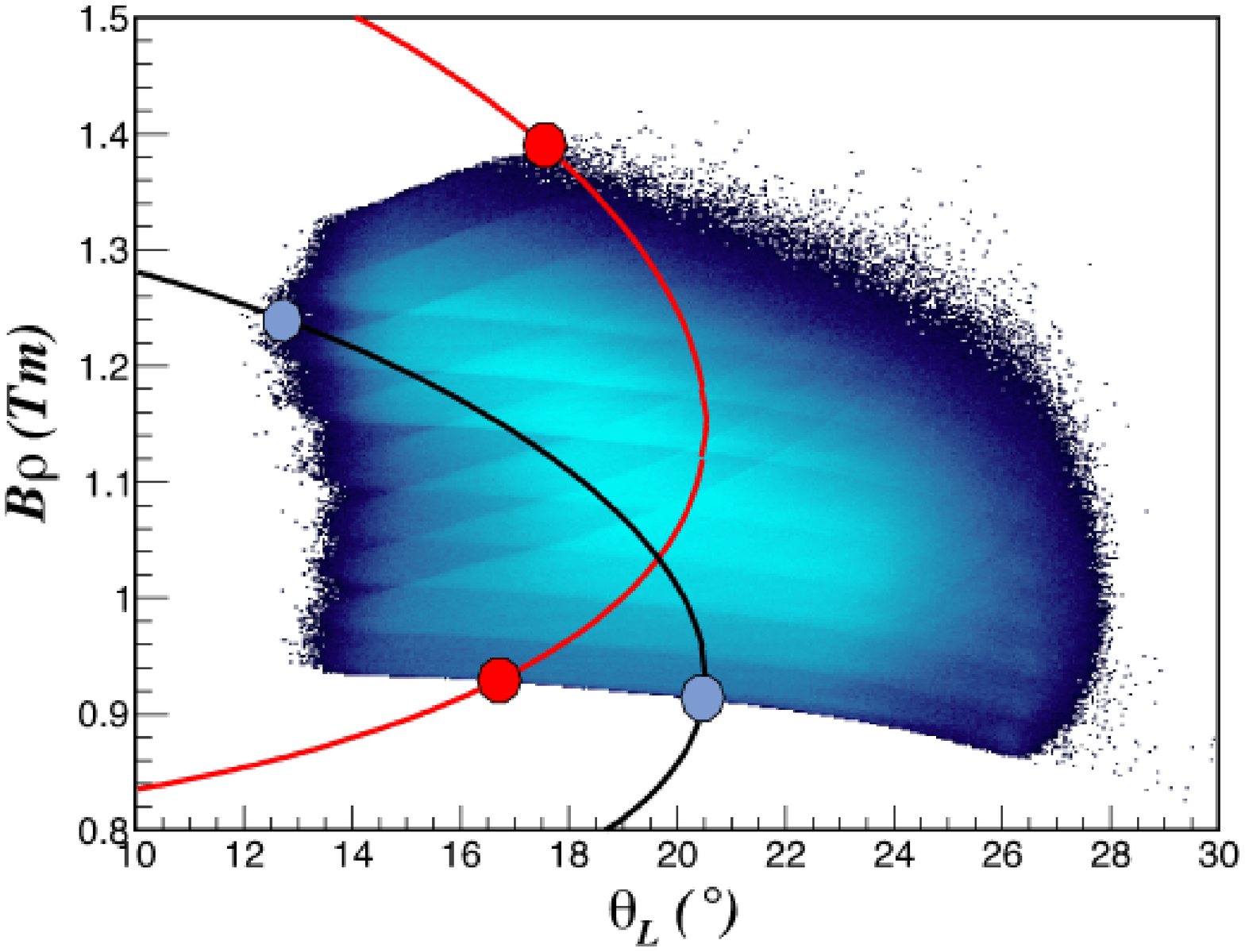}
\caption{\label{fig4} Left panel: shape of the VAMOS $a_{\theta,\delta}$ acceptance. Right panel: The magnetic settings used in the experiment are displayed in $B\rho$ and $\theta_L$ coordinates. The overlapping areas can be distinguished inside the physical distribution. The black and red lines correspond to the kinematics of two charge states of the same isotope. The dots mark the limits of the spectrometer acceptance for each charge state.}
\end{center}
\end{figure*}

\subsection{Fragment yield reconstruction}
\label{section_method}
The magnetic-rigidity distribution of the fission fragments is a convolution of the mass distribution, the ionic charge-state distribution and the velocity distribution, being all of them strongly connected by the kinematics of the fission process and atomic physics. In the laboratory frame, the fission fragments have smaller or larger velocities than that of the beam depending on their direction of emission in the reference frame of the fissioning system, backward or forward, respectively. In order to cover the most of the fragment magnetic-rigidity distribution, the spectrometer was tuned with a series of settings with different reference magnetic rigidities ($B\rho_{ref}=$1.015, 1.055, 1.099, 1.143, 1.190, 1.251, 1.302, and 1.355 Tm) during the experiment. The values of the reference settings were chosen to be closer to each other than the actual spectrometer magnetic-rigidity acceptance in order to have an overlap between two adjacent measurements. In the right panel of Fig. ~\ref{fig4}, the overlap between the different spectrometer settings is shown in the bi-dimensional space defined by the magnetic rigidity and the polar angle $\theta_L$ of the fragments. The number of counts found in the overlap area between adjacent settings was used to normalize the relative beam intensity. As was shown in~\cite{pul08}, the spectrometer acceptance in azimuthal angle $\phi_L$ depends on the polar angle $\theta_L$ and the reduced magnetic rigidity $\delta =\frac{B\rho}{B\rho_{ref}}$. It can be seen in Fig. ~\ref{fig4} that the same value of the magnetic rigidity corresponds to different values of the reduced magnetic rigidity if measured with different reference settings of the spectrometer, and consequently correspond to different acceptance in $\phi_L$. 

\begin{figure}[ht]
\begin{center}
\includegraphics[width=8cm]{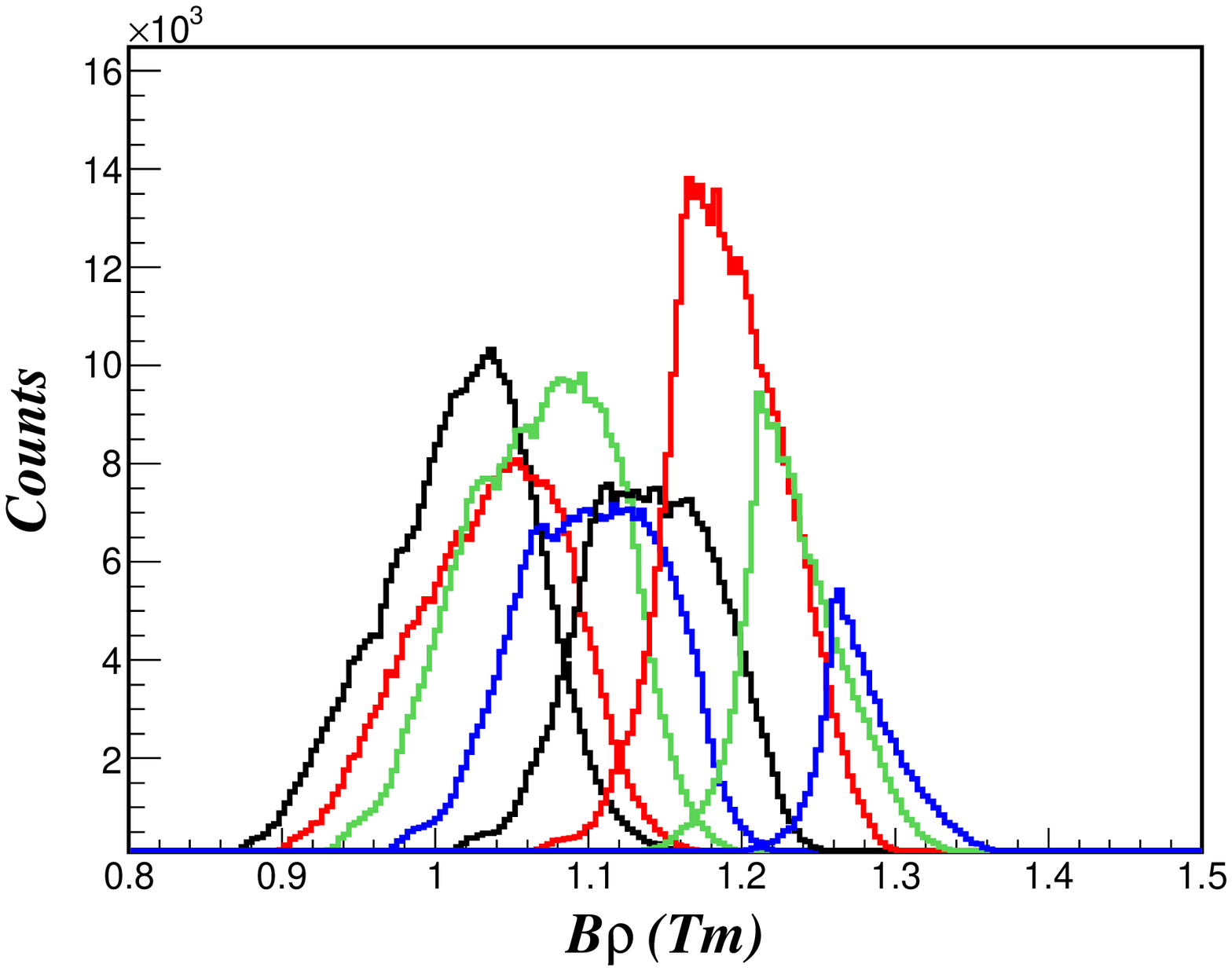}
\includegraphics[width=8cm]{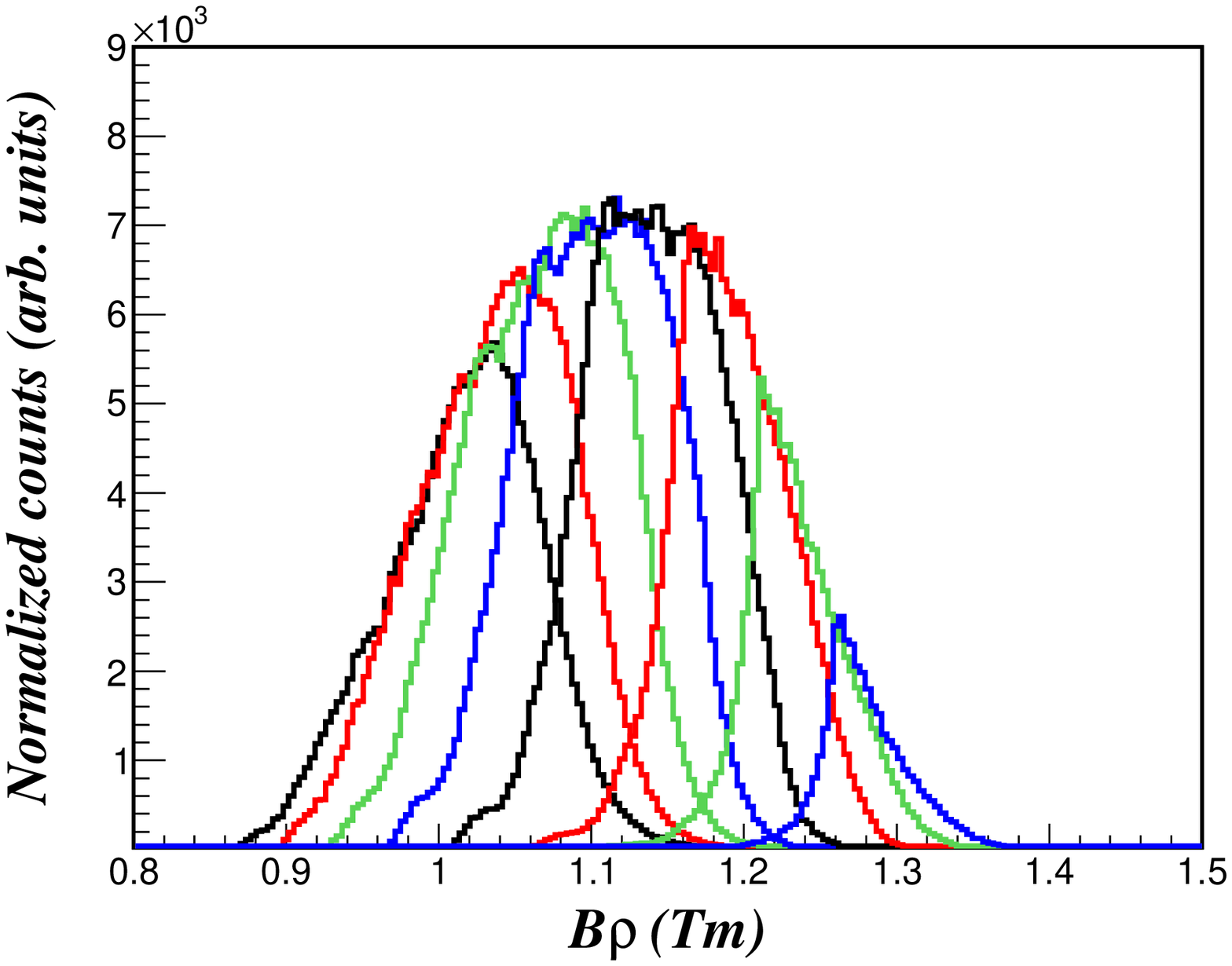}
\caption{\label{fig5} $B\rho$ distributions corresponding to the magnetic settings used in the experiment before (upper panel) and after (lower panel) normalization. Colors (on line) are a guide to separate the individual settings.}
\end{center}
\end{figure}

In order to waive the influence of the different azimuthal-angle acceptance, the normalization between adjacent spectrometer settings was performed for events accepted in a restricted range in $\phi_L$ around 0. Figure \ref{fig5} shows the magnetic-rigidity distributions measured for the different spectrometer settings used in the experiment, before and after applying the calculated normalization factors.

In order to deduce the actual production of each fragment, the number of particles not detected is evaluated by reconstructing the full phase space distributions. In these particular experimental conditions, the momentum distribution is shared between the large $q$ distribution, resulting in a broad magnetic-rigidity distribution. 
The angular acceptance is mainly defined by the size and shape of the magnetic elements (dipole and quadrupoles) and those of the detectors. In the case of VAMOS, the magnetic ensemble forms an entrance of $\theta_L\approx\pm 6^\circ$ in polar angle with respect to the beam direction. The acceptance in azimuthal angle, $\phi_L$, depends on the position of the spectrometer respect to the beam direction. In the present case, with VAMOS rotated 20$^\circ$ with respect to the beam direction, the $\phi_L$ window varies between $\pm 5^\circ$ and $\pm 20^\circ$, depending on $\theta_L$. The momentum acceptance is a $\Delta\delta\approx\pm 7\%$ window in magnetic rigidity. However, as was demonstrated in~\cite{pul08}, the actual azimuthal- and polar-angular acceptances of the spectrometer depend strongly on the value of the particle magnetic-rigidity. A correction factor is applied on an event-by-event basis to correct for the azimuthal angle cuts, depending on the polar angle and magnetic rigidity of the particle. Finally, the polar angular distributions are transformed into the reference frame of the fissioning system in order to reproduce the angular distribution of the fission fragments and deduce the missing fraction of each isotope. These corrections are detailed in the following sections.

\subsubsection{Spectrometer acceptance}

\begin{figure*}[ht]
\begin{center}
\includegraphics[width=8cm]{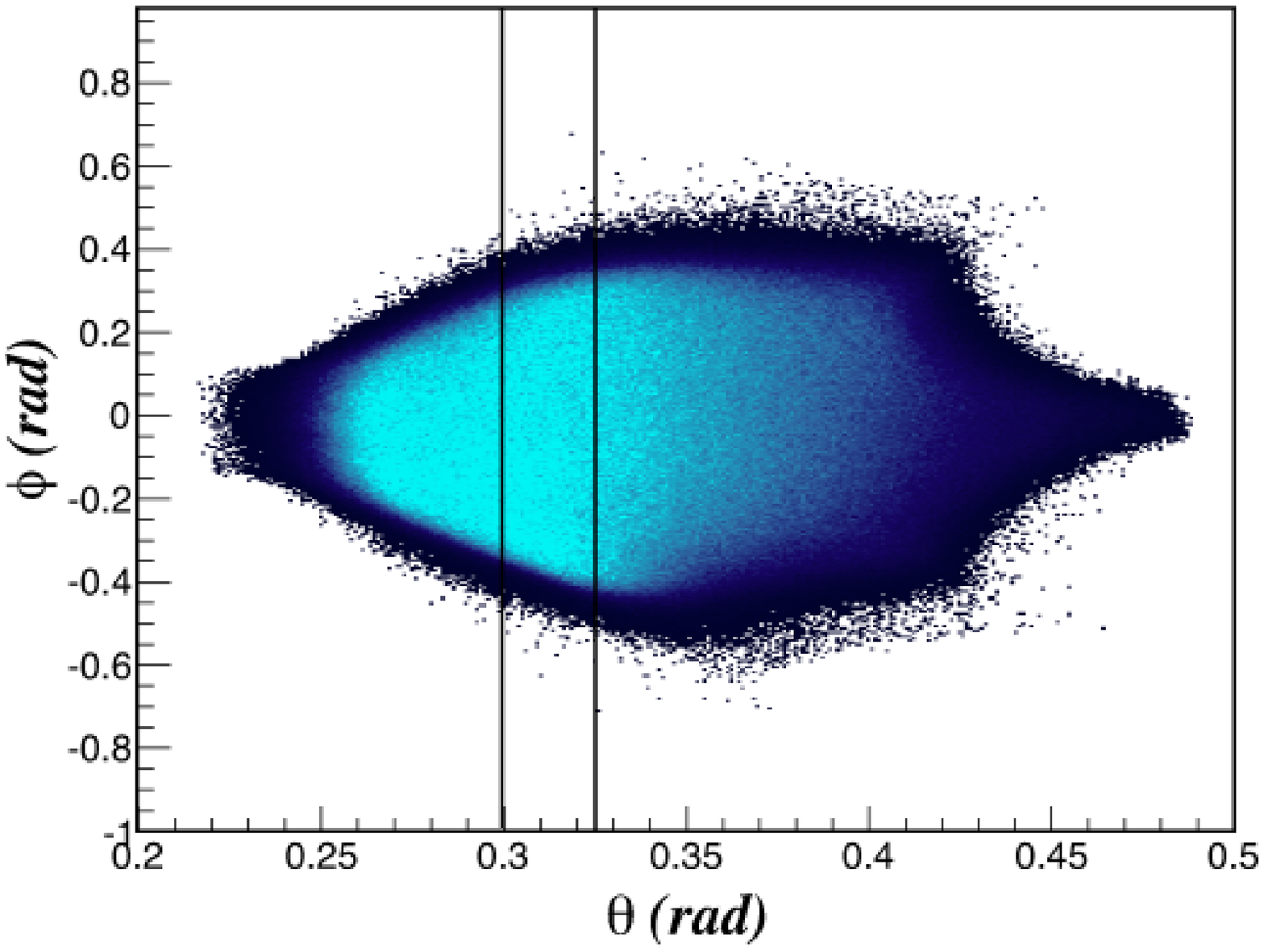}
\includegraphics[width=8cm]{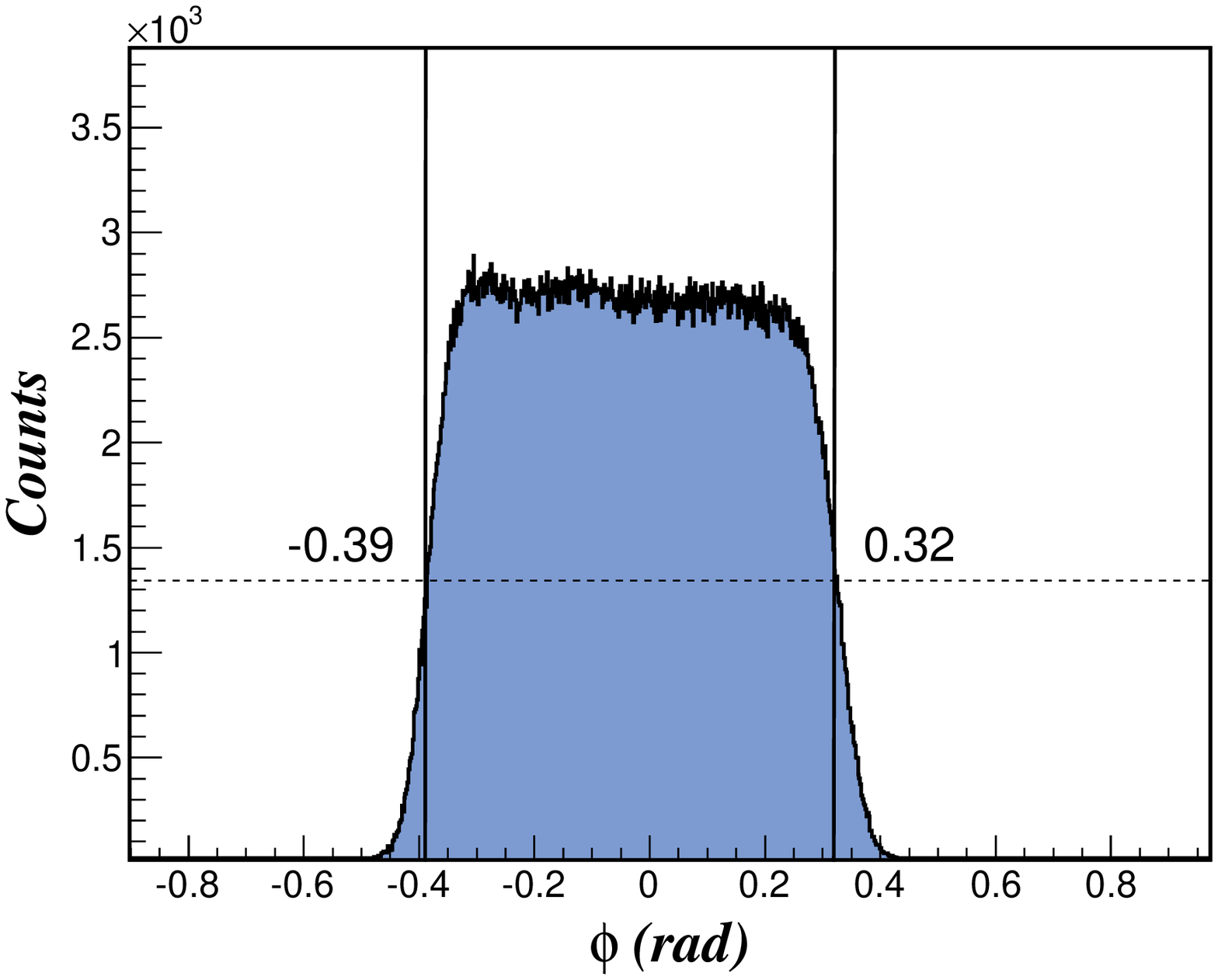}
\includegraphics[width=8cm]{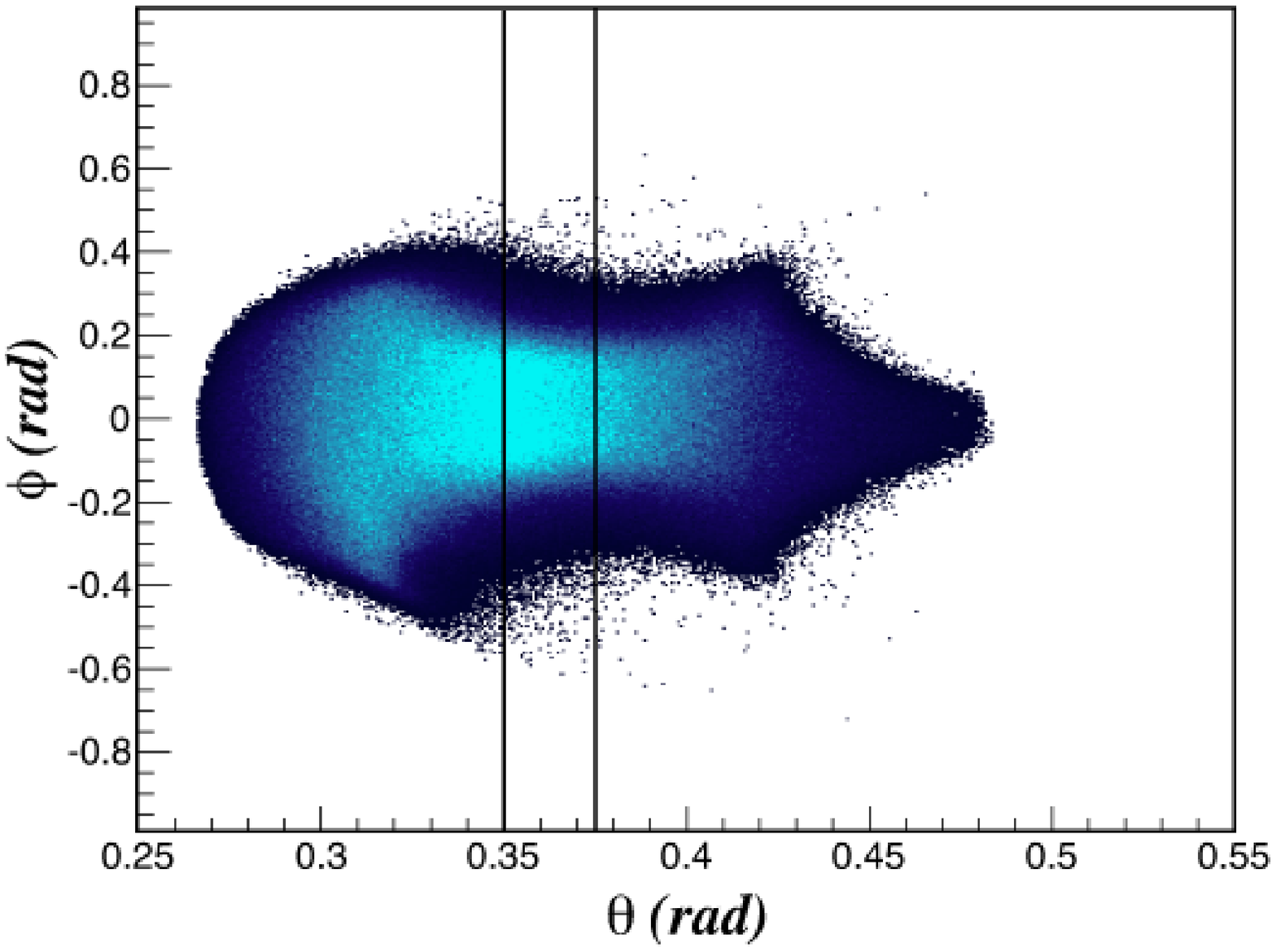}
\includegraphics[width=8cm]{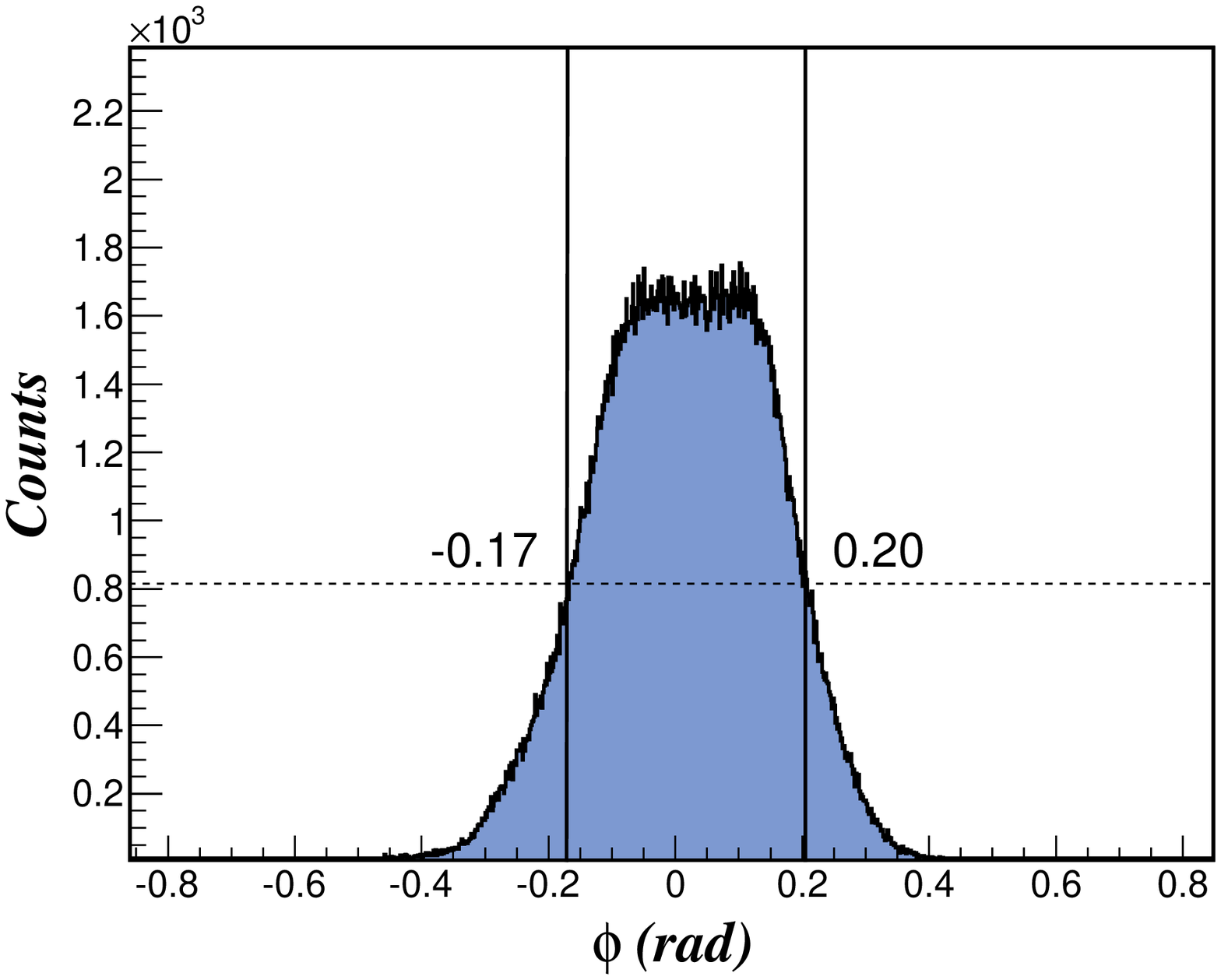}
\caption{\label{fig3} Upper panels: angular distributions for fragments detected in VAMOS with a relative magnetic rigidity between $\delta>0.95$ and $\delta<1.0$. Left panel: $\phi_L$-$\theta_L$ distribution. Right panel: projection in $\phi_L$ of the left panel for a slice between $\theta_L>0.3$ and $\theta_L<0.325$, represented with solid lines in the left panel. The limits of the distribution, defined by its FWHM (dashed line) are shown with solid lines. Lower left panel: $\phi_L$-$\theta_L$ distribution with $\delta>1.0$ and $\delta<1.05$. Lower right panel: projection in $\phi_L$ for a slice between $\theta_L>0.35$ and $\theta_L<0.375$.}
\end{center}
\end{figure*}

Because of the special kinematics of the fission fragments, they cover the complete range of the spectrometer acceptance in magnetic rigidity, and azimuthal and polar angles. A simulation of the reaction based on a fission-fragment kinematics systematics~\cite{boe97} shows that the phase space populated by the fragments is much larger than the acceptance of the spectrometer. The polar-angular distribution is contained between 0$^\circ$ and 25$^\circ$, while the magnetic rigidity spans over 50\% around its average value. Concerning the azimuthal angle, fission fragments are emitted isotropically, covering the full $\Delta\phi_{total}=2\pi$ area. Consequently, the limits in the fragment phase-space induced by the spectrometer acceptance can be measured directly.
Figure \ref{fig3} exemplifies the experimental determination of the $\phi_L$ acceptance. For two values of $d\delta$ slices, the evolution of the limits of $\phi_L$ as a function of $\theta_L$ is represented as a bi-dimensional spectra in the left panels. The distributions in $\phi_L$ obtained for different values of $\theta_L$ are displayed in the right panels as the projections of the slices shown in the bi-dimensional spectra. For each case, the limits in $\phi_L$ transmitted by the spectrometer are defined as the FWHM of the resulting spectra. 
This operation, shown for two different values of $\delta$ in Fig. ~\ref{fig3} is then repeated for values of $\delta$ ranging between 0.85 and 1.14, in steps of 0.01. Steps of 0.5 mrad in $\theta_L$ are covering the corresponding angular distribution to define the projection limits on $\phi_L$ dimension. Finally, for each point of the ($\delta, \theta_L$) map, the acceptance in azimuthal angle $a_{\phi}$ is calculated as the ratio between the detected range and the total $\Delta\phi_{total}=2\pi$:
\begin{equation}
a_{\phi}(\theta_L,\delta)=\frac{\Delta\phi_L(d\theta_L,d\delta)}{2\pi}
\end{equation}

The acceptance $a_{\phi}$ is applied on an event-by-event basis to the measured data. The value of $a_{\phi}$ for each measured $\theta_L$ and $\delta$ is obtained by interpolating the discrete function $a_{\phi}(\theta_L,\delta)$ previously calculated. 

Once the acceptance in $\phi_L$ is corrected, the detection within VAMOS depends solely on the angle $\theta_L$ and the reduced magnetic rigidity $\delta$ of the particle. Figure \ref{fig4} shows the shape of the acceptance in $\delta$ as a function of $\theta_L$. In the right panel of Fig. \ref{fig4}, the different spectrometer settings have been added. The overlapping areas can be distinguished inside the physical distribution. It can be seen that, whereas for small reference settings the fission fragments cover the complete acceptance of the spectrometer, for larger nominal values, the up-right corner of the spectrometer acceptance is not covered. This is a consequence of the fission kinematics: large velocities, and therefore large magnetic rigidities, are associated to small angles. Another feature of the fission kinematics is that, contrary to the case with azimuthal angle, the $\theta_L-\delta$ distribution cannot be assumed to be homogeneous. Therefore, to estimate the losses due to the spectrometer-acceptance limits, it is mandatory to deduce such distribution. For this purpose, the polar angular distribution is calculated in the reference frame of the fissioning system, as explained in the following section.

\subsubsection{\label{secRec}Reconstruction of angular and charge state distributions}

The velocity of the fissioning system $v^*$ in the laboratory reference frame is determined by the momentum and energy conservation laws. In the case of fusion reactions, it is univocally determined from the incoming-beam and target properties. For a transfer reaction, the kinematical properties (angle and velocity) of the actinides produced are deduced from the mass, energy and emission angle of the target-like particle measured in the SPIDER detector. In all the cases, the target thickness (100 $\mu$g/cm$^2$) induces a negligible spread on the kinematic properties of the reaction products. Then, for any reaction, the emission angle of the fragment $\theta_{fiss}$ in the reference frame of the fissioning nucleus is defined as:

\begin{equation}
\tan(\theta_{fiss})=\frac{\sin(\theta_L)}{\gamma^{*}\left[\cos(\theta_L)-\frac{v^*}{v_L}\right]},
\label{tcm}
\end{equation}
where $v_L$ is the measured fragment velocity, and $\gamma^*$ is the Lorentz factor associated to the fissioning system velocity $v^*$.

\begin{figure}[ht]
\begin{center}
\includegraphics[width=8cm]{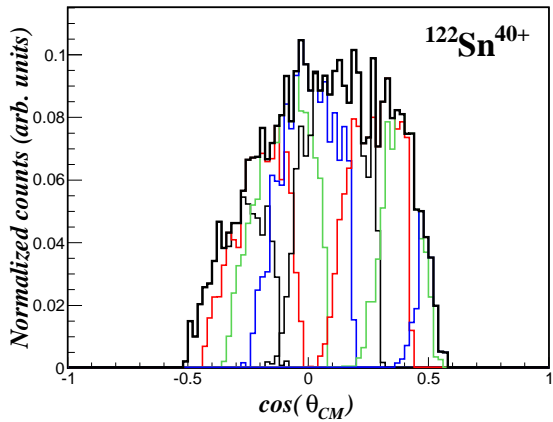}
\caption{\label{figa} Envelope of the angular distributions in the reference frame of the fissioning system of $^{122}$Sn$^{40+}$ ions measured in the different spectrometer settings.}
\end{center}
\end{figure}

\begin{figure*}[ht]
\begin{center}
\includegraphics[width=17cm]{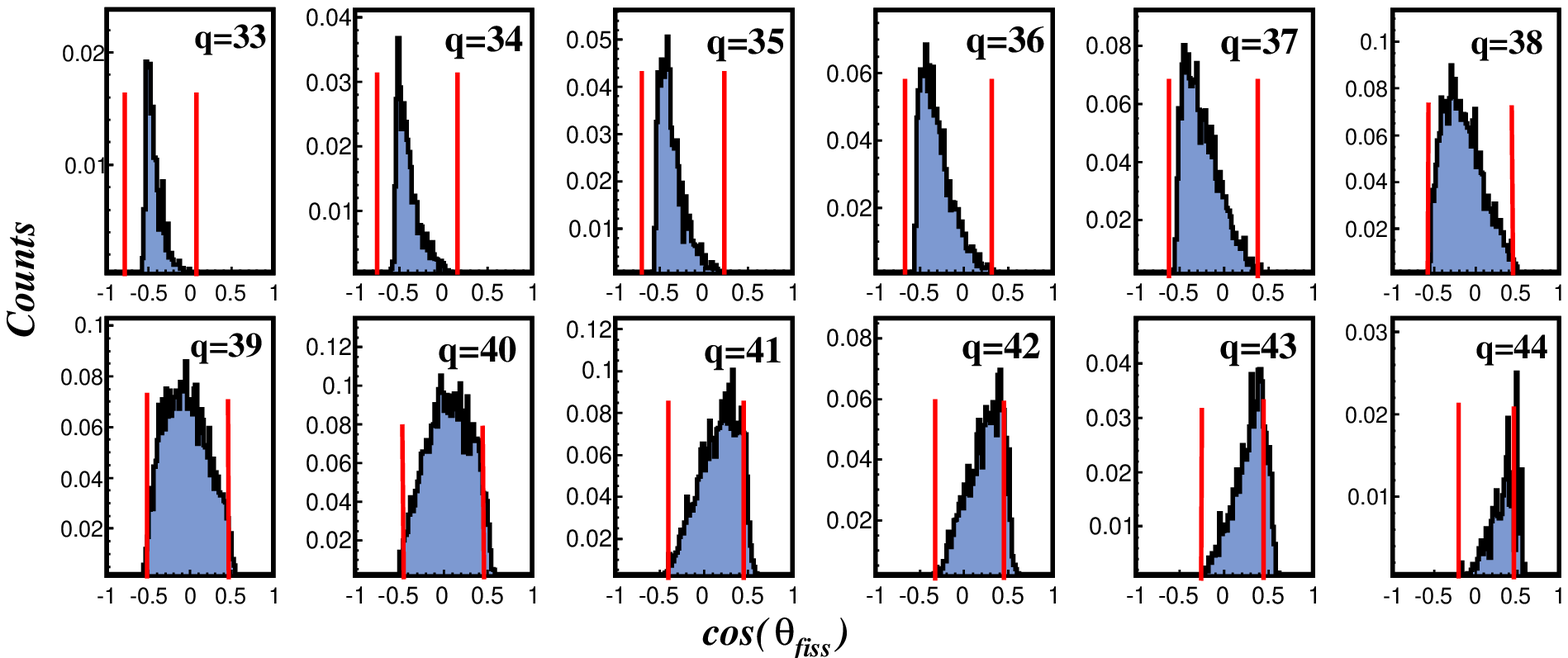}
\caption{\label{figall} Angular distributions in the reference frame of the fissioning system for all significantly contributing charge states of $^{122}$Sn. The integration limits corresponding to a complete transmission within the spectrometer settings for each charge state are shown as vertical bars.}
\end{center}
\end{figure*}

Figure \ref{figa} shows the angular-distribution in the reference frame of the fissioning system for the ion $^{122}$Sn$^{40+}$, measured for the different spectrometer settings, after applying the relative normalization and correcting for the acceptance in $\phi_L$. The resulting angular distribution in Fig. ~\ref{figa} is the convolution of the fission angular-distribution and the probability for the ion to be produced with the charge-state $q=40$. This probability evolves with the velocity in the laboratory reference frame, having a maximum for a particular velocity. Since both $\theta_{fiss}$ and $v_L$ are related through Eq. \ref{tcm}, a similar behaviour is expected in the charge distribution along $\cos(\theta_{fiss})$. For each ($Z$, $A$, $q$) system, the ion kinematics enters into the full transmission ($B\rho$, $\theta_L$) region. However, this region depends on the charge state of the ion, as it is sketched in the right panel of Fig. \ref{fig4}: the solid lines correspond to the kinematics of two different charge states of the same ($Z$, $A$) system, and the dots mark the limits of detection for each one. These limits can be translated to cos($\theta_{fiss}$) in the reference frame of the fissioning system, resulting in a window where each charge state is measured without acceptance restrictions.In principle, the sum of the $q$ distributions corresponds to the yield of the fragment within a region of cos($\theta_{fiss}$) where the charge states are fully transmitted. In order to determine this angular region, the limits for each $q$ distribution are determined. As an example, the limits corresponding to the detected charge states of $^{122}$Sn are shown in Fig. \ref{figall} as solid bars. As it can be seen, in most of the cases the angular distribution of each charge state is measured along the ensemble of magnetic settings chosen during the experiment. However, some distributions are cut in the edges: for low charge states in backward fission, the acceptance is some times larger than the actual cut observed in the distribution. This cut is produced by the slowing down of ions in the ionization chamber, which prevent the fragments with lower velocity (and thus emitted in backward direction) to reach the silicon wall of the focal plane and be fully identified.

Within the cuts produced by the detection efficiency and the spectrometer acceptance, the angular distribution is supposed to be fully transmitted. 

\subsubsection{\label{Yuncer}Isotopic fission-fragments yields and estimated uncertainties}

\begin{figure}[ht]
\begin{center}
\includegraphics[width=8cm]{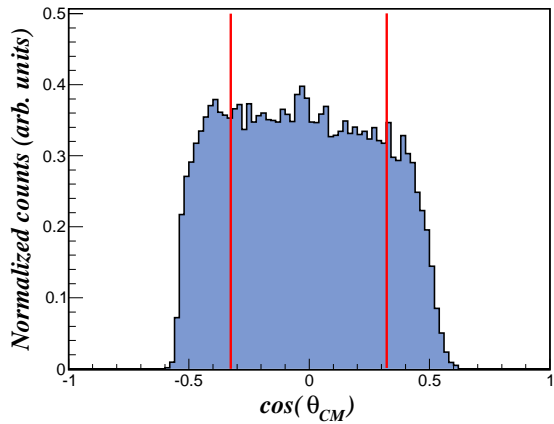}
\caption{\label{sumq} Angular distribution in the reference frame of the fissioning system of $^{122}$Sn summed over its measured charge states. The integration limits corresponding to a complete transmission within the spectrometer settings are shown as vertical bars.}
\end{center}
\end{figure}

To reconstruct the isotopic yields $Y(Z,~A)$, it is necessary to add the contribution of all produced charge-states. This summation is done within the limit angles $\theta_{fiss, min}$ and $\theta_{fiss, max}$, defined as the minimum interval in which all charge-states angular distribution are fully transmitted. In Fig. ~\ref{sumq}, $\theta_{fiss, max}$ is the minimum of the upper limits for the different charge states displayed, while $\theta_{fiss, min}$ is the maximum of the lower limits. In order not to reduce too drastically the integration region, the considered limits were only those that cut the distributions at an amplitude bigger than 10\% of the maximum of each individual distribution. The isotopic yield is then defined as:

\begin{equation}
Y(Z,~A) = f_{\theta}\sum_q \int_{\theta_{fiss, max}}^{\theta_{fiss,min}} \frac{dN(Z,~A,~q)}{d\cos(\theta)} d\cos(\theta)
\label{eqY}
\end{equation}
where $\frac{dN(Z,~A,~q)}{d\cos(\theta)}$ is the angular distribution measured for one ion, see Fig. ~\ref{figa}, corrected event-by-event for the azimuthal acceptance $a_\phi(\theta_L,\delta)$ as discussed in the previous section. 

The summation of all charge states for the particular case of $^{122}$Sn is displayed in Fig. \ref{sumq}, as well as the integration limits settled from Fig. \ref{figall}. These limits define the correction factor $f_\theta$ for the spectrometer acceptance to deduce the yields of each isotope :
\begin{equation}
f_\theta=\frac{2}{\cos(\theta_{fiss, max})-\cos(\theta_{fiss, min})}
\label{fteta}
\end{equation}

This factor is defined under the assumption that the angular distribution in the reference frame of the fissioning system is isotropic. The yields $Y(Z,~A)$ are then multiplied by $f_\theta$ to correct for the spectrometer acceptance, and finally, yields are normalized to 200.

The uncertainties on the final yields originate from three sources; statistics, relative normalization, and $A$ and $Z$ resolutions. The statistical error for each ion is taken into account, and summed quadratically to the normalization error, estimated to be around 0.5\%. 
The $A$ and $Z$ resolutions contribute with a systematic error below 10\%. 
Another possible error source comes from the evaluation of the correction factor $f_\theta$ (Eq.~\ref{fteta}) for the limited angular range of the measurement. This factor is calculated assuming an isotropic angular distribution or, similarly, a constant angular anisotropy. This assumption would be inaccurate in the case of a mass-dependent anisotropy, as it has been observed in~\cite{VoI97}, rendering a dependence of the correction factor $f_\theta$ on the mass partition. This would produce a systematic variation on the measured yield estimated in a maximum of 25\% for the most extreme cases. However, in the present measurement, the angular distributions for each isotope, measured in a range between 60$^\circ$ and 120$^\circ$ in the reference frame of the fissioning system, show no evidence of anisotropy. 

In the present experiment, within the limits in velocity (or emission angle in the reference frame of the fissioning system) defined in Figs. \ref{figall} and~\ref{sumq}, the complete charge-state distribution is measured. This ensures the yields to be independent of possible distortions of the charge-state distributions, which have been shown to be non-Gaussian for isotopes that are produced in short-lived ($\sim$1 ns) isomeric states~\cite{BaS11}. 
This advantage is due to the large acceptance of the VAMOS spectrometer.

\section{Fission-Fragment Yields}

In the present section, the results obtained in fusion-fission and transfer-induced fission reactions are discussed. Most of the statistics correspond to the fusion-fission channel, which counts for approximately 90\% of the cross-section~\cite{bis97}. The fusion-fission events are defined with the detection of a fission fragment at the focal plane of the spectrometer without coincidence of the target-like recoil in the SPIDER telescope. As the angular coverage of the transfer events varies between 60 to 75\%~\cite{carme}, the fusion-fission events have less than 10~\% of contamination from transfer-induced fission. 

\begin{figure}[ht]
\begin{center}
\includegraphics[width=8.5cm]{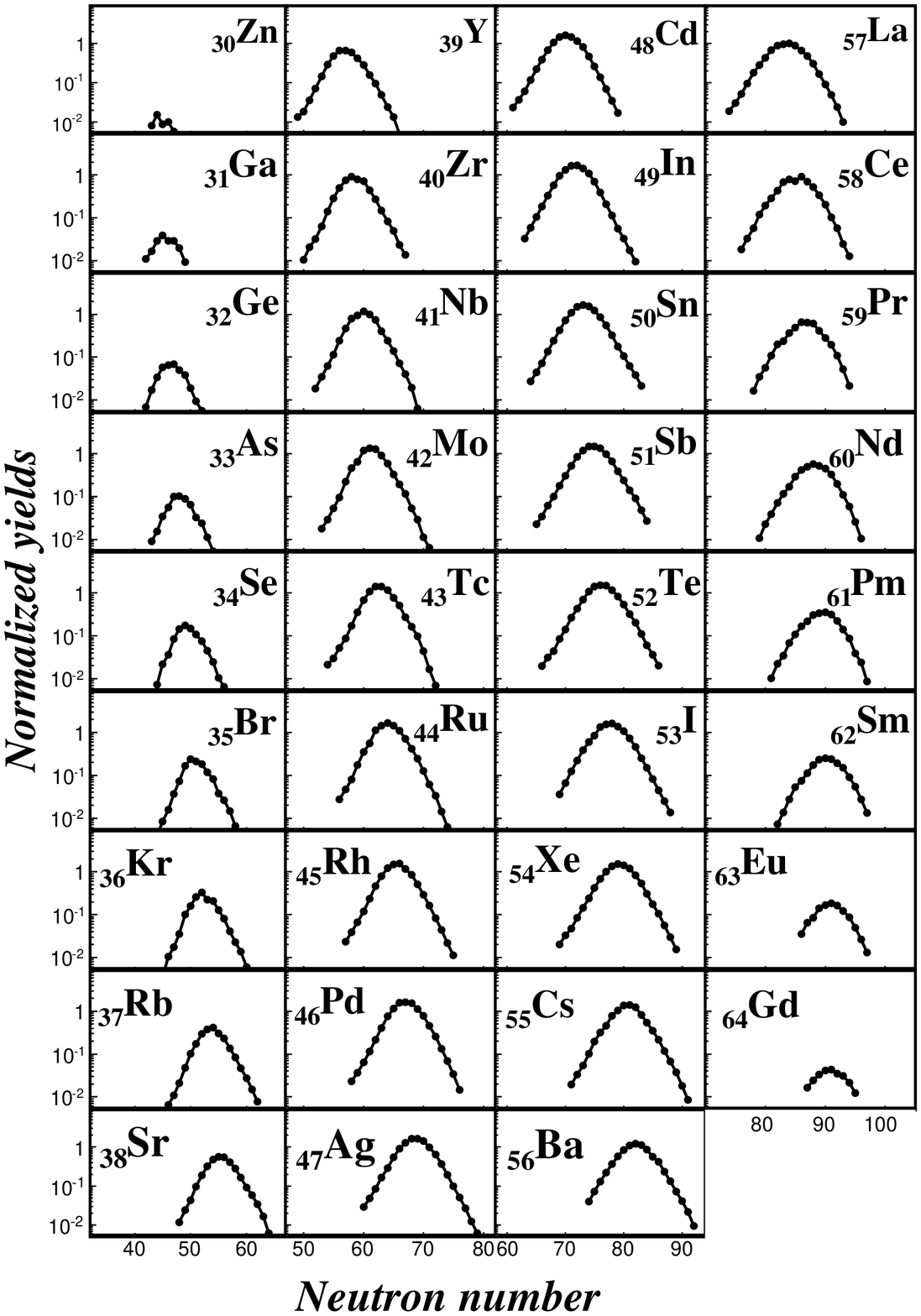}
\caption{\label{iso_cf} Isotopic distributions of fission fragments produced in the fusion-fission reaction of $^{238}$U on $^{12}$C, leading to the compound system$^{250}$Cf with an excitation energy of 45 MeV. Statistical error bars are smaller than the size of the points.}
\end{center}
\end{figure} 

\subsection{Fusion-fission isotopic yields}

The fusion reaction leads to the compound nucleus $^{250}$Cf, produced with an excitation energy of 45 MeV. The isotopic distributions of the corresponding fission yields are displayed in Fig. \ref{iso_cf}. They span from $_{30}$Zn to $_{64}$Gd, over almost 600 isotopes. The isotopic distributions show a typical bell shape and range on two orders of magnitude. The corresponding distributions in mass and atomic number are displayed in Fig. \ref{cf_A_Z} and reviewed in Appendix \ref{app_cf}. For heavier fragments, the expected symmetry of the distributions is not attained due to the experimental resolution and the acceptance limits. As it can be seen in this figure, the mass and atomic-number distributions show a flat top, showing the remaining influence of asymmetric fission at this moderate excitation energy. Such flat-top mass distribution has been observed in the same reaction at a similar energy in~\cite{YaS12}. Compared to the experiments in direct kinematics~\cite{YaS12,HiH92}, the present work allows to investigate also the complete atomic-number distribution (left panel of Fig.~\ref{cf_A_Z}), and therefore the neutron excess of the isotopic distribution. The neutron excess is defined as the average neutron number of one isotopic distribution, divided by the corresponding atomic number: $<N>/Z$. The neutron excess of the fission-fragments reflects the proton-to-neutron equilibration during deformation of the compound nucleus, followed by the evaporation of neutrons post scission, reflecting the sharing of the excitation energy between the two fragments. The resulting neutron excess for the present data on fusion-fission reaction is displayed in left panel of Fig. \ref{cf_A_Z}. It shows a constant value about 1.46, almost independent of the fission-fragment atomic number, while the neutron-over-proton number of the compound nucleus $^{250}$Cf is 1.55. The difference is coherent with a total evaporation (pre and post-scission) of 9 neutrons. The constant trend of the neutron excess along the atomic-number distribution of the fragments is surprising. The Coulomb repulsion within the forming nuclei tends to produce the heavier nuclei more neutron rich than the light ones \cite{arm70}. Consequently, an increase in the neutron excess as a function of the atomic number of the fission fragment is expected. In addition, the contribution of low excitation-energy fission, which appears in the mass and atomic-number distributions, shows no influence in the evolution of the neutron excess. In fission at low excitation energy, the neutron excess is known to exhibit a step behaviour: the heavy fragments being more neutron rich than the light ones (see Ref. \cite{boc89} and Fig. \ref{fig_iso_pu2}). This step behaviour should increase the charge polarization produced by the Coulomb repulsion. The apparent disagreement between the measured data and the expected trend may reside in the role of the excitation energy equilibration between both fragments, which needs to be described in more detail to reproduce the observed data. Such discussion will be the subject of a forthcoming publication.

\begin{figure*}[ht]
\begin{center}
\includegraphics[width=8cm]{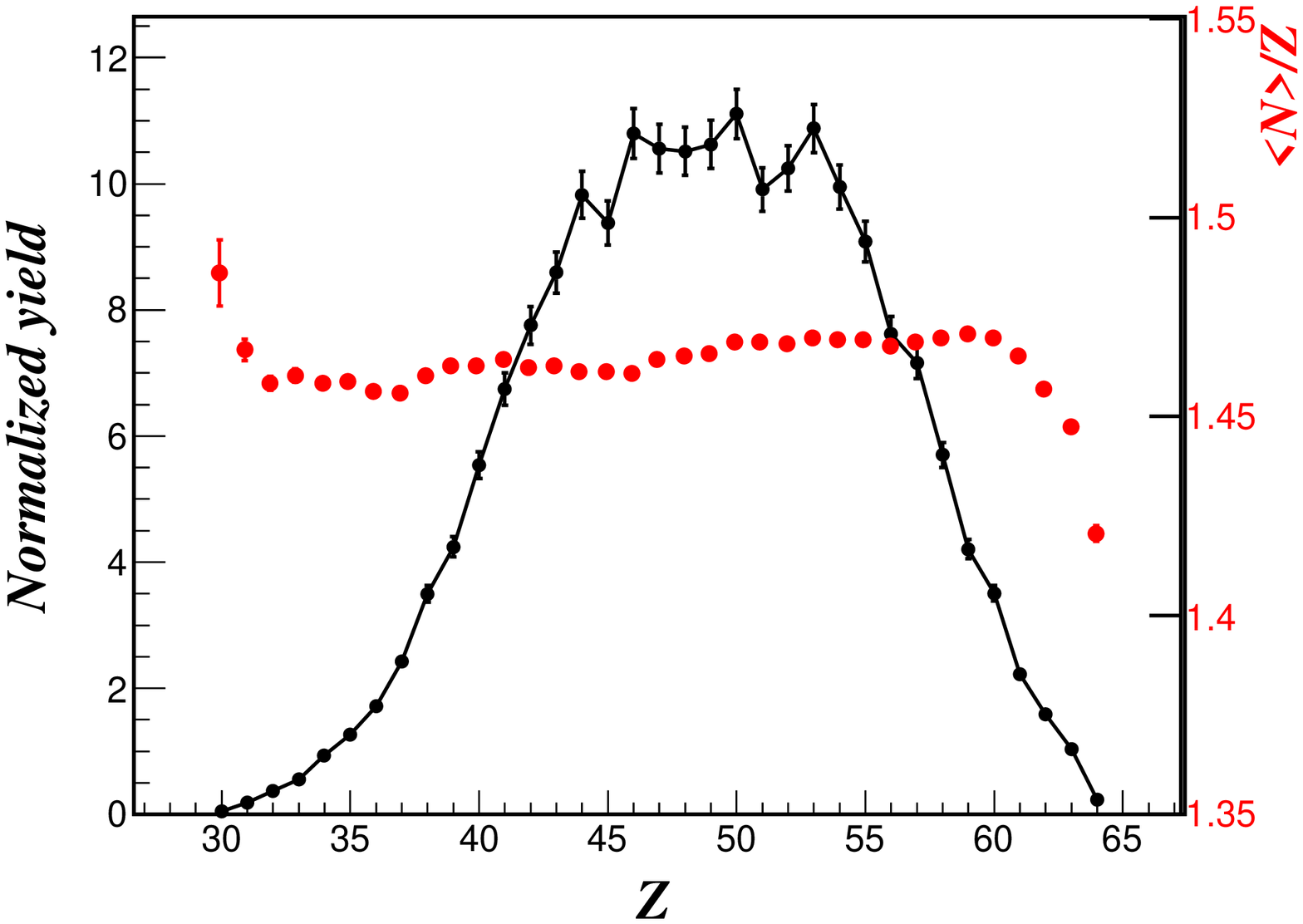}~~~\includegraphics[width=8cm]{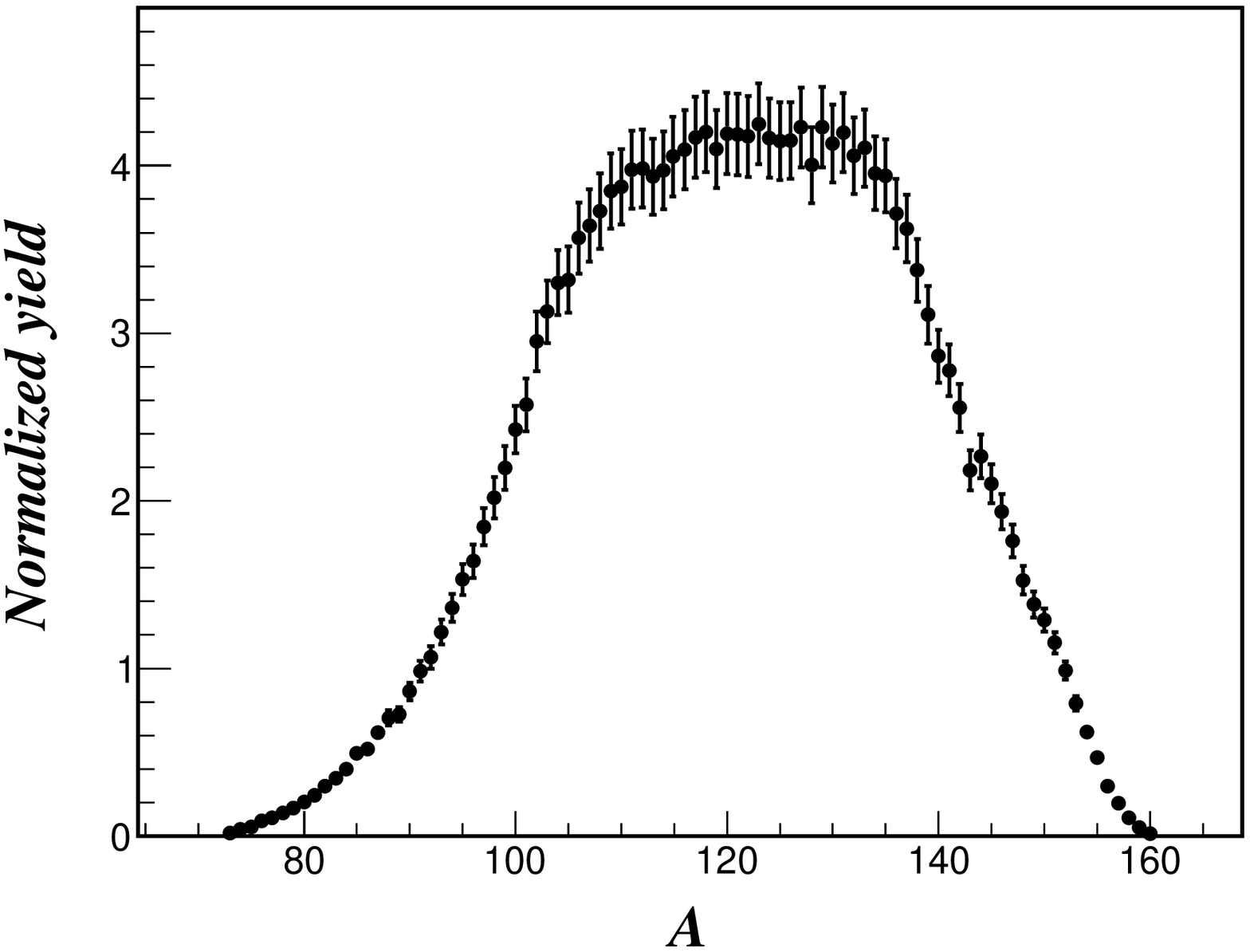}
\hspace{1cm} \caption{\label{cf_A_Z} Atomic-number and mass distributions of fission fragments produced in the fusion-fission reaction of $^{238}$U on $^{12}$C, leading to a compound system$^{250}$Cf with an excitation energy of 45 MeV, are shown in left and right panels, respectively. The neutron excess of the isotopic distributions is superimposed in the left panel. Statistical error bars are displayed.}
\end{center}
\end{figure*}

\subsection{\label{transfer_induced}Transfer-induced fission}

The fission fragments measured in coincidence with a recoil particle detected in SPIDER correspond to transfer-induced fission events. The reconstruction of fission-fragment isotopic yields described in Sec.~\ref{section_method} is done for each of the fissioning systems identified in Fig.~\ref{fig:matrix}. However, the inelastic scattering of the target (leading to the fissioning system $^{238}$U) and the one-proton transfer (leading to the fissioning system $^{239}$Np) channels resulted in too few statistics due to the technical difficulties encountered during the experiment. Results on isotopic yields are given in the following only for the two-proton transfer channel, leading to the formation of the $^{240}$Pu and $^{241}$Pu, with respective yields of 80 and 20\% approximately, as described in Sec.~\ref{section_SPIDER}. 
The radial segmentation of SPIDER gives information on the emission angle of the target-recoil nuclei. Assuming a two-body reaction it is possible to derive the excitation energy gained in the collision:

\begin{align}
\label{eq:eex}
&E^* = T_1 + Q_0 - T_4 - \nonumber \\
&\frac{1}{M_3} \left(M_1T_1 + M_4T_4 - 2~\cos{(\theta_4)} \sqrt{M_1 T_1 M_4 T_4} \right)
\end{align}
where $Q_0$ is the reaction heat, $T_x$, $M_x$ and ${\theta}_x$ are the kinetic energy, the mass and the angle in the laboratory of the particle $x=$1, 2, 3, and 4 representing the beam, the target, the actinide, and the target-recoil nuclei. 

\begin{figure}[h]
\begin{center}
\includegraphics[width=8cm]{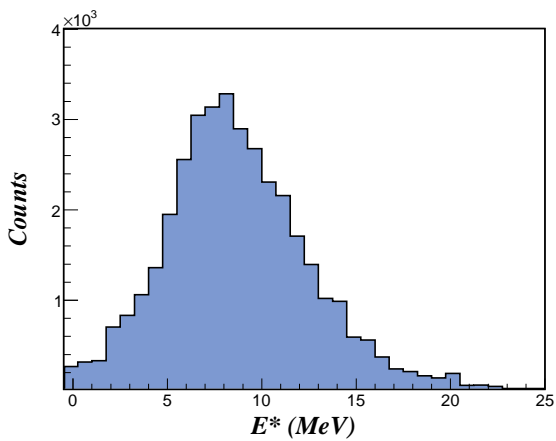}
\caption{\label{figex} Reconstructed excitation-energy spectrum associated to the two-proton transfer channel.}
\end{center}
\end{figure}

As it was already mentioned, it is assumed that the excitation energy produced in the transfer reactions remains in the produced actinides. Precise measurements of the partition of the excitation energy between the fissioning system and the recoil nuclei found that this assumption is valid in around 90 \% of the statistics \cite{carme}. The resolution in total energy is around FWHM$\approx 7$ \% and the angular coverage by each radial segment is around 1$^\circ$, which leads to a resolution in excitation energy that does not exceed FWHM$\approx 1.5$~MeV. Figure~\ref{figex} displays the reconstructed excitation-energy spectrum associated to the two-proton transfer channel ($^{10}$Be as a target recoil identified in Fig.~\ref{fig:matrix}). The excitation-energy spectrum shows an average value of 8.7 MeV, and a FWHM of 6.4 MeV, which corresponds to similar excitation-energy conditions as in the formation of the compound nucleus $^{240}$Pu in fast-neutron induced fission. 

The isotopic yields of almost 300 fission products from $_{34}$Se to $_{60}$Nd, obtained with the method described in Sec.~\ref{section_method}, are listed in Appendix \ref{app_pu} and displayed in Fig. \ref{fig_iso_pu2}, where they are compared with previous data measured in thermal-neutron induced fission at the Lohengrin spectrometer~\cite{sch84,BaS11}. In these works based on direct kinematics, the light fragments are identified in-flight with a spectrometer, while the heavy fragments are identified in mass via time-of-flight techniques, and isotopically via the observation of the $\gamma$-ray emission after $\beta$ decay. The isotopic yields of the light fragments are in a very good agreement between the different techniques, despite the slight difference in excitation energy. The present data show higher yields for the neutron-rich fragments than the thermal-neutron induced fission data. This difference may result from the difference in excitation energy which populates more the symmetric region, or from the contamination of the $^{241}$Pu in the present data. The isotopic yields obtained with $\gamma$-ray spectroscopy after $\beta$ decay show isotopic distributions on limited ranges, due to technical imitations inherent to the half-life of the isotopes and the knowledge of the decay level-scheme \cite{BaS11}.

\begin{figure}[ht]
\begin{center}
\includegraphics[width=8.5cm]{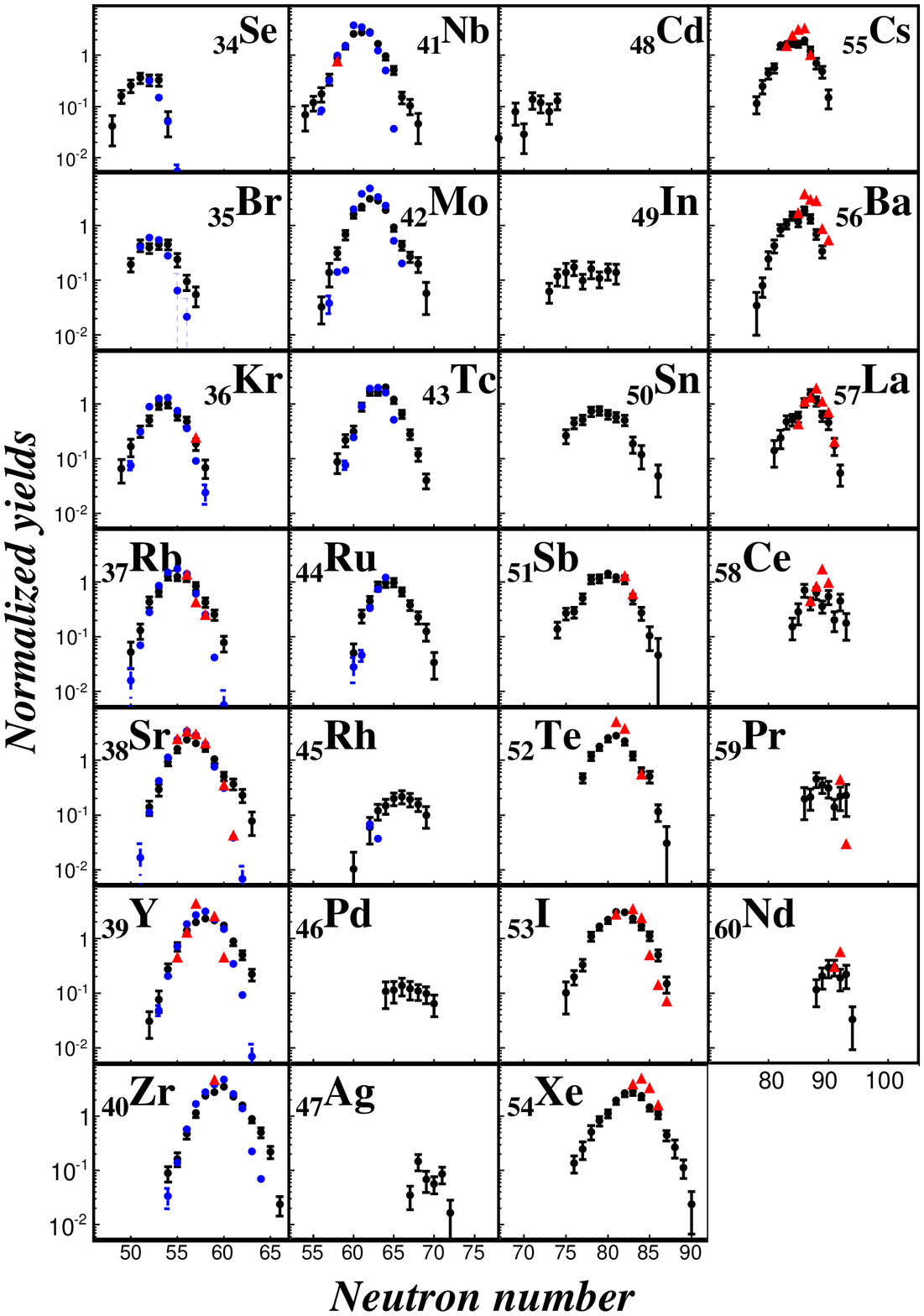}
\caption{\label{fig_iso_pu2} Isotopic fission yields distribution for each of the elements produced in transfer-induced fission leading to the formation of the $^{240}$Pu. The isotopic-yields sum is normalized to 200. Present data (black dots) are compared to thermal-neutron induced fission data measured in direct kinematics from Refs. \cite{sch84} (blue dots) and \cite{BaS11} (red triangles).}
\end{center}
\end{figure}

\begin{figure*}[ht]
\begin{center}
\includegraphics[width=8cm,height=6cm]{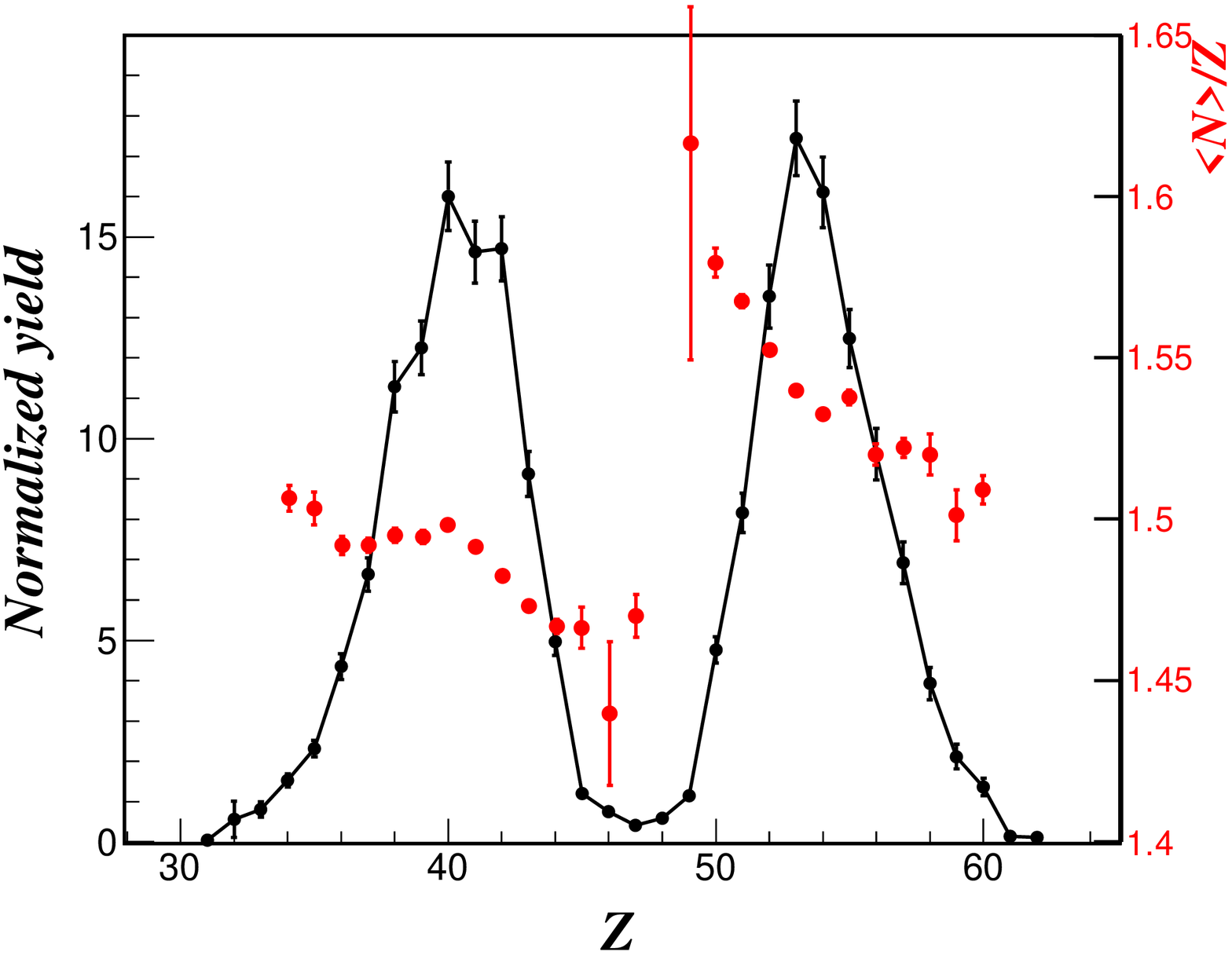}~~~
\includegraphics[width=8cm]{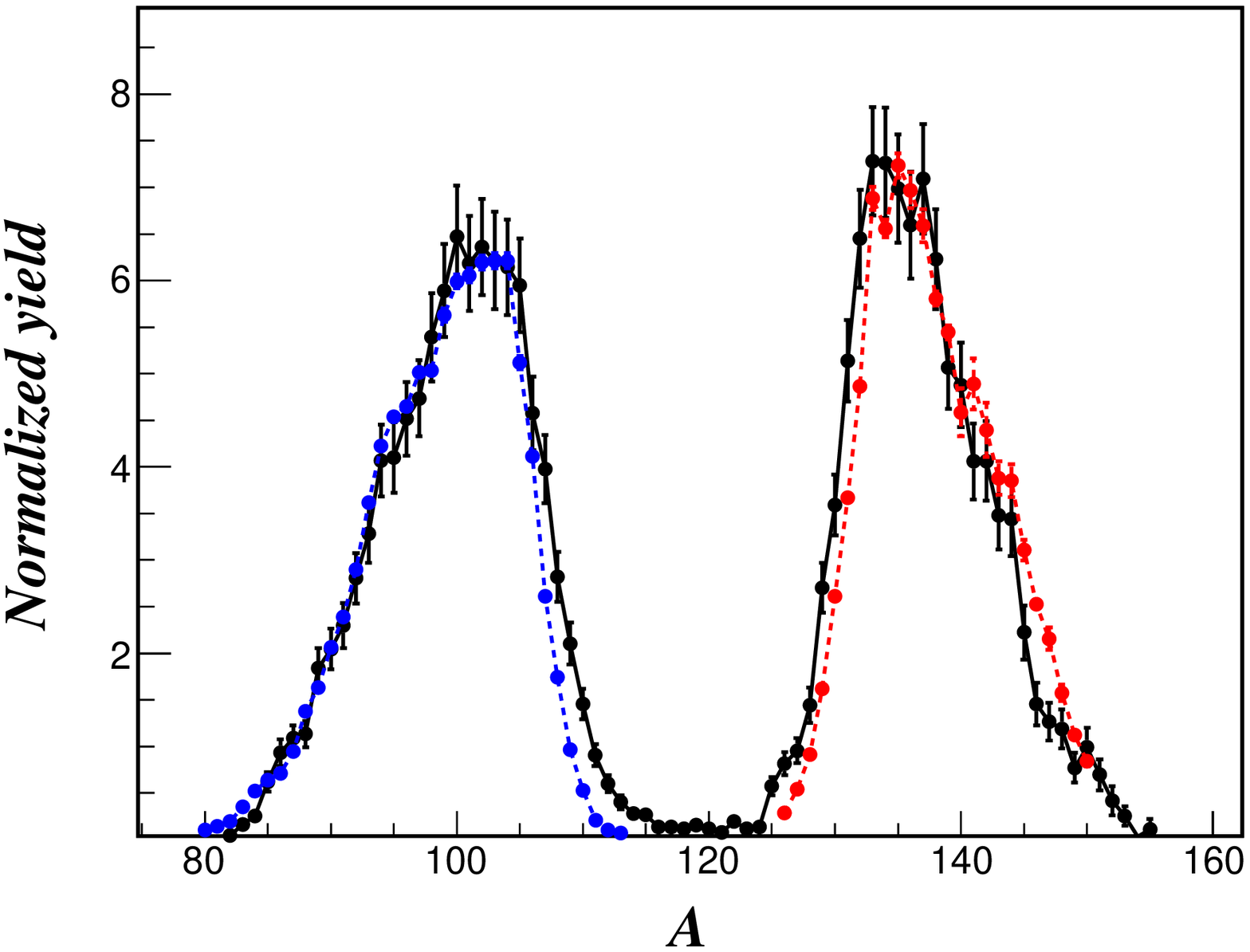}
\caption{\label{fig_ZA_Pu} Atomic-number and mass distributions of fission fragments produced in the transfer-fission reaction of $^{238}$U on $^{12}$C, leading to compound systems $^{240,241}$Pu, are shown in left and right panels, respectively. The neutron excess of the isotopic distributions is superimposed in the left panel. Mass distributions are compared to thermal-neutron induced fission data measured in direct kinematics from Refs. \cite{sch84} (blue dots) and \cite{BaS11} (red dots). Statistical error bars are displayed.}
\end{center}
\end{figure*}

The corresponding atomic-number and mass distributions are displayed in Fig.~\ref{fig_ZA_Pu}. The mass distribution is in good agreement with previous data, showing the validity of the normalization method procedure described in the present work. Compared to the thermal-neutron induced fission, the present data present a more important production in the valley between the two fission-fragment humps. This feature is well-known from fragment mass distributions, where it has been shown that the peak-to-valley ratio decreases with increasing neutron bombarding energy. In the present data, this ratio is about 60, in excellent agreement with the excitation energy of the fissioning nucleus~\cite{glen81}. 

\begin{figure}[ht]
\begin{center}
\includegraphics[width=8cm]{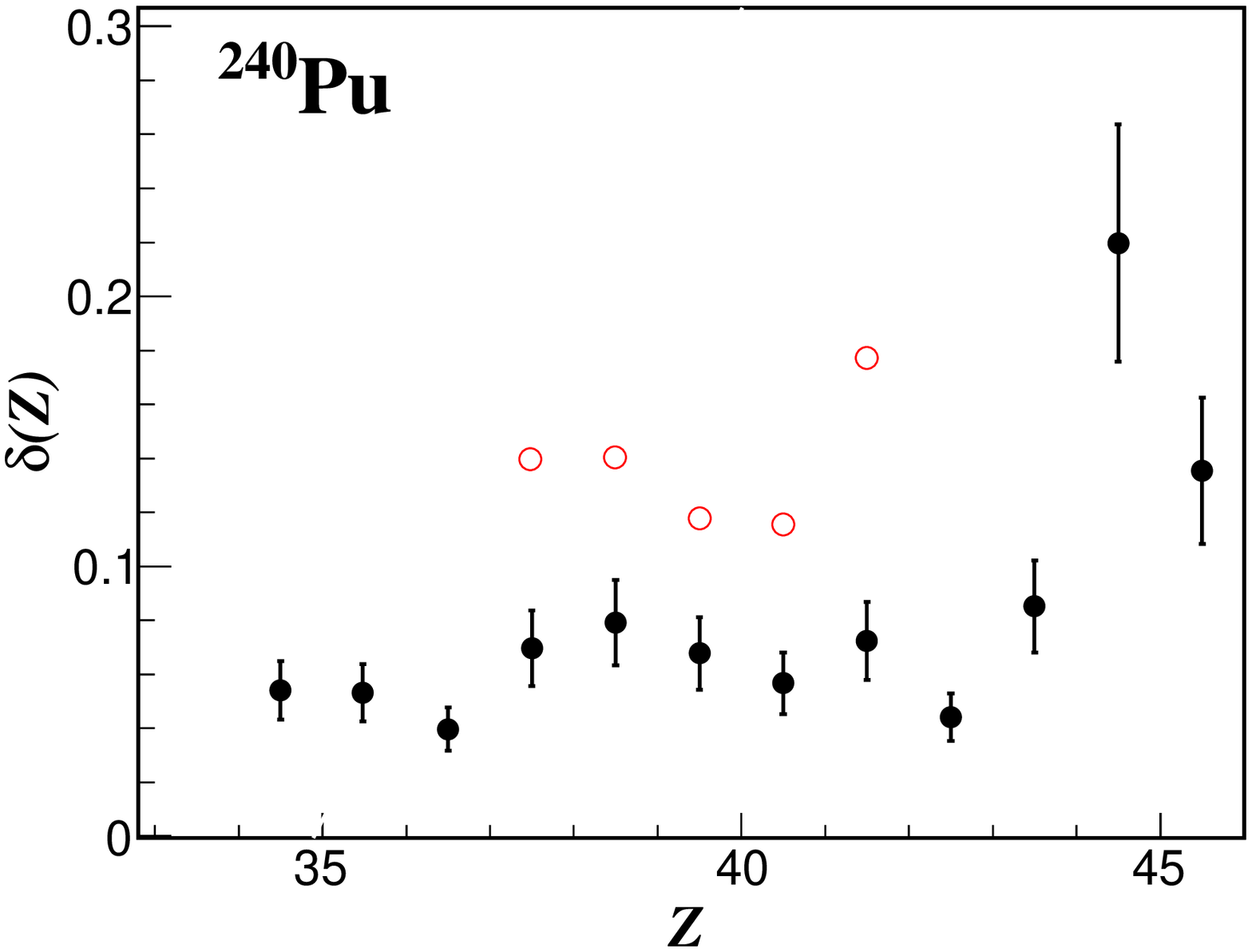}
\caption{\label{delta} Local even-odd effect as a function of the atomic number of the fission fragments for the fissioning nucleus $^{240}$Pu. Present data (open symbols) are compared to thermal-neutron induced fission~\cite{sch84} (full symbols). Statistical error bars are displayed.}
\end{center}
\end{figure}

The atomic-number distribution shows an even-odd staggering, which is not present in the heavy fragments due to the limited resolution in atomic number achieved in the present experiment. In the light-fragment region, the local deviation from a smooth atomic-number distribution may be estimated following the formulation of ~\cite{tracy}:
\begin{equation}
\delta(Z+\frac{3}{2}) = \frac{-1^{Z+1}}{8}\left[L_3-L_0-3\left(L_2-L_1\right)\right],
\label{eq_tracy}
\end{equation}
where $L_i$ is the natural logarithm of the fission yield for atomic number $Z+i$. The resulting local even-odd staggering $\delta(Z)$ is displayed in Fig.~\ref{delta}. The measured even-odd staggering in the present experiment shows a similar trend to the previous data of Schmitt et al.~\cite{sch84}, with a smaller amplitude about 6 to 7 \%, compared to 12 to 14\% in the previous data. This difference reflects the higher excitation energy of the fissioning system in the present experiment compared to data from fission induced by thermal neutrons, which favors the breaking of pairs before scission. In the present data, no evidence of an increase of the even-odd staggering at large asymmetry of the fragment distribution, as expected from the general behaviour described in \cite{caa11}.

\section{Fission kinematics}

\begin{figure}[ht]
\begin{center}
\includegraphics[width=8cm]{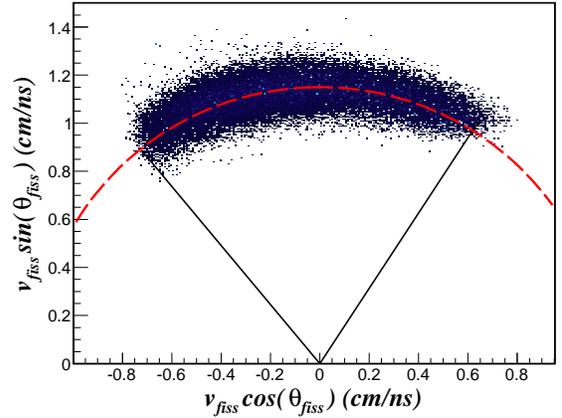}
\caption{\label{velocity sphere} The portion of the velocity sphere relative to the fragment $^{122}$Sn measured in the present experiment. The limits in the measurement correspond to the angular acceptance of the spectrometer and the magnetic rigidity scanning.}
\end{center}
\end{figure}

\begin{figure}
\begin{center}
\includegraphics[width=8.5cm]{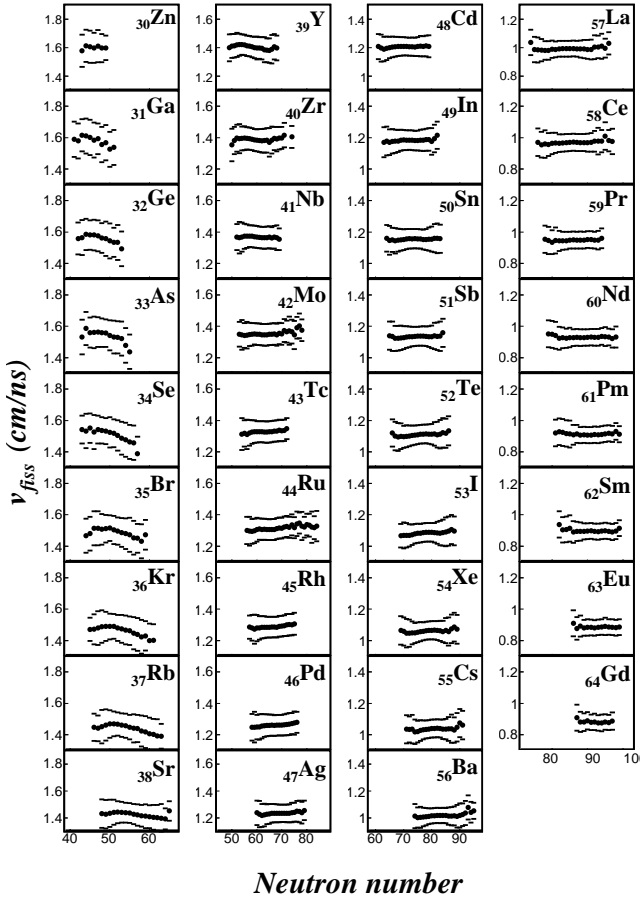}
 \caption{\label{vel_iso_Cf} Fission velocity, $v_{fiss}$, of fission fragments produced in fusion-fission reactions.}
\end{center}
\end{figure}

\begin{figure}
\begin{center}
\includegraphics[width=8cm]{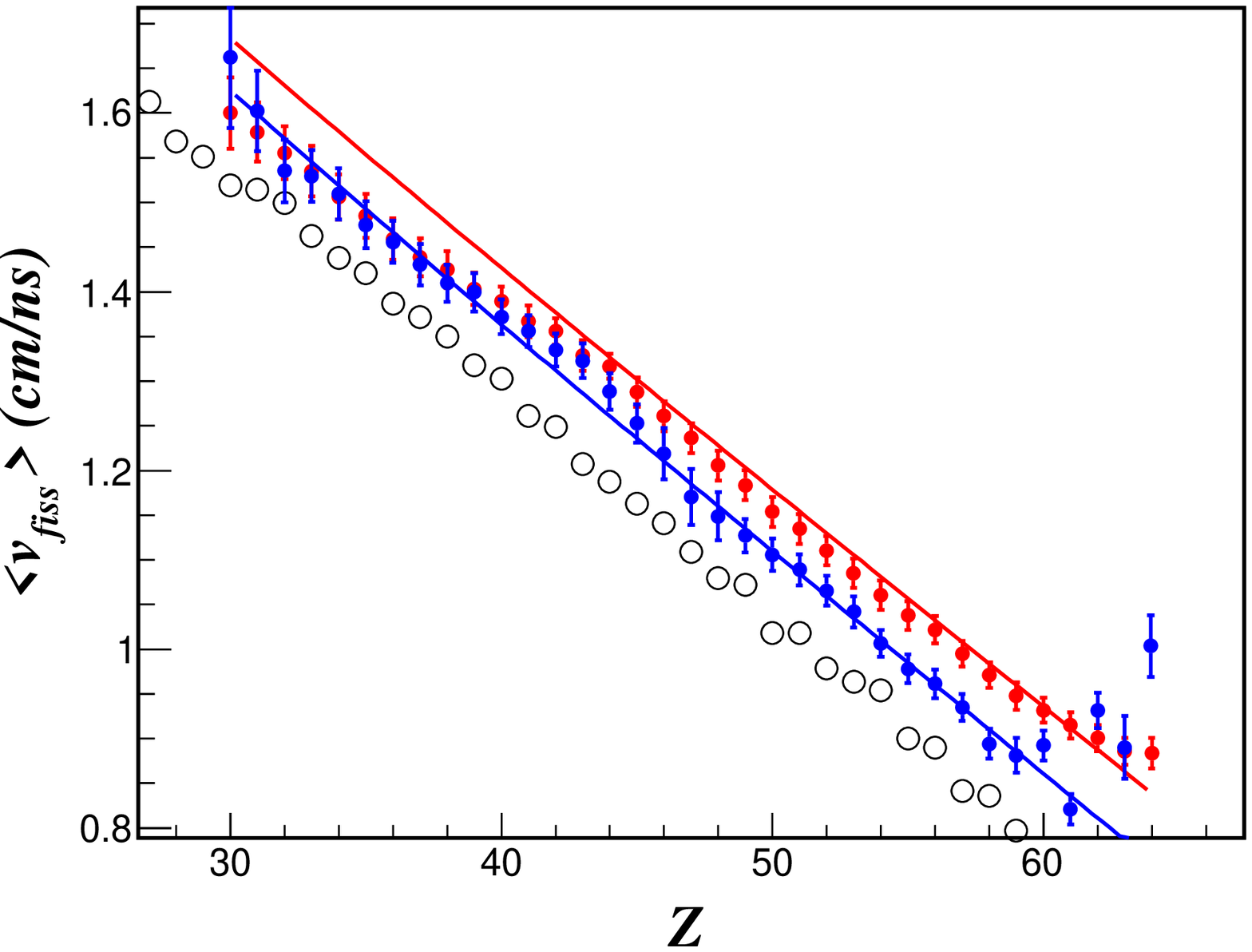}
\caption{\label{vfismoy} Mean fission velocity of the fission fragments in the reference frames of the fissioning systems: $^{250}$Cf in red and $^{240}$Pu in blue. Open symbols correspond to $^{238}$U-induced spallation reaction~\cite{PeB07}. The lines correspond to the estimation of the velocities based on Wilkins prescription~\cite{wilkins, boc08}.}
\end{center}
\end{figure}

The fission kinematics is driven by the strength of the Coulomb repulsion that the fragments experience at the scission point and therefore it gives a hint on the scission configuration in terms of deformation. Because the experimental set-up of the VAMOS focal plane allows for a precise angle reconstruction, the measure of the velocity is performed with a good accuracy in the reference frame of the fissioning nucleus. This is illustrated in Fig.~\ref{velocity sphere}, where the components of the velocity vector of $^{122}$Sn are reconstructed in the reference frame of the $^{250}$Cf fissioning system. The sphere populated by the fission kinematics is recognized within the limits of the spectrometer acceptance. The fission velocity, $v_{fiss}$, of each fragment corresponds to the radius of its particular sphere. Figure \ref{vel_iso_Cf} shows the fission velocity measured for the ensemble of the isotopes produced in the fusion-fission reaction of the present experiment. Similar data for transfer-induced fission were obtained but not shown in the figure for clarity. The bars illustrate the fission-velocity width and not the error on the measurement, which is much smaller. As it can be seen in Fig.~\ref{vel_iso_Cf}, the fragment velocity decreases from 1.6 cm/ns to 0.9 cm/ns as the atomic number of the fragment increases from $_{30}$Zn to $_{64}$Gd isotopes. This is in agreement with what is expected from the momentum conservation during the fission process, where the heavy fragment has a smaller velocity than that of the light fragment. This is well observed when looking at the mean velocity averaged over the isotopic yield as a function of the corresponding atomic number of the fragments, displayed in Fig.~\ref{vfismoy}. In this figure, the two systems $^{240}$Pu and $^{250}$Cf are compared to previous data obtained in spallation reaction of $^{238}$U on deuterium at 1 GeV. The fragments produced in the fission of $^{250}$Cf have a larger velocity than those from $^{240}$Pu, and these are larger than those from $^{238}$U spallation reaction, in agreement with the decreasing Coulomb repulsion. The total kinetic energy can be reproduced by the model of Wilkins et al.~\cite{wilkins}, as shown in Refs.~\cite{boe97,PeB07}. Following the prescription of Wilkins, and considering the momentum conservation, the velocity of the fission fragments at scission can be expressed as:
\begin{align}
&v_{fiss}(Z_1,A_1) = \left( \frac{A_2 Z_1 Z_2 \cdot 2e^2}{A_1 A_{fs}\cdot u} \right)^{1/2}\times \nonumber\\
&\left(\frac{1}{r_0 A_1^{1/3}\left(1+\frac{2}{3}\beta_1\right) + r_0 A_2^{1/3}\left(1+\frac{2}{3}\beta_2\right) + d}\right)^{1/2}
 \label{eq:vfiss}
\end{align}
where $A_i$, $Z_i$ and $\beta_i$ are the mass number, the atomic number and the deformation of the fission fragment $i$, $u$ the unified atomic mass unit, $e$ the elementary charge, $r_0$ the nucleon radius, and $d$ the length of the neck between the two fragments. In the present evaluation, the average mass number $A_1$ and $A_2$ for each atomic number are corrected for the mean neutron evaporation multiplicity, obtained experimentally as explained in Sec.~\ref{sec:neutron_evaporation}. The evolution of the calculated velocity is shown in Fig.~\ref{vfismoy} using the parameters determined in~\cite{boe97}, for Cf (red line) and Pu (blue line) respectively. The prediction follows the mean velocity of the fragments of $^{240}$Pu, while the light fragments of $^{250}$Cf show slightly smaller velocities than expected. This feature has already been reported in~\cite{bernas03}, where the lower velocities of light fission fragments were suggested to result from the contribution of different lighter fissioning systems to the fragment production, inducing smaller velocities. However, in the present experiment, the $^{250}$Cf fissioning system is well defined, as the most part of the scant charge-particle evaporation is ruled out with its detection in SPIDER and only few neutrons are evaporated before scission. A possible contamination of lighter fissioning systems produced in transfer reactions that would not be detected in SPIDER is evaluated to be less than 10\%. With this amount, the contamination from other channels cannot explain the smaller velocities of light fragments. This shows that a deeper investigation on the kinematical properties of the fission fragments is needed.

The same argumentation of the contribution of different fissioning systems was used in~\cite{PeB07} to explain the plateau observed in the evolution of the fission velocities as a function of the fragment mass number for each element (see Fig.~\ref{vel_iso_Cf}). The plateau is unexpected from momentum conservation: a regular decrease would be awaited as the mass of the fragment decreases, considering a constant deformation at scission. 
However, the results obtained in the present reaction, for which the fissioning system is well defined, demonstrates the need for a deeper investigation of the kinematical properties of the fragments that might be linked to a strong variation of the deformation at scission with the neutron excess of the fragments.

\subsection{Post-scission neutron multiplicities}
\label{sec:neutron_evaporation}
The detection of fission fragments happens few hundreds of nanoseconds after the reaction, hence it is not possible to access the fragment distribution at scission. The characterization in mass and atomic number of fragments may be derived from the scission point model~\cite{wilkins}, in which the scission configuration is assumed to be the one that minimizes the total energy of the fissioning nucleus. The potential energy of the fissioning nucleus is calculated as the sum of the potential energy of both fragments with interaction terms relative to Coulomb and nuclear interaction, and, if needed, a centrifugal potential~\cite{Ber79}. Neglecting the shell structure in the expression of the potential energy of the fragments, it is possible to estimate the most probable value of the atomic number $<Z_1>$ for a fragment mass $A_1$, being $Z_1$ the value minimizing the derivative of the total energy with respect to atomic number for a fixed value of mass. This gives the relation~\cite{Ber79}:
\begin{align}
 &<Z_1>\left(\frac{2a_c}{A_1^{1/3}\left(1+\frac{2}{3}\beta_1\right)} + \frac{2a_c}{A_2^{1/3}\left(1+\frac{2}{3}\beta_2\right)} + \frac{4a_a}{A_1}\right.\nonumber\\ 
& \left. + \frac{4a_a}{A_2} - \frac{2e^2}{R}\right) =  Z\left(\frac{2a_c}{A_2^{1/3}\left(1+\frac{2}{3}\beta_2\right)}+\frac{4a_a}{A_2}-\frac{e^2}{R} \right),
 \label{eq:most_probable_Z1}
\end{align}
where $a_a$ and $a_c$ are the liquid-drop parameters for the volume and Coulomb energy terms, $\beta_1$ and $\beta_2$ are the quadrupole deformation at scission of each fragment, and $R$ is the distance between the centers of the two touching deformed fragments and is derived as~\cite{wilkins}:
\begin{equation}
R = r_0 A_1^{1/3}\left(1+\frac{2}{3}\beta_1\right) + r_0 A_2^{1/3}\left(1+\frac{2}{3}\beta_2\right) + d
\end{equation}

\begin{figure}[ht]
\begin{center}
 \includegraphics[width=8cm]{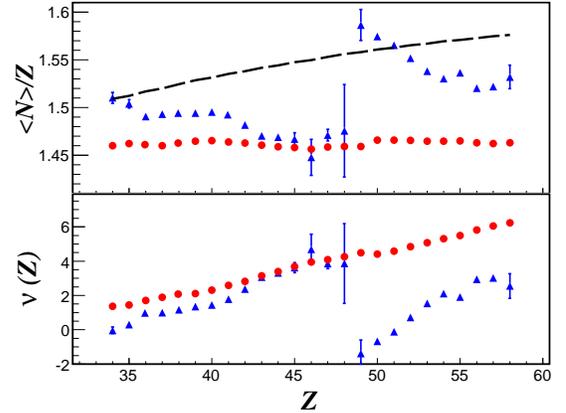}
 \caption{\label{fig:nu_Z} Upper panel: neutron excess $<N>/Z$ expected at the scission point for the $^{240}$Pu (dashed line) compared to the data obtained from the fission of the $^{240,241}$Pu (triangles) and the fission of the $^{250}$Cf (empty circle). Lower panel: difference between the expected $<N>/Z$ value and the measured one for the same system presented above (same symbols).}
\end{center}
\end{figure}

The neutron excess of fragments at scission ${<N>/Z}_{sciss}$ is finally obtained from Eq.~\ref{eq:most_probable_Z1} considering that it is equal to $(A_1-<Z_1>)/<Z_1>$ and that $A_2 = A_{fs} - A_1$ and $Z_2 = Z_{fs} - Z_1$, with $A_{fs}$ and $Z_{fs}$ as the mass and atomic number of the fissioning system. Upper panel of Fig. \ref{fig:nu_Z} shows the calculated ${<N>/Z}_{sciss}$ for the fissioning nucleus $^{240}$Pu as a dashed line. This neutron excess follows naturally a similar trend as the valley of stability, where the heavier nuclei are progressively more neutron rich in order to compensate for the increasing Coulomb repulsion. However, the curve is generally more neutron rich than the valley of stability because the neutron excess of both fragments is bound to the $N/Z$ of the fissioning system. The calculated neutron excess for the fragments of $^{250}$Cf gives very similar results, since both fissioning systems have identical $N/Z$ ratios, and it is not shown for clarity. The experimental neutron excess of the fission fragments from $^{240,241}$Pu and $^{250}$Cf are plotted in the same figure. The difference between the calculated curve and the experimental data can be interepreted as the average multiplicity $\nu(Z)$ of the evaporated neutrons after the separation of fragments, and is displayed for both systems in the lower panel of Fig.~\ref{fig:nu_Z}.

The negative values of the neutron multiplicity can be explained as a consequence of the approximations in the model and from the absence of shell-effect in the calculation of the potential energy. The well known saw-tooth behaviour is observed for the fragments of the $^{240}$Pu fissioning nucleus at an excitation energy of $\sim$ 9 MeV. This pattern is connected to the shell effects in the scission configuration and corresponds to the dissipation of the deformation energy of fragments into intrinsic excitation energy that may lead to the evaporation of neutrons~\cite{wilkins}. In particular, no evaporation is observed for Sn fragments, produced in a spherical shape at scission, while symmetric-mass fragments show a strong deformation associated to the symmetric mode of fission. 
For the fragments produced in fusion-fission reaction at higher excitation energy, a steady increase of the neutron multiplicity with the atomic number of the fragments is observed. This steady increase was already observed in direct kinematic measurements of similar systems~\cite{HiH92}. 

\section{Conclusions}

In the present work, a pioneering method based on inverse kinematics and transfer-induced fission is developed for the investigation of fission-fragment yields. The multi-nucleon transfer reactions give the potential to access different neutron-rich actinides from $^{238}$U to Cm isotopes, with a regime of excitation energy compatible with fast-neutron induced fission. Due to different technical difficulties during the experiment, only the two-proton transfer channel leading to the compound nucleus $^{240}$Pu is investigated. The complete isotopic yields from Se to Nd are accessible for the first time, and their corresponding mass- and atomic-number distributions are in good agreement with previous measurements in direct kinematics, showing the validity of the proposed method. In addition to transfer-fission channels, fusion-fission reactions, leading to the compound nucleus $^{250}$Cf at an excitation energy of 45 MeV, are also presented. The comparison of the two systems sheds light on the evolution of the fission process from a regime strongly influenced by nuclear shell structure to a regime where the macroscopic description of the nucleus prevails, while both regimes are strongly influenced by nuclear dissipation. The neutron excess of the isotopic distributions is used to investigate the influence of nuclear shell structure and the sharing of excitation energy between the fission fragments in both regimes. In addition, the precise measurement of the fragment velocities in the reference frame of the fissioning nucleus allows to get a deep insight into the scission configuration. This wealth of data can be an important step in the development of the description of the fission process, and further work in this direction will be the subject of a forthcoming publication. The present data will be improved in a second iteration of the experiment, already performed, in which a better description of the fissioning systems was achieved~\cite{carme}. In addition, the new experimental setup at the VAMOS spectrometer \cite{rej11} allows for a much more efficient collection of fission fragments. Consequently, the possibilities of systematic studies of the different fissioning systems are expected to increase significantly, including the evolution of the fragment distribution with the excitation energy of the fissioning system. 

The present experimental program has been inspired by the inverse-kinematics experiment performed at relativistic energies at GSI in the 1990s~\cite{sch01}. With respect to this pioneering experiment, the present experimental conditions allow to measure, in addition to the atomic number, the mass of the complete fragment production. The lack of data on the mass of the fragments in the GSI experiment was due to the absence of a spectrometer, while in the present experiment, the large acceptance spectrometer VAMOS shows a sufficiently high resolution to resolve the complete fragment distribution. The potential of the GSI experiment has been impressively improved with a new experimental programme using the ALADIN magnet \cite{sofia}. The relativistic-energy based experimental programme has the advantage to offer a better atomic-number resolution, as the ionic charge-state distributions are much restricted compared to the possible energies at GANIL. In counterpart, the multi-nucleon transfer reaction allows for the investigation of heavier-than-the beam actinides with a precise measurement of their excitation energy.

The present work shows the strong potential that inverse kinematics coupled to a spectrometer brings into the field of fission investigations. The present experimental setup can also serve as a basis for additional systems, such as neutron detection devices. The next generation of fission experiments certainly resides in the fELISE project~\cite{felise} at the future FAIR facility. In order to enrich the variety of fissioning systems for the study with inverse kinematics, the different ISOL projects of exotic-beam production based on fission \cite{spiral2,HIE} should consider the possibility of extracting and accelerating actinides, produced in capture and transfer reactions inside the thick targets, as a unique opportunity to produce heavy and neutron-rich actinide beams.

\section*{Acknowledgements}
This work has been supported by the EURATOM programme under the contract number 44816 and by GEDEPEON (CEA/CNRS/EDF/Areva NP). X.D. and K.-H.S. were in part supported by the Region Basse-Normandie (France). Enlightening discussions with W. Mittig, A. C. Villari, and B. Tamain contributed to characterize the optimal experimental conditions. Also, the excellent support and environment from the GANIL staff in the experimental area is acknowledged.
\appendix
\section{\label{app_cf} Isotopic yields of $^{250}$Cf}
\input{./latex_yields_papier_Cf.tex}
\section{\label{app_pu} Isotopic yields of $^{240,241}$Pu}
\input{./latex_yields_papier_Pu.tex}
\pagebreak

\end{document}

%% file: latex_yields_papier_Cf.tex
\begin{longtable}{| c c c c c c | c c c c c c |}
  \multicolumn{12}{l}
  {Isotopic Yields of $^{250}$Cf} \\
  \hline
 ~ &~ &~ &~ &~ &~ &~ &~ &~ &~ &~ &~\\
 ~ & Z & A & Y(Z,A) & $\epsilon_{stat}$ &~&~& Z & A & Y(Z,A) & $\epsilon_{stat}$ & ~ \\
  \hline
  \endhead
~& 30 & 70 & 0.04 & 0.01 &~&~&
 30 & 71 & 0.05 & 0.01 &~\\~&
 30 & 72 & 0.10 & 0.02 &~&~&
 30 & 73 & 0.18 & 0.03 &~\\~&
 30 & 74 & 0.37 & 0.05 &~&~&
 30 & 75 & 0.33 & 0.04 &~\\~&
 30 & 76 & 0.40 & 0.05 &~&~&
 30 & 77 & 0.30 & 0.04 &~\\~&
 30 & 78 & 0.19 & 0.03 &~&~&
 30 & 79 & 0.13 & 0.03 &~\\~&
 30 & 80 & 0.08 & 0.02 &~&~&
 30 & 81 & 0.04 & 0.02 &~\\~&
 30 & 82 & 0.00 & 0.00 &~&~&
 30 & 85 & 0.03 & 0.02 &~\\~&
 31 & 70 & 0.03 & 0.01 &~&~&
 31 & 71 & 0.03 & 0.01 &~\\~&
 31 & 72 & 0.12 & 0.02 &~&~&
 31 & 73 & 0.25 & 0.04 &~\\~&
 31 & 74 & 0.38 & 0.05 &~&~&
 31 & 75 & 0.69 & 0.08 &~\\~&
 31 & 76 & 0.69 & 0.08 &~&~&
 31 & 77 & 0.93 & 0.10 &~\\~&
 31 & 78 & 0.95 & 0.11 &~&~&
 31 & 79 & 0.68 & 0.08 &~\\~&
 31 & 80 & 0.51 & 0.07 &~&~&
 31 & 81 & 0.39 & 0.06 &~\\~&
 31 & 82 & 0.21 & 0.03 &~&~&
 31 & 83 & 0.12 & 0.03 &~\\~&
 31 & 84 & 0.03 & 0.01 &~&~&
 31 & 85 & 0.02 & 0.01 &~\\~&
 32 & 71 & 0.03 & 0.01 &~&~&
 32 & 72 & 0.05 & 0.01 &~\\~&
 32 & 73 & 0.06 & 0.01 &~&~&
 32 & 74 & 0.15 & 0.02 &~\\~&
 32 & 75 & 0.27 & 0.04 &~&~&
 32 & 76 & 0.75 & 0.09 &~\\~&
 32 & 77 & 1.15 & 0.13 &~&~&
 32 & 78 & 1.57 & 0.17 &~\\~&
 32 & 79 & 1.87 & 0.20 &~&~&
 32 & 80 & 1.63 & 0.17 &~\\~&
 32 & 81 & 1.33 & 0.15 &~&~&
 32 & 82 & 0.86 & 0.10 &~\\~&
 32 & 83 & 0.66 & 0.08 &~&~&
 32 & 84 & 0.33 & 0.05 &~\\~&
 32 & 85 & 0.22 & 0.04 &~&~&
 32 & 86 & 0.09 & 0.02 &~\\~&
 32 & 87 & 0.03 & 0.01 &~&~&
 33 & 73 & 0.04 & 0.01 &~\\~&
 33 & 74 & 0.05 & 0.01 &~&~&
 33 & 75 & 0.07 & 0.02 &~\\~&
 33 & 76 & 0.19 & 0.03 &~&~&
 33 & 77 & 0.39 & 0.05 &~\\~&
 33 & 78 & 0.73 & 0.08 &~&~&
 33 & 79 & 1.29 & 0.14 &~\\~&
 33 & 80 & 2.41 & 0.25 &~&~&
 33 & 81 & 2.64 & 0.27 &~\\~&
 33 & 82 & 2.57 & 0.27 &~&~&
 33 & 83 & 2.23 & 0.23 &~\\~&
 33 & 84 & 1.44 & 0.16 &~&~&
 33 & 85 & 1.08 & 0.12 &~\\~&
 33 & 86 & 0.58 & 0.07 &~&~&
 33 & 87 & 0.28 & 0.04 &~\\~&
 33 & 88 & 0.14 & 0.02 &~&~&
 33 & 89 & 0.05 & 0.01 &~\\~&
 34 & 73 & 0.07 & 0.02 &~&~&
 34 & 75 & 0.05 & 0.01 &~\\~&
 34 & 76 & 0.05 & 0.01 &~&~&
 34 & 77 & 0.13 & 0.02 &~\\~&
 34 & 78 & 0.22 & 0.03 &~&~&
 34 & 79 & 0.35 & 0.04 &~\\~& 
 34 & 78 & 0.01 & 0.00 &~&~&
 34 & 79 & 0.02 & 0.00 &~\\~&
 34 & 80 & 0.04 & 0.00 &~&~&
 34 & 81 & 0.09 & 0.01 &~\\~&
 34 & 82 & 0.15 & 0.02 &~&~&
 34 & 83 & 0.18 & 0.02 &~\\~&
 34 & 84 & 0.16 & 0.02 &~&~&
 34 & 85 & 0.11 & 0.01 &~\\~&
 34 & 86 & 0.08 & 0.01 &~&~&
 34 & 87 & 0.04 & 0.01 &~\\~&
 34 & 88 & 0.02 & 0.00 &~&~&
 34 & 89 & 0.01 & 0.00 &~\\~&
 34 & 90 & 0.01 & 0.00 &~&~&
 35 & 79 & 0.00 & 0.00 &~\\~&
 35 & 80 & 0.01 & 0.00 &~&~&
 35 & 81 & 0.02 & 0.00 &~\\~&
 35 & 82 & 0.04 & 0.00 &~&~&
 35 & 83 & 0.08 & 0.01 &~\\~&
 35 & 84 & 0.17 & 0.02 &~&~&
 35 & 85 & 0.25 & 0.03 &~\\~&
 35 & 86 & 0.22 & 0.02 &~&~&
 35 & 87 & 0.19 & 0.02 &~\\~&
 35 & 88 & 0.12 & 0.01 &~&~&
 35 & 89 & 0.08 & 0.01 &~\\~&
 35 & 90 & 0.04 & 0.00 &~&~&
 35 & 91 & 0.03 & 0.00 &~\\~&
 35 & 92 & 0.02 & 0.00 &~&~&
 35 & 93 & 0.01 & 0.00 &~\\~&
 36 & 81 & 0.00 & 0.00 &~&~&
 36 & 82 & 0.01 & 0.00 &~\\~&
 36 & 83 & 0.02 & 0.00 &~&~&
 36 & 84 & 0.04 & 0.00 &~\\~&
 36 & 85 & 0.11 & 0.01 &~&~&
 36 & 86 & 0.17 & 0.02 &~\\~&
 36 & 87 & 0.27 & 0.03 &~&~&
 36 & 88 & 0.35 & 0.04 &~\\~&
 36 & 89 & 0.24 & 0.03 &~&~&
 36 & 90 & 0.21 & 0.02 &~\\~&
 36 & 91 & 0.13 & 0.01 &~&~&
 36 & 92 & 0.08 & 0.01 &~\\~&
 36 & 93 & 0.04 & 0.00 &~&~&
 36 & 94 & 0.02 & 0.00 &~\\~&
 36 & 95 & 0.01 & 0.00 &~&~&
 36 & 96 & 0.01 & 0.00 &~\\~&
 37 & 83 & 0.01 & 0.00 &~&~&
 37 & 84 & 0.01 & 0.00 &~\\~&
 37 & 85 & 0.02 & 0.00 &~&~&
 37 & 86 & 0.05 & 0.01 &~\\~&
 37 & 87 & 0.11 & 0.01 &~&~&
 37 & 88 & 0.19 & 0.02 &~\\~&
 37 & 89 & 0.32 & 0.04 &~&~&
 37 & 90 & 0.40 & 0.04 &~\\~&
 37 & 91 & 0.44 & 0.05 &~&~&
 37 & 92 & 0.33 & 0.04 &~\\~&
 37 & 93 & 0.24 & 0.03 &~&~&
 37 & 94 & 0.14 & 0.02 &~\\~&
 37 & 95 & 0.08 & 0.01 &~&~&
 37 & 96 & 0.04 & 0.01 &~\\~&
 37 & 97 & 0.03 & 0.00 &~&~&
 37 & 98 & 0.01 & 0.00 &~\\~&
 37 & 99 & 0.01 & 0.00 &~&~&
 38 & 86 & 0.01 & 0.00 &~\\~&
 38 & 87 & 0.03 & 0.00 &~&~&
 38 & 88 & 0.05 & 0.01 &~\\~&
 38 & 89 & 0.10 & 0.01 &~&~&
 38 & 90 & 0.20 & 0.02 &~\\~&
 38 & 91 & 0.34 & 0.04 &~&~&
 38 & 92 & 0.51 & 0.06 &~\\~&
 38 & 93 & 0.60 & 0.07 &~&~&
 38 & 94 & 0.57 & 0.06 &~\\~&
 38 & 95 & 0.43 & 0.05 &~&~&
 38 & 96 & 0.29 & 0.03 &~\\~&
 38 & 97 & 0.16 & 0.02 &~&~&
 38 & 98 & 0.09 & 0.01 &~\\~&
 38 & 99 & 0.05 & 0.01 &~&~&
 38 & 100 & 0.03 & 0.00 &~\\~&
 38 & 101 & 0.02 & 0.00 &~&~&
 38 & 102 & 0.01 & 0.00 &~\\~&
 39 & 88 & 0.01 & 0.00 &~&~&
 39 & 89 & 0.02 & 0.00 &~\\~&
 39 & 90 & 0.04 & 0.00 &~&~&
 39 & 91 & 0.08 & 0.01 &~\\~&
 39 & 92 & 0.16 & 0.02 &~&~&
 39 & 93 & 0.31 & 0.03 &~\\~&
 39 & 94 & 0.52 & 0.06 &~&~&
 39 & 95 & 0.70 & 0.08 &~\\~&
 39 & 96 & 0.72 & 0.08 &~&~&
 39 & 97 & 0.64 & 0.07 &~\\~&
 39 & 98 & 0.44 & 0.05 &~&~&
 39 & 99 & 0.29 & 0.03 &~\\~&
 39 & 100 & 0.15 & 0.02 &~&~&
 39 & 101 & 0.09 & 0.01 &~\\~&
 39 & 102 & 0.05 & 0.01 &~&~&
 39 & 103 & 0.02 & 0.00 &~\\~&
 39 & 104 & 0.01 & 0.00 &~&~&
 39 & 105 & 0.00 & 0.00 &~\\~&
 40 & 90 & 0.01 & 0.00 &~&~&
 40 & 91 & 0.02 & 0.00 &~\\~&
 40 & 92 & 0.03 & 0.00 &~&~&
 40 & 93 & 0.07 & 0.01 &~\\~&
 40 & 94 & 0.15 & 0.02 &~&~&
 40 & 95 & 0.30 & 0.03 &~\\~&
 40 & 96 & 0.53 & 0.06 &~&~&
 40 & 97 & 0.79 & 0.09 &~\\~&
 40 & 98 & 0.98 & 0.11 &~&~&
 40 & 99 & 0.86 & 0.10 &~\\~&
 40 & 100 & 0.77 & 0.09 &~&~&
 40 & 101 & 0.45 & 0.05 &~\\~&
 40 & 102 & 0.27 & 0.03 &~&~&
 40 & 103 & 0.14 & 0.02 &~\\~&
 40 & 104 & 0.08 & 0.01 &~&~&
 40 & 105 & 0.05 & 0.01 &~\\~&
 40 & 106 & 0.02 & 0.00 &~&~&
 40 & 107 & 0.01 & 0.00 &~\\~&
 41 & 93 & 0.02 & 0.00 &~&~&
 41 & 94 & 0.04 & 0.00 &~\\~&
 41 & 95 & 0.06 & 0.01 &~&~&
 41 & 96 & 0.12 & 0.01 &~\\~&
 41 & 97 & 0.25 & 0.03 &~&~&
 41 & 98 & 0.49 & 0.05 &~\\~&
 41 & 99 & 0.82 & 0.09 &~&~&
 41 & 100 & 1.01 & 0.11 &~\\~&
 41 & 101 & 1.24 & 0.14 &~&~&
 41 & 102 & 1.03 & 0.11 &~\\~&
 41 & 103 & 0.75 & 0.08 &~&~&
 41 & 104 & 0.40 & 0.05 &~\\~&
 41 & 105 & 0.24 & 0.03 &~&~&
 41 & 106 & 0.13 & 0.01 &~\\~&
 41 & 107 & 0.07 & 0.01 &~&~&
 41 & 108 & 0.04 & 0.00 &~\\~&
 41 & 109 & 0.02 & 0.00 &~&~&
 41 & 110 & 0.01 & 0.00 &~\\~&
 42 & 95 & 0.02 & 0.00 &~&~&
 42 & 96 & 0.03 & 0.00 &~\\~&
 42 & 97 & 0.05 & 0.01 &~&~&
 42 & 98 & 0.10 & 0.01 &~\\~&
 42 & 99 & 0.23 & 0.03 &~&~&
 42 & 100 & 0.48 & 0.05 &~\\~&
 42 & 101 & 0.66 & 0.07 &~&~&
 42 & 102 & 1.23 & 0.14 &~\\~&
 42 & 103 & 1.39 & 0.15 &~&~&
 42 & 104 & 1.34 & 0.15 &~\\~&
 42 & 105 & 0.93 & 0.10 &~&~&
 42 & 106 & 0.59 & 0.07 &~\\~&
 42 & 107 & 0.32 & 0.04 &~&~&
 42 & 108 & 0.19 & 0.02 &~\\~&
 42 & 109 & 0.11 & 0.01 &~&~&
 42 & 110 & 0.05 & 0.01 &~\\~&
 42 & 111 & 0.03 & 0.00 &~&~&
 42 & 112 & 0.01 & 0.00 &~\\~&
 42 & 113 & 0.01 & 0.00 &~&~&
 43 & 97 & 0.02 & 0.00 &~\\~&
 43 & 98 & 0.03 & 0.00 &~&~&
 43 & 99 & 0.05 & 0.01 &~\\~&
 43 & 100 & 0.09 & 0.01 &~&~&
 43 & 101 & 0.17 & 0.02 &~\\~&
 43 & 102 & 0.36 & 0.04 &~&~&
 43 & 103 & 0.69 & 0.08 &~\\~&
 43 & 104 & 1.12 & 0.12 &~&~&
 43 & 105 & 1.47 & 0.16 &~\\~&
 43 & 106 & 1.46 & 0.16 &~&~&
 43 & 107 & 1.22 & 0.14 &~\\~&
 43 & 108 & 0.79 & 0.09 &~&~&
 43 & 109 & 0.53 & 0.06 &~\\~&
 43 & 110 & 0.27 & 0.03 &~&~&
 43 & 111 & 0.16 & 0.02 &~\\~&
 43 & 112 & 0.09 & 0.01 &~&~&
 43 & 113 & 0.04 & 0.01 &~\\~&
 43 & 114 & 0.02 & 0.00 &~&~&
 43 & 115 & 0.01 & 0.00 &~\\~&
 44 & 100 & 0.03 & 0.00 &~&~&
 44 & 101 & 0.05 & 0.01 &~\\~&
 44 & 102 & 0.08 & 0.01 &~&~&
 44 & 103 & 0.18 & 0.02 &~\\~&
 44 & 104 & 0.36 & 0.04 &~&~&
 44 & 105 & 0.56 & 0.06 &~\\~&
 44 & 106 & 1.14 & 0.13 &~&~&
 44 & 107 & 1.51 & 0.17 &~\\~&
 44 & 108 & 1.73 & 0.19 &~&~&
 44 & 109 & 1.49 & 0.17 &~\\~&
 44 & 110 & 1.13 & 0.13 &~&~&
 44 & 111 & 0.70 & 0.08 &~\\~&
 44 & 112 & 0.41 & 0.05 &~&~&
 44 & 113 & 0.24 & 0.03 &~\\~&
 44 & 114 & 0.12 & 0.01 &~&~&
 44 & 115 & 0.06 & 0.01 &~\\~&
 44 & 116 & 0.03 & 0.00 &~&~&
 44 & 117 & 0.01 & 0.00 &~\\~&
 44 & 118 & 0.01 & 0.00 &~&~&
 45 & 102 & 0.02 & 0.00 &~\\~&
 45 & 103 & 0.04 & 0.00 &~&~&
 45 & 104 & 0.06 & 0.01 &~\\~&
 45 & 105 & 0.12 & 0.01 &~&~&
 45 & 106 & 0.23 & 0.03 &~\\~&
 45 & 107 & 0.47 & 0.05 &~&~&
 45 & 108 & 0.81 & 0.09 &~\\~&
 45 & 109 & 1.25 & 0.14 &~&~&
 45 & 110 & 1.52 & 0.17 &~\\~&
 45 & 111 & 1.63 & 0.18 &~&~&
 45 & 112 & 1.22 & 0.14 &~\\~&
 45 & 113 & 0.88 & 0.10 &~&~&
 45 & 114 & 0.52 & 0.06 &~\\~&
 45 & 115 & 0.29 & 0.03 &~&~&
 45 & 116 & 0.16 & 0.02 &~\\~&
 45 & 117 & 0.08 & 0.01 &~&~&
 45 & 118 & 0.04 & 0.01 &~\\~&
 45 & 119 & 0.02 & 0.00 &~&~&
 45 & 120 & 0.01 & 0.00 &~\\~&
 46 & 104 & 0.02 & 0.00 &~&~&
 46 & 105 & 0.03 & 0.00 &~\\~&
 46 & 106 & 0.06 & 0.01 &~&~&
 46 & 107 & 0.11 & 0.01 &~\\~&
 46 & 108 & 0.21 & 0.02 &~&~&
 46 & 109 & 0.41 & 0.05 &~\\~&
 46 & 110 & 0.77 & 0.09 &~&~&
 46 & 111 & 1.21 & 0.13 &~\\~&
 46 & 112 & 1.66 & 0.18 &~&~&
 46 & 113 & 1.67 & 0.19 &~\\~&
 46 & 114 & 1.64 & 0.18 &~&~&
 46 & 115 & 1.18 & 0.13 &~\\~&
 46 & 116 & 0.79 & 0.09 &~&~&
 46 & 117 & 0.48 & 0.05 &~\\~&
 46 & 118 & 0.28 & 0.03 &~&~&
 46 & 119 & 0.14 & 0.02 &~\\~&
 46 & 120 & 0.08 & 0.01 &~&~&
 46 & 121 & 0.04 & 0.00 &~\\~&
 46 & 122 & 0.02 & 0.00 &~&~&
 47 & 107 & 0.03 & 0.00 &~\\~&
 47 & 108 & 0.05 & 0.01 &~&~&
 47 & 109 & 0.08 & 0.01 &~\\~&
 47 & 110 & 0.16 & 0.02 &~&~&
 47 & 111 & 0.28 & 0.03 &~\\~&
 47 & 112 & 0.52 & 0.06 &~&~&
 47 & 113 & 0.92 & 0.10 &~\\~&
 47 & 114 & 1.30 & 0.14 &~&~&
 47 & 115 & 1.66 & 0.18 &~\\~&
 47 & 116 & 1.66 & 0.18 &~&~&
 47 & 117 & 1.43 & 0.16 &~\\~&
 47 & 118 & 1.01 & 0.11 &~&~&
 47 & 119 & 0.65 & 0.07 &~\\~&
 47 & 120 & 0.40 & 0.04 &~&~&
 47 & 121 & 0.21 & 0.02 &~\\~&
 47 & 122 & 0.11 & 0.01 &~&~&
 47 & 123 & 0.06 & 0.01 &~\\~&
 47 & 124 & 0.03 & 0.00 &~&~&
 47 & 125 & 0.01 & 0.00 &~\\~&
 47 & 126 & 0.01 & 0.00 &~&~&
 48 & 109 & 0.02 & 0.00 &~\\~&
 48 & 110 & 0.03 & 0.00 &~&~&
 48 & 111 & 0.06 & 0.01 &~\\~&
 48 & 112 & 0.11 & 0.01 &~&~&
 48 & 113 & 0.21 & 0.02 &~\\~&
 48 & 114 & 0.39 & 0.04 &~&~&
 48 & 115 & 0.70 & 0.08 &~\\~&
 48 & 116 & 1.10 & 0.12 &~&~&
 48 & 117 & 1.50 & 0.17 &~\\~&
 48 & 118 & 1.66 & 0.19 &~&~&
 48 & 119 & 1.50 & 0.17 &~\\~&
 48 & 120 & 1.21 & 0.14 &~&~&
 48 & 121 & 0.84 & 0.09 &~\\~&
 48 & 122 & 0.52 & 0.06 &~&~&
 48 & 123 & 0.31 & 0.03 &~\\~&
 48 & 124 & 0.17 & 0.02 &~&~&
 48 & 125 & 0.09 & 0.01 &~\\~&
 48 & 126 & 0.05 & 0.01 &~&~&
 48 & 127 & 0.02 & 0.00 &~\\~&
 49 & 112 & 0.03 & 0.00 &~&~&
 49 & 113 & 0.05 & 0.01 &~\\~&
 49 & 114 & 0.10 & 0.01 &~&~&
 49 & 115 & 0.17 & 0.02 &~\\~&
 49 & 116 & 0.31 & 0.03 &~&~&
 49 & 117 & 0.55 & 0.06 &~\\~&
 49 & 118 & 0.93 & 0.10 &~&~&
 49 & 119 & 1.27 & 0.14 &~\\~&
 49 & 120 & 1.62 & 0.18 &~&~&
 49 & 121 & 1.64 & 0.18 &~\\~&
 49 & 122 & 1.38 & 0.15 &~&~&
 49 & 123 & 1.04 & 0.12 &~\\~&
 49 & 124 & 0.68 & 0.08 &~&~&
 49 & 125 & 0.40 & 0.04 &~\\~&
 49 & 126 & 0.22 & 0.03 &~&~&
 49 & 127 & 0.13 & 0.01 &~\\~&
 49 & 128 & 0.06 & 0.01 &~&~&
 49 & 129 & 0.04 & 0.00 &~\\~&
 49 & 130 & 0.02 & 0.00 &~&~&
 49 & 131 & 0.01 & 0.00 &~\\~&
 50 & 114 & 0.02 & 0.00 &~&~&
 50 & 115 & 0.04 & 0.00 &~\\~&
 50 & 116 & 0.06 & 0.01 &~&~&
 50 & 117 & 0.13 & 0.01 &~\\~&
 50 & 118 & 0.23 & 0.03 &~&~&
 50 & 119 & 0.41 & 0.05 &~\\~&
 50 & 120 & 0.73 & 0.08 &~&~&
 50 & 121 & 1.11 & 0.12 &~\\~&
 50 & 122 & 1.48 & 0.16 &~&~&
 50 & 123 & 1.69 & 0.19 &~\\~&
 50 & 124 & 1.58 & 0.18 &~&~&
 50 & 125 & 1.27 & 0.14 &~\\~&
 50 & 126 & 0.90 & 0.10 &~&~&
 50 & 127 & 0.60 & 0.07 &~\\~&
 50 & 128 & 0.37 & 0.04 &~&~&
 50 & 129 & 0.22 & 0.02 &~\\~&
 50 & 130 & 0.12 & 0.01 &~&~&
 50 & 131 & 0.07 & 0.01 &~\\~&
 50 & 132 & 0.04 & 0.01 &~&~&
 50 & 133 & 0.03 & 0.00 &~\\~&
 51 & 116 & 0.02 & 0.00 &~&~&
 51 & 117 & 0.03 & 0.00 &~\\~&
 51 & 118 & 0.05 & 0.01 &~&~&
 51 & 119 & 0.09 & 0.01 &~\\~&
 51 & 120 & 0.16 & 0.02 &~&~&
 51 & 121 & 0.29 & 0.03 &~\\~&
 51 & 122 & 0.50 & 0.06 &~&~&
 51 & 123 & 0.81 & 0.09 &~\\~&
 51 & 124 & 1.14 & 0.13 &~&~&
 51 & 125 & 1.41 & 0.16 &~\\~&
 51 & 126 & 1.44 & 0.16 &~&~&
 51 & 127 & 1.29 & 0.14 &~\\~&
 51 & 128 & 0.95 & 0.11 &~&~&
 51 & 129 & 0.69 & 0.08 &~\\~&
 51 & 130 & 0.42 & 0.05 &~&~&
 51 & 131 & 0.27 & 0.03 &~\\~&
 51 & 132 & 0.16 & 0.02 &~&~&
 51 & 133 & 0.10 & 0.01 &~\\~&
 51 & 134 & 0.05 & 0.01 &~&~&
 51 & 135 & 0.03 & 0.00 &~\\~&
 52 & 118 & 0.02 & 0.00 &~&~&
 52 & 119 & 0.03 & 0.00 &~\\~&
 52 & 120 & 0.04 & 0.00 &~&~&
 52 & 121 & 0.07 & 0.01 &~\\~&
 52 & 122 & 0.13 & 0.02 &~&~&
 52 & 123 & 0.25 & 0.03 &~\\~&
 52 & 124 & 0.41 & 0.05 &~&~&
 52 & 125 & 0.68 & 0.08 &~\\~&
 52 & 126 & 1.03 & 0.11 &~&~&
 52 & 127 & 1.38 & 0.15 &~\\~&
 52 & 128 & 1.47 & 0.16 &~&~&
 52 & 129 & 1.45 & 0.16 &~\\~&
 52 & 130 & 1.11 & 0.12 &~&~&
 52 & 131 & 0.82 & 0.09 &~\\~&
 52 & 132 & 0.54 & 0.06 &~&~&
 52 & 133 & 0.34 & 0.04 &~\\~&
 52 & 134 & 0.21 & 0.02 &~&~&
 52 & 135 & 0.12 & 0.01 &~\\~&
 52 & 136 & 0.07 & 0.01 &~&~&
 52 & 137 & 0.04 & 0.00 &~\\~&
 52 & 138 & 0.02 & 0.00 &~&~&
 53 & 122 & 0.03 & 0.00 &~\\~&
 53 & 123 & 0.06 & 0.01 &~&~&
 53 & 124 & 0.12 & 0.01 &~\\~&
 53 & 125 & 0.21 & 0.02 &~&~&
 53 & 126 & 0.38 & 0.04 &~\\~&
 53 & 127 & 0.59 & 0.07 &~&~&
 53 & 128 & 0.84 & 0.09 &~\\~&
 53 & 129 & 1.31 & 0.15 &~&~&
 53 & 130 & 1.52 & 0.17 &~\\~&
 53 & 131 & 1.62 & 0.18 &~&~&
 53 & 132 & 1.39 & 0.15 &~\\~&
 53 & 133 & 1.07 & 0.12 &~&~&
 53 & 134 & 0.71 & 0.08 &~\\~&
 53 & 135 & 0.45 & 0.05 &~&~&
 53 & 136 & 0.26 & 0.03 &~\\~&
 53 & 137 & 0.15 & 0.02 &~&~&
 53 & 138 & 0.09 & 0.01 &~\\~&
 53 & 139 & 0.05 & 0.01 &~&~&
 53 & 140 & 0.03 & 0.00 &~\\~&
 53 & 141 & 0.02 & 0.00 &~&~&
 54 & 123 & 0.02 & 0.00 &~\\~&
 54 & 124 & 0.03 & 0.00 &~&~&
 54 & 125 & 0.04 & 0.01 &~\\~&
 54 & 126 & 0.08 & 0.01 &~&~&
 54 & 127 & 0.13 & 0.01 &~\\~&
 54 & 128 & 0.24 & 0.03 &~&~&
 54 & 129 & 0.39 & 0.04 &~\\~&
 54 & 130 & 0.66 & 0.07 &~&~&
 54 & 131 & 0.99 & 0.11 &~\\~&
 54 & 132 & 1.32 & 0.15 &~&~&
 54 & 133 & 1.50 & 0.17 &~\\~&
 54 & 134 & 1.41 & 0.16 &~&~&
 54 & 135 & 1.14 & 0.13 &~\\~&
 54 & 136 & 0.80 & 0.09 &~&~&
 54 & 137 & 0.52 & 0.06 &~\\~&
 54 & 138 & 0.31 & 0.03 &~&~&
 54 & 139 & 0.18 & 0.02 &~\\~&
 54 & 140 & 0.10 & 0.01 &~&~&
 54 & 141 & 0.06 & 0.01 &~\\~&
 54 & 142 & 0.03 & 0.00 &~&~&
 54 & 143 & 0.02 & 0.00 &~\\~&
 55 & 126 & 0.02 & 0.00 &~&~&
 55 & 127 & 0.03 & 0.00 &~\\~&
 55 & 128 & 0.05 & 0.01 &~&~&
 55 & 129 & 0.09 & 0.01 &~\\~&
 55 & 130 & 0.19 & 0.02 &~&~&
 55 & 131 & 0.31 & 0.03 &~\\~&
 55 & 132 & 0.46 & 0.05 &~&~&
 55 & 133 & 0.78 & 0.09 &~\\~&
 55 & 134 & 1.05 & 0.12 &~&~&
 55 & 135 & 1.33 & 0.15 &~\\~&
 55 & 136 & 1.36 & 0.15 &~&~&
 55 & 137 & 1.21 & 0.13 &~\\~&
 55 & 138 & 0.83 & 0.09 &~&~&
 55 & 139 & 0.56 & 0.06 &~\\~&
 55 & 140 & 0.35 & 0.04 &~&~&
 55 & 141 & 0.22 & 0.02 &~\\~&
 55 & 142 & 0.12 & 0.01 &~&~&
 55 & 143 & 0.07 & 0.01 &~\\~&
 55 & 144 & 0.04 & 0.00 &~&~&
 55 & 145 & 0.02 & 0.00 &~\\~&
 55 & 146 & 0.01 & 0.00 &~&~&
 56 & 130 & 0.03 & 0.00 &~\\~&
 56 & 131 & 0.07 & 0.01 &~&~&
 56 & 132 & 0.12 & 0.01 &~\\~&
 56 & 133 & 0.21 & 0.02 &~&~&
 56 & 134 & 0.36 & 0.04 &~\\~&
 56 & 135 & 0.55 & 0.06 &~&~&
 56 & 136 & 0.80 & 0.09 &~\\~&
 56 & 137 & 1.06 & 0.12 &~&~&
 56 & 138 & 1.17 & 0.13 &~\\~&
 56 & 139 & 1.04 & 0.12 &~&~&
 56 & 140 & 0.78 & 0.09 &~\\~&
 56 & 141 & 0.54 & 0.06 &~&~&
 56 & 142 & 0.40 & 0.04 &~\\~&
 56 & 143 & 0.23 & 0.03 &~&~&
 56 & 144 & 0.13 & 0.02 &~\\~&
 56 & 145 & 0.07 & 0.01 &~&~&
 56 & 146 & 0.04 & 0.01 &~\\~&
 56 & 147 & 0.02 & 0.00 &~&~&
 56 & 148 & 0.01 & 0.00 &~\\~&
 57 & 131 & 0.02 & 0.00 &~&~&
 57 & 132 & 0.03 & 0.00 &~\\~&
 57 & 133 & 0.05 & 0.01 &~&~&
 57 & 134 & 0.08 & 0.01 &~\\~&
 57 & 135 & 0.17 & 0.02 &~&~&
 57 & 136 & 0.27 & 0.03 &~\\~&
 57 & 137 & 0.42 & 0.05 &~&~&
 57 & 138 & 0.66 & 0.07 &~\\~&
 57 & 139 & 0.86 & 0.10 &~&~&
 57 & 140 & 0.95 & 0.11 &~\\~&
 57 & 141 & 1.01 & 0.11 &~&~&
 57 & 142 & 0.88 & 0.10 &~\\~&
 57 & 143 & 0.65 & 0.07 &~&~&
 57 & 144 & 0.47 & 0.05 &~\\~&
 57 & 145 & 0.30 & 0.03 &~&~&
 57 & 146 & 0.17 & 0.02 &~\\~&
 57 & 147 & 0.09 & 0.01 &~&~&
 57 & 148 & 0.05 & 0.01 &~\\~&
 57 & 149 & 0.02 & 0.00 &~&~&
 57 & 150 & 0.01 & 0.00 &~\\~&
 58 & 134 & 0.02 & 0.00 &~&~&
 58 & 135 & 0.03 & 0.00 &~\\~&
 58 & 136 & 0.05 & 0.01 &~&~&
 58 & 137 & 0.11 & 0.01 &~\\~&
 58 & 138 & 0.17 & 0.02 &~&~&
 58 & 139 & 0.25 & 0.03 &~\\~&
 58 & 140 & 0.41 & 0.05 &~&~&
 58 & 141 & 0.62 & 0.07 &~\\~&
 58 & 142 & 0.73 & 0.08 &~&~&
 58 & 143 & 0.67 & 0.08 &~\\~&
 58 & 144 & 0.84 & 0.09 &~&~&
 58 & 145 & 0.65 & 0.07 &~\\~&
 58 & 146 & 0.47 & 0.05 &~&~&
 58 & 147 & 0.31 & 0.04 &~\\~&
 58 & 148 & 0.18 & 0.02 &~&~&
 58 & 149 & 0.10 & 0.01 &~\\~&
 58 & 150 & 0.05 & 0.01 &~&~&
 58 & 151 & 0.02 & 0.00 &~\\~&
 58 & 152 & 0.01 & 0.00 &~&~&
 59 & 137 & 0.02 & 0.00 &~\\~&
 59 & 138 & 0.03 & 0.00 &~&~&
 59 & 139 & 0.06 & 0.01 &~\\~&
 59 & 140 & 0.10 & 0.01 &~&~&
 59 & 141 & 0.18 & 0.02 &~\\~&
 59 & 142 & 0.23 & 0.03 &~&~&
 59 & 143 & 0.34 & 0.04 &~\\~&
 59 & 144 & 0.46 & 0.05 &~&~&
 59 & 145 & 0.61 & 0.07 &~\\~&
 59 & 146 & 0.60 & 0.07 &~&~&
 59 & 147 & 0.59 & 0.07 &~\\~&
 59 & 148 & 0.39 & 0.04 &~&~&
 59 & 149 & 0.26 & 0.03 &~\\~&
 59 & 150 & 0.19 & 0.02 &~&~&
 59 & 151 & 0.09 & 0.01 &~\\~&
 59 & 152 & 0.05 & 0.01 &~&~&
 59 & 153 & 0.02 & 0.00 &~\\~&
 60 & 139 & 0.01 & 0.00 &~&~&
 60 & 140 & 0.02 & 0.00 &~\\~&
 60 & 141 & 0.04 & 0.00 &~&~&
 60 & 142 & 0.06 & 0.01 &~\\~&
 60 & 143 & 0.11 & 0.01 &~&~&
 60 & 144 & 0.16 & 0.02 &~\\~&
 60 & 145 & 0.26 & 0.03 &~&~&
 60 & 146 & 0.36 & 0.04 &~\\~&
 60 & 147 & 0.45 & 0.05 &~&~&
 60 & 148 & 0.51 & 0.06 &~\\~&
 60 & 149 & 0.47 & 0.05 &~&~&
 60 & 150 & 0.42 & 0.05 &~\\~&
 60 & 151 & 0.29 & 0.03 &~&~&
 60 & 152 & 0.17 & 0.02 &~\\~&
 60 & 153 & 0.09 & 0.01 &~&~&
 60 & 154 & 0.05 & 0.01 &~\\~&
 60 & 155 & 0.02 & 0.00 &~&~&
 60 & 156 & 0.01 & 0.00 &~\\~&
 61 & 142 & 0.01 & 0.00 &~&~&
 61 & 143 & 0.02 & 0.00 &~\\~&
 61 & 144 & 0.03 & 0.00 &~&~&
 61 & 145 & 0.06 & 0.01 &~\\~&
 61 & 146 & 0.10 & 0.01 &~&~&
 61 & 147 & 0.15 & 0.02 &~\\~&
 61 & 148 & 0.20 & 0.02 &~&~&
 61 & 149 & 0.28 & 0.03 &~\\~&
 61 & 150 & 0.30 & 0.03 &~&~&
 61 & 151 & 0.33 & 0.04 &~\\~&
 61 & 152 & 0.29 & 0.03 &~&~&
 61 & 153 & 0.19 & 0.02 &~\\~&
 61 & 154 & 0.12 & 0.01 &~&~&
 61 & 155 & 0.07 & 0.01 &~\\~&
 61 & 156 & 0.03 & 0.00 &~&~&
 61 & 157 & 0.02 & 0.00 &~\\~&
 61 & 158 & 0.01 & 0.00 &~&~&
 62 & 144 & 0.01 & 0.00 &~\\~&
 62 & 145 & 0.01 & 0.00 &~&~&
 62 & 146 & 0.03 & 0.00 &~\\~&
 62 & 147 & 0.05 & 0.01 &~&~&
 62 & 148 & 0.07 & 0.01 &~\\~&
 62 & 149 & 0.10 & 0.01 &~&~&
 62 & 150 & 0.17 & 0.02 &~\\~&
 62 & 151 & 0.21 & 0.02 &~&~&
 62 & 152 & 0.23 & 0.03 &~\\~&
 62 & 153 & 0.21 & 0.02 &~&~&
 62 & 154 & 0.18 & 0.02 &~\\~&
 62 & 155 & 0.14 & 0.02 &~&~&
 62 & 156 & 0.09 & 0.01 &~\\~&
 62 & 157 & 0.05 & 0.01 &~&~&
 62 & 158 & 0.02 & 0.00 &~\\~&
 62 & 159 & 0.01 & 0.00 &~&~&
 63 & 149 & 0.03 & 0.00 &~\\~&
 63 & 150 & 0.05 & 0.01 &~&~&
 63 & 151 & 0.07 & 0.01 &~\\~&
 63 & 152 & 0.12 & 0.01 &~&~&
 63 & 153 & 0.15 & 0.02 &~\\~&
 63 & 154 & 0.18 & 0.02 &~&~&
 63 & 155 & 0.15 & 0.02 &~\\~&
 63 & 156 & 0.11 & 0.01 &~&~&
 63 & 157 & 0.08 & 0.01 &~\\~&
 63 & 158 & 0.05 & 0.01 &~&~&
 63 & 159 & 0.02 & 0.00 &~\\~&
 63 & 160 & 0.01 & 0.00 &~&~&
 64 & 151 & 0.01 & 0.00 &~\\~&
 64 & 152 & 0.02 & 0.00 &~&~&
 64 & 153 & 0.03 & 0.01 &~\\~&
 64 & 154 & 0.04 & 0.01 &~&~&
 64 & 155 & 0.04 & 0.01 &~\\~&
 64 & 156 & 0.03 & 0.00 &~&~&
 64 & 157 & 0.03 & 0.00 &~\\~&
 64 & 158 & 0.02 & 0.00 &~&~&
 64 & 159 & 0.01 & 0.00 &~\\
 \hline
\caption{Relative isotopic yields of $^{250}$Cf, normalized to 200, and the associated statistical errors. A systematic error of 10\% needs to be considered as well.}
\end{longtable}

%% file: latex_yields_papier_Pu.tex
\begin{longtable}{| c c c c c c | c c c c c c |}
\multicolumn{12}{l}
{Isotopic Yields of $^{240,241}$Pu}\\
\hline
~ &~ &~ &~ &~ &~ &~ &~ &~ &~ &~ &~\\
~ & Z & A & Y(Z,A) & $\epsilon_{stat}$ & ~ & ~ & Z & A & Y(Z,A) & $\epsilon_{stat}$ & ~ \\
\hline
\endhead
~& 34 & 82 & 0.04 & 0.02 &~&~&
 34 & 83 & 0.16 & 0.04 &~\\~&
 34 & 84 & 0.25 & 0.07 &~&~&
 34 & 85 & 0.36 & 0.08 &~\\~&
 34 & 86 & 0.33 & 0.08 &~&~&
 34 & 87 & 0.33 & 0.08 &~\\~&
 34 & 88 & 0.05 & 0.03 &~&~&
 35 & 85 & 0.20 & 0.05 &~\\~&
 35 & 86 & 0.44 & 0.10 &~&~&
 35 & 87 & 0.39 & 0.08 &~\\~&
 35 & 88 & 0.45 & 0.09 &~&~&
 35 & 89 & 0.45 & 0.09 &~\\~&
 35 & 90 & 0.24 & 0.07 &~&~&
 35 & 91 & 0.09 & 0.03 &~\\~&
 35 & 92 & 0.05 & 0.02 &~&~&
 36 & 85 & 0.07 & 0.03 &~\\~&
 36 & 86 & 0.17 & 0.06 &~&~&
 36 & 87 & 0.32 & 0.07 &~\\~&
 36 & 88 & 0.50 & 0.10 &~&~&
 36 & 89 & 0.95 & 0.17 &~\\~&
 36 & 90 & 1.00 & 0.17 &~&~&
 36 & 91 & 0.61 & 0.11 &~\\~&
 36 & 92 & 0.47 & 0.09 &~&~&
 36 & 93 & 0.19 & 0.05 &~\\~&
 36 & 94 & 0.07 & 0.03 &~&~&
 37 & 87 & 0.05 & 0.03 &~\\~&
 37 & 88 & 0.13 & 0.04 &~&~&
 37 & 89 & 0.44 & 0.10 &~\\~&
 37 & 90 & 0.66 & 0.12 &~&~&
 37 & 91 & 1.27 & 0.20 &~\\~&
 37 & 92 & 1.26 & 0.20 &~&~&
 37 & 93 & 1.22 & 0.19 &~\\~&
 37 & 94 & 0.84 & 0.14 &~&~&
 37 & 95 & 0.43 & 0.08 &~\\~&
 37 & 96 & 0.25 & 0.05 &~&~&
 37 & 97 & 0.08 & 0.02 &~\\~&
 38 & 90 & 0.14 & 0.04 &~&~&
 38 & 91 & 0.29 & 0.07 &~\\~&
 38 & 92 & 0.94 & 0.16 &~&~&
 38 & 93 & 1.61 & 0.24 &~\\~&
 38 & 94 & 2.35 & 0.33 &~&~&
 38 & 95 & 2.04 & 0.29 &~\\~&
 38 & 96 & 1.69 & 0.25 &~&~&
 38 & 97 & 1.04 & 0.17 &~\\~&
 38 & 98 & 0.51 & 0.09 &~&~&
 38 & 99 & 0.37 & 0.08 &~\\~&
 38 & 100 & 0.23 & 0.06 &~&~&
 38 & 101 & 0.08 & 0.04 &~\\~&
 39 & 91 & 0.03 & 0.02 &~&~&
 39 & 92 & 0.08 & 0.03 &~\\~&
 39 & 93 & 0.27 & 0.07 &~&~&
 39 & 94 & 0.71 & 0.13 &~\\~&
 39 & 95 & 1.41 & 0.22 &~&~&
 39 & 96 & 1.98 & 0.28 &~\\~&
 39 & 97 & 2.30 & 0.32 &~&~&
 39 & 98 & 2.13 & 0.30 &~\\~&
 39 & 99 & 1.73 & 0.26 &~&~&
 39 & 100 & 0.89 & 0.15 &~\\~&
 39 & 101 & 0.50 & 0.09 &~&~&
 39 & 102 & 0.22 & 0.05 &~\\~&
 40 & 94 & 0.09 & 0.03 &~&~&
 40 & 95 & 0.16 & 0.05 &~\\~&
 40 & 96 & 0.48 & 0.10 &~&~&
 40 & 97 & 1.14 & 0.19 &~\\~&
 40 & 98 & 2.37 & 0.34 &~&~&
 40 & 99 & 2.78 & 0.38 &~\\~&
 40 & 100 & 3.50 & 0.46 &~&~&
 40 & 101 & 2.27 & 0.32 &~\\~&
 40 & 102 & 1.58 & 0.24 &~&~&
 40 & 103 & 0.89 & 0.15 &~\\~&
 40 & 104 & 0.50 & 0.11 &~&~&
 40 & 105 & 0.22 & 0.06 &~\\~&
 40 & 106 & 0.02 & 0.01 &~&~&
 41 & 95 & 0.07 & 0.04 &~\\~&
 41 & 96 & 0.12 & 0.04 &~&~&
 41 & 97 & 0.18 & 0.05 &~\\~&
 41 & 98 & 0.34 & 0.08 &~&~&
 41 & 99 & 0.88 & 0.16 &~\\~&
 41 & 100 & 1.53 & 0.24 &~&~&
 41 & 101 & 2.56 & 0.36 &~\\~&
 41 & 102 & 2.74 & 0.37 &~&~&
 41 & 103 & 2.79 & 0.38 &~\\~&
 41 & 104 & 1.65 & 0.24 &~&~&
 41 & 105 & 0.96 & 0.16 &~\\~&
 41 & 106 & 0.51 & 0.10 &~&~&
 41 & 107 & 0.15 & 0.04 &~\\~&
 41 & 108 & 0.10 & 0.03 &~&~&
 41 & 109 & 0.05 & 0.03 &~\\~&
 42 & 98 & 0.03 & 0.02 &~&~&
 42 & 99 & 0.14 & 0.06 &~\\~&
 42 & 100 & 0.32 & 0.08 &~&~&
 42 & 101 & 0.68 & 0.13 &~\\~&
 42 & 102 & 1.60 & 0.25 &~&~&
 42 & 103 & 2.23 & 0.32 &~\\~&
 42 & 104 & 3.06 & 0.41 &~&~&
 42 & 105 & 2.85 & 0.39 &~\\~&
 42 & 106 & 1.92 & 0.27 &~&~&
 42 & 107 & 0.93 & 0.15 &~\\~&
 42 & 108 & 0.43 & 0.08 &~&~&
 42 & 109 & 0.27 & 0.06 &~\\~&
 42 & 110 & 0.20 & 0.06 &~&~&
 42 & 111 & 0.06 & 0.03 &~\\~&
 43 & 101 & 0.09 & 0.03 &~&~&
 43 & 102 & 0.22 & 0.05 &~\\~&
 43 & 103 & 0.31 & 0.08 &~&~&
 43 & 104 & 0.88 & 0.16 &~\\~&
 43 & 105 & 1.67 & 0.26 &~&~&
 43 & 106 & 1.67 & 0.25 &~\\~&
 43 & 107 & 1.99 & 0.29 &~&~&
 43 & 108 & 1.21 & 0.19 &~\\~&
 43 & 109 & 0.65 & 0.11 &~&~&
 43 & 110 & 0.28 & 0.06 &~\\~&
 43 & 111 & 0.12 & 0.03 &~&~&
 43 & 112 & 0.04 & 0.01 &~\\~&
 44 & 104 & 0.05 & 0.02 &~&~&
 44 & 105 & 0.24 & 0.07 &~\\~&
 44 & 106 & 0.45 & 0.09 &~&~&
 44 & 107 & 0.84 & 0.15 &~\\~&
 44 & 108 & 0.96 & 0.16 &~&~&
 44 & 109 & 0.99 & 0.18 &~\\~&
 44 & 110 & 0.67 & 0.12 &~&~&
 44 & 111 & 0.38 & 0.08 &~\\~&
 44 & 112 & 0.23 & 0.06 &~&~&
 44 & 113 & 0.13 & 0.04 &~\\~&
 44 & 114 & 0.03 & 0.02 &~&~&
 45 & 105 & 0.01 & 0.01 &~\\~&
 45 & 107 & 0.06 & 0.03 &~&~&
 45 & 108 & 0.12 & 0.04 &~\\~&
 45 & 109 & 0.15 & 0.04 &~&~&
 45 & 110 & 0.20 & 0.05 &~\\~&
 45 & 111 & 0.22 & 0.06 &~&~&
 45 & 112 & 0.19 & 0.05 &~\\~&
 45 & 113 & 0.16 & 0.04 &~&~&
 45 & 114 & 0.10 & 0.04 &~\\~&
 46 & 110 & 0.11 & 0.05 &~&~&
 46 & 111 & 0.12 & 0.05 &~\\~&
 46 & 112 & 0.14 & 0.05 &~&~&
 46 & 113 & 0.12 & 0.04 &~\\~&
 46 & 114 & 0.11 & 0.04 &~&~&
 46 & 115 & 0.10 & 0.03 &~\\~&
 46 & 116 & 0.06 & 0.03 &~&~&
 47 & 114 & 0.04 & 0.02 &~\\~&
 47 & 115 & 0.15 & 0.05 &~&~&
 47 & 116 & 0.07 & 0.03 &~\\~&
 47 & 117 & 0.06 & 0.02 &~&~&
 47 & 118 & 0.09 & 0.03 &~\\~&
 47 & 119 & 0.02 & 0.01 &~&~&
 48 & 115 & 0.02 & 0.02 &~\\~&
 48 & 117 & 0.08 & 0.04 &~&~&
 48 & 118 & 0.03 & 0.02 &~\\~&
 48 & 119 & 0.14 & 0.05 &~&~&
 48 & 120 & 0.12 & 0.04 &~\\~&
 48 & 121 & 0.08 & 0.03 &~&~&
 48 & 122 & 0.13 & 0.04 &~\\~&
 49 & 122 & 0.06 & 0.02 &~&~&
 49 & 123 & 0.12 & 0.04 &~\\~&
 49 & 124 & 0.14 & 0.06 &~&~&
 49 & 125 & 0.17 & 0.05 &~\\~&
 49 & 126 & 0.10 & 0.03 &~&~&
 49 & 127 & 0.16 & 0.05 &~\\~&
 49 & 128 & 0.11 & 0.03 &~&~&
 49 & 129 & 0.15 & 0.05 &~\\~&
 49 & 130 & 0.14 & 0.05 &~&~&
 50 & 125 & 0.26 & 0.07 &~\\~&
 50 & 126 & 0.45 & 0.10 &~&~&
 50 & 127 & 0.51 & 0.10 &~\\~&
 50 & 128 & 0.72 & 0.14 &~&~&
 50 & 129 & 0.75 & 0.14 &~\\~&
 50 & 130 & 0.64 & 0.12 &~&~&
 50 & 131 & 0.57 & 0.11 &~\\~&
 50 & 132 & 0.50 & 0.11 &~&~&
 50 & 133 & 0.19 & 0.06 &~\\~&
 50 & 134 & 0.12 & 0.05 &~&~&
 50 & 136 & 0.05 & 0.03 &~\\~&
 51 & 125 & 0.14 & 0.04 &~&~&
 51 & 126 & 0.27 & 0.07 &~\\~&
 51 & 127 & 0.28 & 0.06 &~&~&
 51 & 128 & 0.51 & 0.10 &~\\~&
 51 & 129 & 1.13 & 0.19 &~&~&
 51 & 130 & 1.16 & 0.19 &~\\~&
 51 & 131 & 1.43 & 0.22 &~&~&
 51 & 132 & 1.22 & 0.20 &~\\~&
 51 & 133 & 1.11 & 0.19 &~&~&
 51 & 134 & 0.50 & 0.10 &~\\~&
 51 & 135 & 0.27 & 0.07 &~&~&
 51 & 136 & 0.10 & 0.05 &~\\~&
 51 & 137 & 0.05 & 0.05 &~&~&
 52 & 129 & 0.48 & 0.10 &~\\~&
 52 & 130 & 1.19 & 0.21 &~&~&
 52 & 131 & 1.79 & 0.28 &~\\~&
 52 & 132 & 2.54 & 0.37 &~&~&
 52 & 133 & 2.84 & 0.40 &~\\~&
 52 & 134 & 2.22 & 0.32 &~&~&
 52 & 135 & 1.21 & 0.20 &~\\~&
 52 & 136 & 0.61 & 0.12 &~&~&
 52 & 137 & 0.50 & 0.12 &~\\~&
 52 & 138 & 0.11 & 0.04 &~&~&
 52 & 139 & 0.03 & 0.03 &~\\~&
 53 & 128 & 0.10 & 0.06 &~&~&
 53 & 129 & 0.20 & 0.06 &~\\~&
 53 & 130 & 0.33 & 0.08 &~&~&
 53 & 131 & 1.10 & 0.21 &~\\~&
 53 & 132 & 1.67 & 0.26 &~&~&
 53 & 133 & 2.21 & 0.33 &~\\~&
 53 & 134 & 3.03 & 0.43 &~&~&
 53 & 135 & 3.01 & 0.42 &~\\~&
 53 & 136 & 2.34 & 0.36 &~&~&
 53 & 137 & 1.69 & 0.27 &~\\~&
 53 & 138 & 1.11 & 0.20 &~&~&
 53 & 139 & 0.50 & 0.11 &~\\~&
 53 & 140 & 0.15 & 0.05 &~&~&
 54 & 130 & 0.14 & 0.05 &~\\~&
 54 & 131 & 0.25 & 0.09 &~&~&
 54 & 132 & 0.51 & 0.15 &~\\~&
 54 & 133 & 0.82 & 0.16 &~&~&
 54 & 134 & 1.11 & 0.19 &~\\~&
 54 & 135 & 1.96 & 0.32 &~&~&
 54 & 136 & 2.69 & 0.40 &~\\~&
 54 & 137 & 2.85 & 0.41 &~&~&
 54 & 138 & 2.36 & 0.36 &~\\~&
 54 & 139 & 1.48 & 0.23 &~&~&
 54 & 140 & 1.08 & 0.19 &~\\~&
 54 & 141 & 0.44 & 0.10 &~&~&
 54 & 142 & 0.26 & 0.10 &~\\~&
 54 & 143 & 0.11 & 0.04 &~&~&
 54 & 144 & 0.02 & 0.02 &~\\~&
 55 & 133 & 0.11 & 0.04 &~&~&
 55 & 134 & 0.25 & 0.07 &~\\~&
 55 & 135 & 0.44 & 0.10 &~&~&
 55 & 136 & 0.56 & 0.12 &~\\~&
 55 & 137 & 1.57 & 0.27 &~&~&
 55 & 138 & 1.65 & 0.27 &~\\~&
 55 & 139 & 1.75 & 0.28 &~&~&
 55 & 140 & 1.67 & 0.28 &~\\~&
 55 & 141 & 1.94 & 0.31 &~&~&
 55 & 142 & 1.21 & 0.22 &~\\~&
 55 & 143 & 0.70 & 0.16 &~&~&
 55 & 144 & 0.48 & 0.13 &~\\~&
 55 & 145 & 0.15 & 0.06 &~&~&
 56 & 134 & 0.03 & 0.02 &~\\~&
 56 & 135 & 0.08 & 0.03 &~&~&
 56 & 136 & 0.25 & 0.09 &~\\~&
 56 & 137 & 0.43 & 0.11 &~&~&
 56 & 138 & 0.85 & 0.20 &~\\~&
 56 & 139 & 1.07 & 0.21 &~&~&
 56 & 140 & 1.51 & 0.27 &~\\~&
 56 & 141 & 1.16 & 0.20 &~&~&
 56 & 142 & 1.86 & 0.33 &~\\~&
 56 & 143 & 1.33 & 0.24 &~&~&
 56 & 144 & 0.70 & 0.14 &~\\~&
 56 & 145 & 0.34 & 0.09 &~&~&
 57 & 138 & 0.14 & 0.07 &~\\~&
 57 & 139 & 0.24 & 0.09 &~&~&
 57 & 140 & 0.47 & 0.13 &~\\~&
 57 & 141 & 0.52 & 0.13 &~&~&
 57 & 142 & 0.58 & 0.12 &~\\~&
 57 & 143 & 1.06 & 0.19 &~&~&
 57 & 144 & 1.52 & 0.29 &~\\~&
 57 & 145 & 1.12 & 0.22 &~&~&
 57 & 146 & 0.60 & 0.13 &~\\~&
 57 & 147 & 0.45 & 0.11 &~&~&
 57 & 148 & 0.17 & 0.06 &~\\~&
 57 & 149 & 0.05 & 0.02 &~&~&
 58 & 142 & 0.15 & 0.06 &~\\~&
 58 & 143 & 0.28 & 0.12 &~&~&
 58 & 144 & 0.72 & 0.19 &~\\~&
 58 & 145 & 0.41 & 0.10 &~&~&
 58 & 146 & 0.64 & 0.16 &~\\~&
 58 & 147 & 0.36 & 0.09 &~&~&
 58 & 148 & 0.54 & 0.16 &~\\~&
 58 & 149 & 0.20 & 0.08 &~&~&
 58 & 150 & 0.45 & 0.15 &~\\~&
 58 & 151 & 0.18 & 0.09 &~&~&
 59 & 145 & 0.20 & 0.12 &~\\~&
 59 & 146 & 0.21 & 0.09 &~&~&
 59 & 147 & 0.46 & 0.14 &~\\~&
 59 & 148 & 0.36 & 0.11 &~&~&
 59 & 149 & 0.31 & 0.10 &~\\~&
 59 & 150 & 0.14 & 0.05 &~&~&
 59 & 151 & 0.22 & 0.10 &~\\~&
 59 & 152 & 0.23 & 0.13 &~&~&
 60 & 148 & 0.12 & 0.06 &~\\~&
 60 & 149 & 0.21 & 0.09 &~&~&
 60 & 150 & 0.30 & 0.10 &~\\~&
 60 & 151 & 0.30 & 0.10 &~&~&
 60 & 152 & 0.19 & 0.08 &~\\~&
 60 & 153 & 0.22 & 0.10 &~&~&
 60 & 154 & 0.03 & 0.02 &~\\
 \hline
 \caption{Relative isotopic yields of $^{240}$Pu, normalized to 200, and the associated statistical errors. A systematic error of 10\% needs to be considered as well.}
\end{longtable}